\documentclass[a4paper, 11pt]{article}
\usepackage[left=2cm,right=2cm,top=2cm,bottom=2cm]{geometry}
\usepackage{lineno,hyperref}
\usepackage{amsmath}
\usepackage{amssymb}
\usepackage{dsfont}
\usepackage{nicefrac}
\usepackage{here}
\usepackage{multirow}
\usepackage{diagbox}
\usepackage{amssymb}
\usepackage{caption}
\usepackage{subcaption}
\usepackage{color}
\usepackage{titlesec}
\usepackage{appendix}
\usepackage{tikz}
\usepackage{xcolor}

\newcommand\GP[1]{{\color{black}{#1}}}
\newcommand\MM[1]{{\color{black}{#1}}}
\newcommand\LM[1]{{\color{black}{#1}}}

\newcommand*\dif{\mathop{}\!\mathrm{d}}

\begin{document}

\title{Spatial coupling of an explicit temporal adaptive integration scheme with an implicit time integration scheme}
\author{Laurent Muscat, Guillaume Puigt, Marc Montagnac, Pierre Brenner}
\maketitle


\begin{abstract}

The Reynolds-Averaged Navier-Stokes equations and the Large-Eddy Simulation equations
can be coupled using a transition function to switch from a set of equations applied
in some areas of a domain to the other set in the other part of the domain.
Following this idea, different time integration schemes can be coupled. In this context, 
we developed a hybrid time integration scheme
that spatially couples the explicit scheme of Heun and the implicit scheme of
Crank and Nicolson using a dedicated transition function.
This scheme is linearly stable and second-order accurate.
In this paper, an extension of this hybrid scheme is introduced to deal with a
temporal adaptive procedure.
The idea is to treat the time integration procedure with unstructured grids as
it is performed with Cartesian grids with local mesh refinement.
Depending on its characteristic size,
each mesh cell is assigned a rank. And for two cells from two consecutive ranks, the ratio of the
associated time steps for time marching the solutions is $2$. As a consequence,
the cells with the lowest rank iterate more than the other ones to reach the same physical time.
In a finite-volume context, a key ingredient
is to keep the conservation property for the interfaces that separate two cells of different ranks.
After introducing the different schemes, the paper recalls briefly the coupling procedure, and
details the extension to the temporal adaptive procedure.
The new time integration scheme is validated with the propagation of 1D wave packet,
the Sod's tube, and the transport of a bi-dimensional vortex in an uniform flow.
\end{abstract}



\section{Introduction \label{sec:intro}}

The Reynolds-Averaged Navier-Stokes (RANS)
equations account for the mean effects of the turbulence on the main conservative quantities.
Of course, the RANS method is unable to represent unsteady effects of the turbulence on
the flow. But this method has a relatively low CPU cost, and it is accurate enough for the
computation of boundary layers for example. For this reason of low CPU cost, it is today
one of the preferred technique for use in an industrial context.
RANS equations are generally time-marched using an
implicit time formulation for fast convergence to the steady-state solution.
For unsteady RANS equations, the implicit time integration enables large time steps.
Implicit time integration requires to solve large linear system of equations,
and then tends to be expensive in terms of CPU cost.

Large Eddy Simulation (LES) solves some of the turbulence effects,
and is increasingly being considered in an
industrial context. The principle of LES is to introduce a model for the
smallest turbulent scales that have a universal nature, and to capture the largest turbulence scales.
In practice, the separation between the tractable and the modeled spectra is defined by the
mesh spacing and the numerical scheme. Then, LES equations are generally integrated using explicit schemes
that exhibit good spectral properties of dissipation and dispersion and can attain any order of accuracy. However,
the associated time steps are limited by the Courant-Friedrichs-Lewy (CFL) number.


In~\cite{Muscat_JCP_XX_2018} was introduced the AION time integration scheme
that spatially couples the explicit Heun's scheme and the implicit Crank-Nicolson' scheme.
The principle of the AION scheme is to blend both schemes using a transition function $\omega$.
Grid cells can be declared as explicit, implicit, or hybrid depending on their values of $\omega$, and
are associated with their own time integration scheme.
Explicit and implicit cells are respectively handled by the Heun's and Crank-Nicolson' schemes, while
a hybrid scheme is in charge of the other grid cells. An important property of the transition
function is to allow a quick switch between the explicit and implicit cells,
keeping the number of hybrid cells as low as possible.
In addition, attention was paid on the stability analysis to avoid wave amplification for
any wavenumber~\cite{Muscat_JCP_XX_2018}. This paper deals with the adaptation of the AION scheme to a
temporal adaptive procedure.

The Adaptive Mesh Refinement (AMR) method allows more grid cells to be placed
in regions of interest to better capture the flow physics while keeping a small number
of cells in zones of low interest.
Many efforts were dedicated to the spatial adaptation, especially for Cartesian grids.
For unsteady simulations, it can be of great interest to couple the spatial adaptive method with
a temporal adaptive approach. Indeed, a standard unsteady explicit computation is constrained by
the CFL condition over the whole computational domain, and the maximum stable time step is generally
associated with the smallest cells. It follows that the time integration of the largest cells is
performed by a fraction of the maximal local allowable time step, and this leads to useless computations.
The principle of temporal adaptive schemes is to allow the mesh cells to integrate
the solution using their own time step according to the local CFL condition. This type of
local time-stepping method is included in
FLUSEPA\footnotemark[1] solver~\cite{Brenner_AFOSR_1996}. This solver, developed by ArianeGroup, is used
in the certification process of launchers, when complex physical
phenomena occur.

Concerning AMR, the main difficulty lies in the treatment of flux at the interface between cells of
different grid sizes. Sanders and Osher~\cite{Osher_MC_1983} proposed a one-dimensional
local time-stepping algorithm and
later on, Dawson~\cite{Dawson_1995_NMPD} extended the procedure to multi-dimensional problems. The procedure
is built in order to keep a space time conservation: for an interface between large and small cells, the sum of
fluxes on small cells is exactly the one on the larger cell. However, the procedure leads to a first-order time scheme.
%
An extension to the second order of accuracy was proposed more recently. Dawson and Kirby~\cite{Dawson_2001_SIAM}
extended the procedure by means of Total Variation Diminishing (TVD) Runge-Kutta schemes.
Constantinescu and Sandu~\cite{Constantinescu_2007_JSC} developed a set of multirate
Runge-Kutta integrators called Partitioned Runge-Kutta. They are adapted to automatic mesh refinement, and respect
strong mathematical properties (Strong Stability Preserving). Later on, the same authors extended the procedure
to the multirate explicit Adams scheme~\cite{Sandu_2009_JSP}.

\begin{figure}[!htbp]
\begin{center}
\includegraphics [width=10cm]{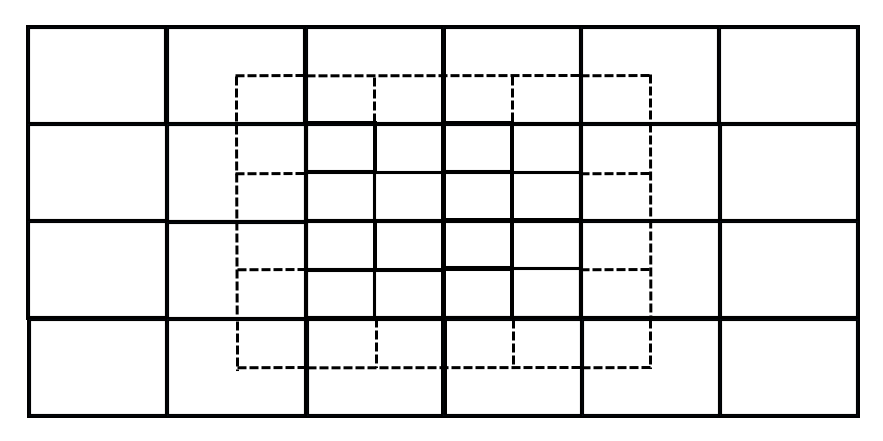}
\caption{Two grids of different level, dotted cells correspond to ghost cells for the finest grid.
\label{fig:grids}}
\end{center}
\end{figure}

\begin{figure}[!ht]
\begin{center}
\includegraphics [width=12cm]{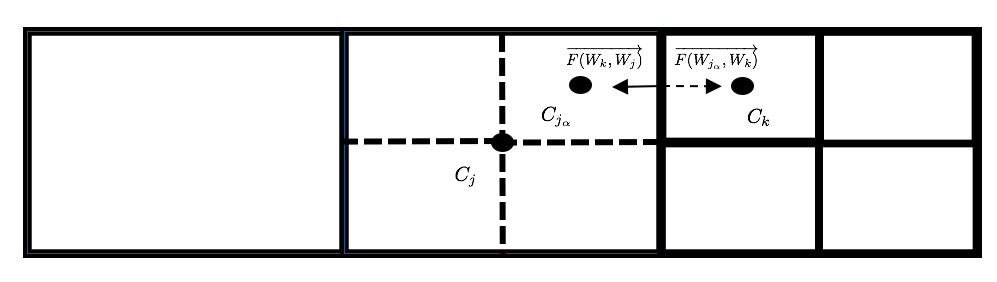}
\caption{Example of inter-level flux conservation issue.
\label{fig:flux_level}}
\end{center}
\end{figure}

Berger and Oliger~\cite{Berger_JCP_1984} and Berger and Colella~\cite{Berger_JCP_1989} proposed a different approach
for playing with both space and time refinements.
Starting from the initial grid level indexed
$l=0$, several grid levels $l=1$,..., $l=l_{max}$ are introduced by local refinement (see Fig.~\ref{fig:grids}).
For transferring information from coarse to fine grids, ghost cells are introduced at the level $l$, and the values inside them are interpolated
from the coarse grid at level $l-1$. It must be highlighted that these ghost cells can be found on boundary conditions or
inside the computational domain, and serve as local boundary conditions for the grid level.
The interest of such method lies in its simple implementation. 
The procedure consists of time integrating the mesh by means of the time classes, starting from a coarse grid ($l=0$) and
finishing by the most refined level ($l=l_{max}$). On any grid level, the following steps are considered:
\begin{itemize}
\item{time integrate cells of level $l$ with local time step ($\Delta t_l$). If $l>0$, the boundary information is collected from the next coarser level
$l-1$ by interpolation of the flux.}
\item{synchronise cells of levels $l$ and $l+1$, and interpolate correction to finest cells (of level $[l+2,...,l_{max}]$).}
\end{itemize}
Hence the adaptive scheme allows the time integrations of the different level of grids by means of their own time steps, and
more time integration steps are needed on the refined levels than on the coarser one. The synchronization step must be seen as a
correction that makes the procedure consistent. Nevertheless, the procedure must be defined carefully since
synchronization and interpolation drive order of accuracy and conservation.
For instance, let's take the example introduced in Fig.~\ref{fig:flux_level}.
The principle of ``internal'' ghost-cell $C_{j_{\alpha}}$ is to serve as an additional variable. The ghost cell value and
the gradient can be interpolated from cell-centered values in the surrounding cells. However, this can lead to a non
conservative procedure since the interface state computed from the ghost cells (using state and gradients for instance)
and the one from the large cell $C_j$ may be different.

Several methods were introduced to limit the impact of loss of accuracy. Bell {\it et al.}~\cite{Bell_1987_CFDC}
introduced a correction on the state $W$ thanks to a passive scalar, which allows a fast convergence for quasi-steady state.
But they noticed that the conservation of flux is not guaranteed.
Bell~\cite{Bell_2005_Springer} also applied a correction seen as a kind of "fixed" Dirichlet boundary condition but
involving an inconsistency at interfaces between grids.

The temporal adaptive scheme in FLUSEPA follows some basic principles introduced by Kleb {\it et al.}~\cite{KLEB_1992_AIAA}.
Their procedure is based on the explicit Euler time integration, and leads to a first-order time-accurate solution.
Obtaining a second-order time integration is a prerequisite, and the approach shares some ideas with Krivodonova's work~\cite{Krivodonova_2010_APP},
even if the procedure differs in the way the time integration is performed on classes holding the smallest cells. The first
specific point is the use of the second order explicit scheme proposed by Heun. The second point of importance is the extension to the adaptive
time treatment for Heun's scheme. The last key ingredient is the coupling of the adaptive time integration with the hybrid AION
scheme presented in~\cite{Muscat_JCP_XX_2018}. The remainder of this paper unfolds as follows.
In Sec.~\ref{sec:1b_Equations}, the standard form of the Navier-Stokes
is recalled and the explicit, implicit and hybrid schemes chosen for the present study are introduced.
Our explicit, implicit and hybrid basic schemes (Heun's, Crank-Nicolson's and AION schemes) are defined in Sec.~\ref{sec:BasicTI}.
In Sec.~\ref{sec:TA_Heun}, the temporal adaptive method implemented in FLUSEPA is introduced.
It must be underlined that the adaptive version
of the Heun's scheme was briefly described in~\cite{Brenner_AIAA_1993}, and the full
description of the scheme implemented in FLUSEPA is also
one objective of the paper.
Mathematical analysis including order of accuracy, stability and spectral behaviour is then presented.
In Sec.~\ref{sec:TA_AION}, the procedure is extended to our coupled explicit / implicit time integrator called AION.
Before concluding, Sec.~\ref{sec:Validation} is dedicated to the validation of
the adaptive AION scheme for 1D and 2D test cases with temporal adaptive approach.

\section{Discretization of the Navier-Stokes Equations}\label{sec:1b_Equations}

The Navier-Stokes system of equations is written in the following compact conservation form,
\begin{equation}
\frac{\partial W}{\partial t} + \nabla \cdot F(W) + \nabla \cdot G(W,\nabla W)  = 0,
\label{eq:euler}
\end{equation}
with $W$ the vector of conservative variables, its gradient $\nabla W$,
the convection flux $F$ and the diffusion flux $G$.
The $N$ non-overlapping rigid stationary cells $\Omega_j$ map the computational domain $\Omega$, and
Eq.~\eqref{eq:euler} is integrated over every mesh cell.
Applying the Gauss relation that ties the volume integrals of the divergence terms to the interface fluxes gives the weak form,
\begin{equation}
\frac{\dif}{\dif t} \iiint_{\Omega_{j}}  {W}  \dif \Omega = -\iint_{A_{j}} F(W) \cdot \vec{n} \dif S   -\iint_{A_{j}}   {G(W, \nabla W)} \cdot \vec{n} \dif S,
\label{eq:weakform}
\end{equation}
where $\Omega_{j}$ is the $j$-th control volume with its border $A_{j}$, and $\vec{n}$ is the outgoing unit normal
vector.
The averaged conservative variables are defined as
\begin{equation}
\overline{W}_j= \frac{1}{|{\Omega_j}|} \iiint_{\Omega_j} W \dif \Omega.
\label{eq:averaged}
\end{equation}
This relation~\eqref{eq:averaged} allows to rewrite the conservation laws~\eqref{eq:weakform} discretized with a finite-volume formulation in the following
differential form,
\begin{equation}
  \frac{\dif \overline{W}_j}{\dif t}=R(\overline{W}_j),\label{eq:compact_form}
\end{equation}
where $R(\overline{W}_j)$ is the residual computed using the averaged quantities $\overline{W}_j$ in cell $j$.
For any cell, the residual is the sum of the flux over the whole
boundary of a cell.

In this paper, convection and diffusion fluxes are discretized by means of a $k$-exact
formulation coupled with successive corrections~\cite{Pont_JCP_2017,Haider_2014_SpringerBook}.
The principle of the formulation is to define a local polynomial approximation of the unknowns.
The successive corrections are designed to avoid geometrical reconstructions for parallel computations.
Starting from the solution, the Taylor expansion of the unknowns defines the local polynomial reconstruction,
and the order of accuracy is defined by the error term in the Taylor approximation.
The process for defining the coefficients of the Taylor expansion can be found in~\cite{Pont_JCP_2017,Haider_2014_SpringerBook}.

Finally, the finite-volume formulation of Eq.~\eqref{eq:compact_form} can be expressed as the following Cauchy problem,
\begin{equation}
\left\{
\begin{aligned}
&\frac{\dif \overline{W}_j(t)}{\dif t} = R\big({\overline{W}}_j(t)\big), \quad \forall t \in R^+,\\
&\overline{W}_j(0)  =W_j^{0},  \quad \forall j \in\{1,...,N\},
\end{aligned}
\right .
\label{Cauchy1}
\end{equation}
where ${W_j^{0}}_{1\le j\le N}$ denotes the initial solution in the mesh cells.
For the sake of clarity, the averaging symbol will be dropped,
and ${W}$ will represent the averaged quantities over the control volumes.

\section{Heun, Crank-Nicolson, and AION methods for time integration\label{sec:BasicTI}}

Heun's, {Crank-Nicolson's} and AION schemes, which are considered as standard time integration schemes in
this study, are presented below.

\subsection{Heun's scheme}

Heun's explicit scheme~\cite{Heun_ZMathPhys_1900} is a second-order accurate predictor-corrector method.
The state $W^{n+1}$ is first predicted with a forward Euler scheme, and then corrected by a standard trapezoidal rule,
\begin{equation}
  \left\{
  \begin{array}{ll}
  \hbox{Predictor {stage}:} & \displaystyle \widehat{W}= W^n + \Delta t R(W^n), \\
  \vspace*{-3mm} &  \\
  \hbox{Corrector {stage}:} & \displaystyle W^{n+1}=W^n+\frac{\Delta t}{2} \big( R(W^n) + R(\widehat{W}) \big). \\
  \end{array}
  \right.
  \label{eq:HEUN}
\end{equation}

\subsection{Crank-Nicolson's scheme (IRK2)}

Crank-Nicolson's implicit scheme~\cite{Crank_1947_PCP} is a
second-order accurate method defined as
\begin{equation}
\begin{aligned}
W^{n+1}=W^{n}+\frac{\Delta t}{2} \big( R(W^n) + R(W^{n+1}) \big).
\label{IRK2_f}
\end{aligned}
\end{equation}
Eq.~\eqref{IRK2_f} is solved with an iterative Newton method since the residual $R$ is nonlinear in $W$.
{This scheme belongs to the class of second-order Implicit Runge-Kutta schemes, and
in the following, it will also be denoted IRK2 for conciseness.}

\subsection{The AION scheme}\label{sec:aion}

The AION scheme couples Heun's and Crank-Nicolson's schemes by means of a hybrid scheme
that enables a smooth transition between the latter two schemes. The AION scheme
is defined to ensure a unique flux on each interface, leading to a conservative formulation.
The full scheme description is available in~\cite{Muscat_JCP_XX_2018} and the key points are recalled below
for the sake of clarity.

The cell status $\omega_j$ is used to smoothly switch from a time-explicit ($\omega_j = 1$)
to a time-hybrid ($0.6<\omega_j < 1$) and then to a time-implicit ($\omega_j \leq 0.6$) scheme.
The cell status is named Heun for $\omega_j=1$, IRK2 for $\omega_j \leq 0.6$, and hybrid otherwise.

\begin{figure}[!htbp]
    \begin{center}
    \includegraphics [width=7cm]{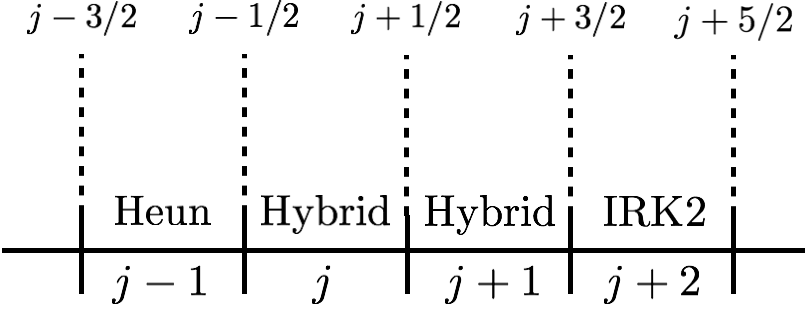}
    \caption{1D example to explain the flux conservation property of the AOIN scheme}
    \label{fig:flux_conserv}
    \end{center}
\end{figure}

The flux formula used for an interface separating two cells depends on the status of these cells as shown in Tab.~\ref{tab:flux-faces}.
The flux formula type is determined using the cell status provided on Fig.~\ref{fig:flux_conserv}.

\begin{table}[!htbp]
\caption{Flux formula depending on neighbour cells. The table is "symmetrical":
the flux between two cells is independent of the direction of information propagation.\label{tab:flux-faces}}
\begin{center}
\begin{tabular}[b]{|l|c|c|c|c|}
\hline
\backslashbox{Left cell status}{Right cell status} & Heun & Hybrid & IRK2 \\
\hline
Heun & $F^{Heun}$ & $F^{Heun}$ & $\times$ \\
\hline
Hybrid & $F^{Heun}$& $F^{Hybrid}$ & $F^{IRK2}$ \\
\hline
IRK2 & $\times$ & $F^{IRK2}$ & $F^{IRK2}$ \\
\hline
\end{tabular}
\end{center}
\end{table}
According to Tab.~\ref{tab:flux-faces}, for the 1D example on Fig.~\ref{fig:flux_conserv},
the AION scheme is expressed as:
\begin{equation}
\left\{
\begin{array}{ll}
\hbox{Predictor stage:} & \displaystyle \widehat{W_j}=W_j^{n}+ \Delta t R(W_j^{n})\\
\vspace*{-3mm} &  \\
\hbox{Corrector stage:} & \displaystyle \left \{
    \begin{array}{ll}
    \displaystyle W^{n+1}_{j-1}=W^n_{j-1}+ \frac{\Delta t}{2}
    \big( F_{j-\frac{1}{2}}^{n}+ \widehat{F}_{j-\frac{1}{2}} - F_{j-\frac{3}{2}}^{n}- \widehat{F}_{j-\frac{3}{2}} \big)  \\
    \vspace*{-3mm} &  \\
    \displaystyle W^{n+1}_j=W^n_j+ \Delta t \big( F_{j+\frac{1}{2}}^{Hybrid} - \frac{1}{2} (F_{j-\frac{1}{2}}^{n}+\widehat{F}_{j-\frac{1}{2}}) \big)  \\
    \vspace*{-3mm} &  \\
    \displaystyle W^{n+1}_{j+1}=W^n_{j+1}+ \Delta t \big( \frac{1}{2}(F_{j+\frac{3}{2}}^{n}+ F_{j+\frac{3}{2}}^{n+1}) -F_{j+\frac{1}{2}}^{Hybrid}     \big)  \\
    \vspace*{-3mm} &  \\
    \displaystyle W^{n+1}_{j+2}=W^n_{j+2}+ \frac{\Delta t}{2} (F_{j+\frac{5}{2}}^{n}+ F_{j+\frac{5}{2}}^{n+1}- F_{j+\frac{3}{2}}^{n}- F_{j+\frac{3}{2}}^{n+1})  .
    \end{array}
\right.
\end{array}
\right.\label{eq:HCS2}
\end{equation}
The reconstructed primitive states (vector $V$) at an interface between cells with indices $j$ and $j+1$, which are used in the computation of the hybrid flux $F^{Hybrid}$,
are defined as
\begin{equation}
    \begin{array}{ll}
V^L=& \omega_j  \displaystyle \bigg[ \frac{V^n_j+\widehat{V}_j}{2}\bigg]+\frac{1}{2}(\widetilde{\nabla V})_j^{n} \cdot \overrightarrow{C_jC_f} +
\bigg(\omega_j-\frac{1}{2}\bigg)(\widetilde{\widehat{\nabla V}})_j \cdot \overrightarrow{C_jC_f} +\\
&(1-\omega_j)  \displaystyle \bigg[ V^{n+1}_j+(\widetilde{\nabla V})_j^{n+1}\cdot \overrightarrow{C_jC_f} - \frac{1-\omega_j}{2}(\widetilde{\Delta_t V})^{n+1}_j \bigg], \\
V^R=& \omega_{j+1}  \displaystyle \bigg[ \frac{V^n_{j+1}+\widehat{V}_{j+1}}{2}\bigg]+\frac{1}{2}(\widetilde{\nabla V})_{j+1}^{n}\cdot \overrightarrow{C_{j+1}C_f}
+ \bigg(\omega_{j+1} - \frac{1}{2}\bigg)(\widetilde{\widehat{\nabla V}})_{j+1} \cdot \overrightarrow{C_{j+1}C_f}+\\
&(1-\omega_{j+1})  \displaystyle \bigg[ V^{n+1}_{j+1}+(\widetilde{\nabla V})_{j+1}^{n+1}\cdot \overrightarrow{C_{j+1}C_f} - \frac{1-\omega_i}{2}(\widetilde{\Delta_t V})^{n+1}_{j+1} \bigg],\\
\end{array}
\label{eq:AION_NHRM}
\end{equation}
where $C_j$, $C_{j+1}$ and $C_f$ represent the centers of cell $j$ and $j+1$ and the interface center, respectively.

{\bf Remark:} A situation not adressed in Tab.~\ref{tab:flux-faces} nor in Fig.~\ref{fig:flux_conserv} must be described. 
For any hybrid cell (according to the value $\omega_j$) with hybrid flux for all the faces, the predictor state $\widehat{W}_j$ is 
time-integrated by accounting for $\omega_j$: $\widehat{W}_j=W_j^{n}+\omega_j{\Delta t} R(W_j^{n})$.

The cell status $\omega_j$ is locally adapted to the flow and the stability constraint. First,
the global time step is chosen to be the maximum allowable time step on a part $\cal{D}$
of the whole computational domain without source of stiff terms,
\begin{equation}
\Delta t = \min_{j \in \cal{D}} \hat{\nu} \frac{h_j}{\| \vec{v_j}\| + c_j},
\end{equation}
where $\hat{\nu} < 1$, $h_j$ is a reference length scale, $\vec{v_j}$ the velocity vector,
 and $c_j$ the speed of sound in
cell $j$.
Then, the parameter $\omega_j$ is defined as
\begin{equation}
\begin{aligned}
\omega_j = \min\big( 1, \frac{1}{\nu_j} \big),\hbox{ with }\nu_j=\frac{(\| \vec{v_j}\| + c_j)\,\Delta t}{h_j}.
\end{aligned}
\end{equation}

In Eq.~\eqref{eq:AION_NHRM}, the term ${(\widetilde{\Delta_t V})}^{n+1}_{j}$ comes
from a generalization of the TVD property to the coupled space/time reconstruction.
All details regarding
this specific term and the proof of the TVD property are provided in previous paper ~\cite{Muscat_JCP_XX_2018}.
It can be seen as a time slope limiter.
This term is given by a Newton's algorithm which at any step $s$ reads,
\begin{equation}
(\widetilde{\Delta_t V})^{n+1,s}_{j} = \left\{
\begin{array}{ll}
\displaystyle \max\left[0, \min\left(\min_j [\beta (V_j^{n+1,s} - V_j^{n}) +
(\widetilde{\Delta_t V})^{n+1,s-1}_{j}], V_j^{n+1,s} - V_j^{n}\right)\right] & \\
& \hspace*{-3cm}\hbox{ if } V_j^{n+1,s} - V_j^{n} \ge 0,\\
& \vspace*{-0mm} \\
\displaystyle \min\left[0, \max\left(\max_j [\beta (V_j^{n+1,s} - V_j^{n}) +
(\widetilde{\Delta_t V})^{n+1,s-1}_{j}], V_j^{n+1,s} - V_j^{n}\right)\right] & \\
& \hspace*{-3cm}\hbox{ if } V_j^{n+1,s} - V_j^{n} < 0.\\
\end{array}
\right.
\end{equation}
The parameter $\beta$ is computed from the local CFL number $\nu_j$ and $\omega_j$ at each step $s$,
\begin{equation}
\beta = \frac{2 \big(1-\omega_j^s\nu_j^s(2-\omega_j^s \nu_j^s)\big)}{\nu_j^2(1-\omega_i^s)^2}
\label{eq:def_beta}
\end{equation}
This method is second-order accurate in both space and time in fully explicit,
fully implicit and explicit-implicit domains.

The next section is dedicated to the time-adaptive extension of the Heun' scheme.

\section{Temporal Adaptive Heun's scheme\label{sec:TA_Heun}}

The time-adaptive extension of Heun's scheme is fully described below. It starts from the very brief
introduction in~\cite{Brenner_AIAA_1993}.

\subsection{Time class for the mesh cells}

The local time step of each cell $j$ is computed, $\Delta \tau_j = \mbox{CFL} \frac{h_j}{\| \vec{v_j}\| + c_j}$
where $h_j$ is a reference length scale, $\vec{v_j}$ the velocity vector and $c_j$ the speed of sound in
cell $j$. The minimum value of the local time step, $\Delta t_{\min}$, enables to define the
time class of rank $K$ with
\begin{equation}
K=\left\lfloor \frac{\ln\Big(\frac{\Delta \tau_j}{\Delta t_{\min}}\Big)}{\ln(2)} \right\rfloor,
\end{equation}
from which the time step of the class denoted $K$ is deduced,
\begin{equation}
\Delta t_K = 2^K \Delta t_{\min}.\label{eq:timestepclass}
\end{equation}
From Eq.~\eqref{eq:timestepclass}, it is clear that the class of rank $0$ is associated with the time step $\Delta t_{\min}$.
If the solution of the partial differential equation is integrated in time in the ascending order of class rank, the associated time steps are
$2 \Delta t_{\min}$, $4 \Delta t_{\min}$, and so on. In order to attain the same physical time, cells of class $K$ must be
 integrated one more time than cells of class $K+1$.
In practice, it is also required that the difference in class rank between two cells sharing the same interface is not greater than one.
Therefore, two adjacent cells are integrated with the same time step if they are in the same class,
or with different time steps with a ratio of 2. There are other approaches with temporal adaptive methodology,
where each cell is time integrated with its own maximal allowable time step~\cite{UNFER_JCP_2013,TOUMI_CRM_2015,Semiletov_IJA_2014}.

In order to obtain time synchronization, attention is paid on the treatment of the flux between cells that belong to
classes with different ranks since it is the key point to keep a conservative treatment~\cite{Berger_JCP_1984, Berger_JCP_1989}.

\subsection{Update of the Solution\label{sec:TA_method}}

When two cells sharing a common interface belong to two different time classes, the cell with the lower rank
is iterated one more time than cells with the higher rank. It is then necessary to elaborate a specific
computation of the flux on the interface between these two cells to perform time synchronization with conservation,
while maintaining the local scheme accuracy.

\begin{figure}[h]
\begin{center}
\begin{tikzpicture}[scale=4.5]
    \draw [->,black] (-1.05,0.) -- (1,0.);
    \draw (1.1,0) node {$x$} ;
    \draw [black] (-0.8,-0.02) -- (-0.8, 0.02);
    \fill (canvas cs:x=-0.7cm,y=0cm) circle (0.2mm);
    \draw [black] (-0.6,-0.02) -- (-0.6, 0.02);
    \fill (canvas cs:x=-0.5cm,y=0cm) circle (0.2mm);
    \draw [black] (-0.4,-0.02) -- (-0.4, 0.02);
    \fill (canvas cs:x=-0.3cm,y=0cm) circle (0.2mm);
    \draw [black] (-0.2,-0.02) -- (-0.2, 0.02);
    \fill (canvas cs:x=-0.1cm,y=0cm) circle (0.2mm);
    \draw [black] (-0. ,-0.02) -- (-0. , 0.02);
    \fill (canvas cs:x=0.2cm,y=0cm) circle (0.2mm);
    \draw [black] (0.4 ,-0.02) -- (0.4 , 0.02);
    \fill (canvas cs:x=0.6cm,y=0cm) circle (0.2mm);
    \draw [black] (0.8 ,-0.02) -- (0.8 , 0.02);
    \draw[->,black] (-1,-0.05) -- (-1.0,0.5);
    \draw (-1.0,0.6) node {$t$} ;
    \draw [dashed] (-1.04,0.2) -- (0.9,0.2);
    \draw (-1.15,0.2) node {$\Delta t$} ;
    \draw [dashed] (-1.04,0.4) -- (0.9,0.4);
    \draw (-1.15,0.4) node {$2\Delta t$} ;
    \draw [dotted] (0.0,0.7) -- (0.0,-0.3);
    \draw (-0.5,-0.2) node {$\text{cells of class 0}$} ;
    \draw (0.5,-0.2) node {$\text{cells of class 1}$} ;
    \draw (0.2,-0.07) node {$\beta+1$} ;
    \draw (-0.1,-0.07) node {$\beta$} ;
    \draw (-0.3,-0.07) node {$\beta-1$} ;
\end{tikzpicture}
\caption{1D configuration consisting of a mesh composed of two classes. Initial solution is represented by the black circles. The key point will be
the definition of the interface $F_{\beta+1/2}$
\label{fig:TA0}}
\end{center}
\end{figure}
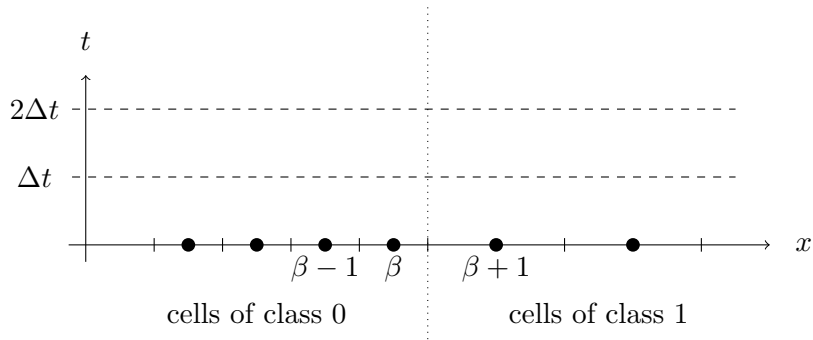

The method is illustrated on a one-dimensional example with cells of class $K=0$ and $K=1$ in Fig.~\ref{fig:TA0}
with space as the abscissa and time as the ordinate. The dashed lines
represent the time at which the solution must be computed starting from the initial solution represented by
\begin{tikzpicture}[scale=4.5]
    \fill (canvas cs:x=-0.0cm,y=0cm) circle (0.2mm);
\end{tikzpicture} symbols.
The first-order time-accurate solutions using the predictor stage of Heun's scheme will be represented by $\widehat{\times}$.

Starting from the initial time $t=0$, the final time reached by all cells is
equal to $2 \Delta t$, which is the time step of the class with rank $K=1$ and twice the time step
of the cells with rank $K=0$.
In the following, $F_{i}^{k \Delta t}$ will represent the flux at the interface $i$ at
the time $k \Delta t$. If $k=0$, the superscript will be simply $0$. In the standard finite volume approximation,
the local residual of a cell of index $j$ is simply $R_j=F_{j+1/2}-F_{j-1/2}$.

Heun's scheme is an explicit second-order time integrator. Since it contains a predictor-corrector procedure~\cite{Heun_ZMathPhys_1900},
cells with class rank $1$  will undergo only two stages to reach $t = 2 \Delta t$:
\begin{itemize}
    \item[a-1. ] $\displaystyle \widehat{W}^{2 \Delta t} = W^0 + 2 \Delta t R(W^0)$
    \item[b-1. ] $\displaystyle W^{2 \Delta t} = W^0 + \frac{2 \Delta t}{2} \big( R(W^0) + R(\widehat{W}^{2 \Delta t}) \big)$,
\end{itemize}
but cells with class rank $0$ will undergo four stages:
\begin{itemize}
    \item[a-0. ] $\displaystyle \widehat{W}^{\Delta t} = W^0 + \Delta t R(W^0)$
    \item[b-0. ] $\displaystyle W^{\Delta t} = W^0 + \frac{\Delta t}{2} \big( R(W^0) + R(\widehat{W}^{\Delta t}) \big)$
    \item[c-0. ] $\displaystyle \widehat{W}^{2 \Delta t} = W^{\Delta t} + \Delta t R(W^{\Delta t})$
    \item[d-0. ] $\displaystyle W^{2 \Delta t} = W^{\Delta t} + \frac{\Delta t}{2} \big( R(W^{\Delta t}) + R(\widehat{W}^{2\Delta t}) \big)$.
\end{itemize}

\begin{figure}[h]
\begin{center}
\begin{tikzpicture}[scale=4.5]
    \draw [->,black] (-1.05,0.) -- (1,0.);
    \draw (1.1,0) node {$x$} ;
    \draw [black] (-0.8,-0.02) -- (-0.8, 0.02);
    \fill (canvas cs:x=-0.7cm,y=0cm) circle (0.2mm);
    \draw [black] (-0.6,-0.02) -- (-0.6, 0.02);
    \fill (canvas cs:x=-0.5cm,y=0cm) circle (0.2mm);
    \draw [black] (-0.4,-0.02) -- (-0.4, 0.02);
    \fill (canvas cs:x=-0.3cm,y=0cm) circle (0.2mm);
    \draw [black] (-0.2,-0.02) -- (-0.2, 0.02);
    \fill (canvas cs:x=-0.1cm,y=0cm) circle (0.2mm);
    \draw [black] (-0. ,-0.02) -- (-0. , 0.02);
    \fill (canvas cs:x=0.2cm,y=0cm) circle (0.2mm);
    \draw [black] (0.4 ,-0.02) -- (0.4 , 0.02);
    \fill (canvas cs:x=0.6cm,y=0cm) circle (0.2mm);
    \draw [black] (0.8 ,-0.02) -- (0.8 , 0.02);
    \draw[->,black] (-1,-0.05) -- (-1.0,0.5);
    \draw (-1.0,0.6) node {$t$} ;
    \draw [dashed] (-1.04,0.2) -- (0.9,0.2);
    \draw (-1.15,0.2) node {$\Delta t$} ;
    \draw [dashed] (-1.04,0.4) -- (0.9,0.4);
    \draw (-1.15,0.4) node {$2\Delta t$} ;
    \draw [dotted] (0.0,0.7) -- (0.0,-0.3);
    \draw (-0.5,-0.2) node {$\text{cells of class 0}$} ;
    \draw (0.5,-0.2) node {$\text{cells of class 1}$} ;
    \draw (0.2,-0.07) node {$\beta+1$} ;
    \draw (-0.1,-0.07) node {$\beta$} ;
    \draw (-0.3,-0.07) node {$\beta-1$} ;
    \draw (-0.7,0.21) node {$\widehat{\times}$} ;
    \draw (-0.5,0.21) node {$\widehat{\times}$} ;
    \draw (-0.3,0.21) node {$\widehat{\times}$} ;
    \draw (-0.1,0.21) node {$\widehat{\times}$} ;
    \draw (0.2,0.41) node {$\widehat{\times}$} ;
    \draw (0.6,0.41) node {$\widehat{\times}$} ;
\end{tikzpicture}
\caption{Initial solution and predicted states using stages a-0 and a-1.
\label{fig:TA0bis}}
\end{center}
\end{figure}
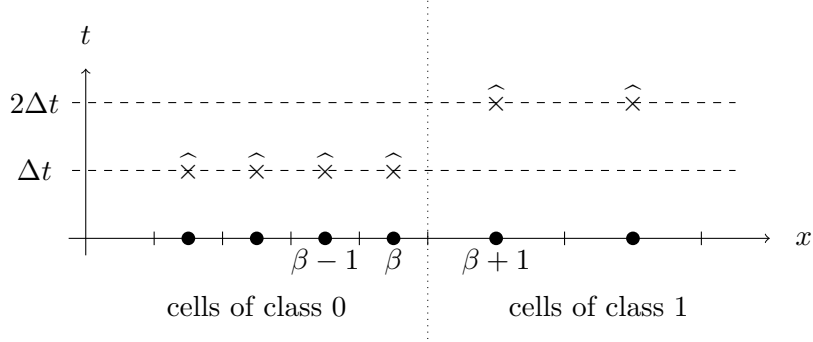

$\bullet$ {\bf Step 1:}\\
The residual $R(W^0)$ is computed using the initial solution, and the predicted states are obtained for all cells (stages a-0 and a-1), as shown in Fig~\ref{fig:TA0bis}.
At this time, it should be highlighted that the stage b-1 needs
the residual using $\widehat{W}^{2\Delta t}$ available only in the cells of class 1 (and not in class 0),
and the stage b-0 needs the residual using $\widehat{W}^{\Delta t}$ available in the cells of class 0 (and not in class 1).

\begin{figure}[!htbp]
\begin{center}
\begin{tikzpicture}[scale=4.5]
    \draw [->,black] (-1.05,0.) -- (1,0.);
    \draw (1.1,0) node {$x$} ;
    \draw [black] (-0.8,-0.02) -- (-0.8, 0.02);
    \fill (canvas cs:x=-0.7cm,y=0cm) circle (0.2mm);
    \draw [black] (-0.6,-0.02) -- (-0.6, 0.02);
    \fill (canvas cs:x=-0.5cm,y=0cm) circle (0.2mm);
    \draw [black] (-0.4,-0.02) -- (-0.4, 0.02);
    \fill (canvas cs:x=-0.3cm,y=0cm) circle (0.2mm);
    \draw [black] (-0.2,-0.02) -- (-0.2, 0.02);
    \fill (canvas cs:x=-0.1cm,y=0cm) circle (0.2mm);
    \draw [black] (-0. ,-0.02) -- (-0. , 0.02);
    \fill (canvas cs:x=0.2cm,y=0cm) circle (0.2mm);
    \draw [black] (0.4 ,-0.02) -- (0.4 , 0.02);
    \fill (canvas cs:x=0.6cm,y=0cm) circle (0.2mm);
    \draw [black] (0.8 ,-0.02) -- (0.8 , 0.02);
    \draw[->,black] (-1,-0.05) -- (-1.0,0.5);
    \draw (-1.0,0.6) node {$t$} ;
    \draw [dashed] (-1.04,0.2) -- (0.9,0.2);
    \draw (-1.15,0.2) node {$\Delta t$} ;
    \draw [dashed] (-1.04,0.4) -- (0.9,0.4);
    \draw (-1.15,0.4) node {$2\Delta t$} ;
    \draw [dotted] (0.0,0.7) -- (0.0,-0.3);
    \draw (-0.5,-0.2) node {$\text{cells of class 0}$} ;
    \draw (0.5,-0.2) node {$\text{cells of class 1}$} ;
    \draw (0.2,-0.07) node {$\beta+1$} ;
    \draw (-0.1,-0.07) node {$\beta$} ;
    \draw (-0.3,-0.07) node {$\beta-1$} ;
    \draw (-0.7,0.21) node {$\widehat{\times}$} ;
    \draw (-0.5,0.21) node {$\widehat{\times}$} ;
    \draw (-0.3,0.21) node {$\widehat{\times}$} ;
    \draw (-0.1,0.21) node {$\widehat{\times}$} ;
    \draw (-0.3,0.41) node {$\widehat{\times}$} ;
    \draw (-0.1,0.41) node {$\widehat{\times}$} ;
    \draw ( 0.6,0.41) node {$\widehat{\times}$} ;
    \draw ( 0.2,0.41) node {$\widehat{\times}$} ;

    \draw [dashed] [->] (-0.8,0.7) -- (-0.3,0.5);
    \draw [dashed] [->] (-0.8,0.7) -- (-0.1,0.5);
    \draw (-0.8,0.7) node [left] {$\text{extrapolated}$} ;
    \draw [dashed] [->] (0.8,0.7) -- (0.6,0.5);
    \draw [dashed] [->] (0.8,0.7) -- (0.2,0.5);
    \draw (0.8,0.7) node [right] {$\text{predicted}$} ;
\end{tikzpicture}
\caption{Predicted or extrapolated states used for the computation of the flux $F_{\beta+1/2}$ at time $2 \Delta t$
\label{fig:TA1}}
\end{center}
\end{figure}
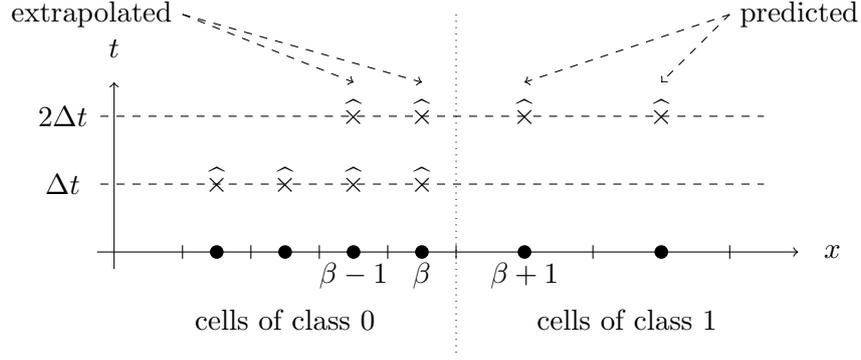

$\bullet$ {\bf Step 2:}\\
The residual $R(\widehat{W}^{2 \Delta t})$ needed by the stage b-1 needs the predicted states on cells of rank 0 on the required stencil to compute
the interface flux (Fig.~\ref{fig:TA1}). For the required cells with class rank 0, the estimated states are
\begin{equation}
\begin{aligned}
\widehat{W}^{2\Delta t}_{\beta}&=W^0_{\beta}+2\Delta t R(W^0),\\
\widehat{W}^{2\Delta t}_{\beta-1}&=W^0_{\beta-1}+2\Delta t R(W^0).
\end{aligned}
\label{eq:StateExtrapolation}
\end{equation}
These states are called extrapolated states since $2\Delta t$ violates the CFL stability condition for these cells ($2\Delta t$ is the time step of
cells from class of rank $1$).

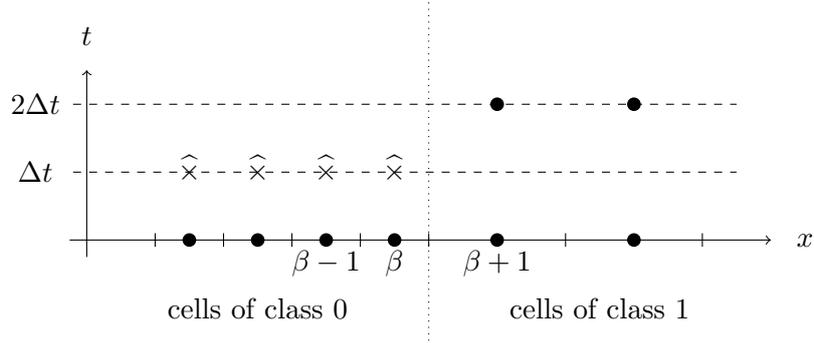
\begin{figure}[h]
\begin{center}
\begin{tikzpicture}[scale=4.5]
    \draw [->,black] (-1.05,0.) -- (1,0.);
    \draw (1.1,0) node {$x$} ;
    \draw [black] (-0.8,-0.02) -- (-0.8, 0.02);
    \fill (canvas cs:x=-0.7cm,y=0cm) circle (0.2mm);
    \draw [black] (-0.6,-0.02) -- (-0.6, 0.02);
    \fill (canvas cs:x=-0.5cm,y=0cm) circle (0.2mm);
    \draw [black] (-0.4,-0.02) -- (-0.4, 0.02);
    \fill (canvas cs:x=-0.3cm,y=0cm) circle (0.2mm);
    \draw [black] (-0.2,-0.02) -- (-0.2, 0.02);
    \fill (canvas cs:x=-0.1cm,y=0cm) circle (0.2mm);
    \draw [black] (-0. ,-0.02) -- (-0. , 0.02);
    \fill (canvas cs:x=0.2cm,y=0cm) circle (0.2mm);
    \draw [black] (0.4 ,-0.02) -- (0.4 , 0.02);
    \fill (canvas cs:x=0.6cm,y=0cm) circle (0.2mm);
    \draw [black] (0.8 ,-0.02) -- (0.8 , 0.02);
    \draw[->,black] (-1,-0.05) -- (-1.0,0.5);
    \draw (-1.0,0.6) node {$t$} ;
    \draw [dashed] (-1.04,0.2) -- (0.9,0.2);
    \draw (-1.15,0.2) node {$\Delta t$} ;
    \draw [dashed] (-1.04,0.4) -- (0.9,0.4);
    \draw (-1.15,0.4) node {$2\Delta t$} ;
    \draw [dotted] (0.0,0.7) -- (0.0,-0.3);
    \draw (-0.5,-0.2) node {$\text{cells of class 0}$} ;
    \draw (0.5,-0.2) node {$\text{cells of class 1}$} ;
    \draw (0.2,-0.07) node {$\beta+1$} ;
    \draw (-0.1,-0.07) node {$\beta$} ;
    \draw (-0.3,-0.07) node {$\beta-1$} ;
    \draw (-0.7,0.21) node {$\widehat{\times}$} ;
    \draw (-0.5,0.21) node {$\widehat{\times}$} ;
    \draw (-0.3,0.21) node {$\widehat{\times}$} ;
    \draw (-0.1,0.21) node {$\widehat{\times}$} ;
    \fill (canvas cs:x=0.2cm,y=0.4cm) circle (0.2mm);
    \fill (canvas cs:x=0.6cm,y=0.4cm) circle (0.2mm);
\end{tikzpicture}
\caption{Update of the solution for cells of class rank 1 at time $2 \Delta t$
\label{fig:TA1bis}}
\end{center}
\end{figure}

For cells of class rank $1$,
once the flux $F_{\beta+1/2}(\widehat{W}^{2\Delta t})$ is computed, the final state of stage b-1
can be computed at $t=2\Delta t$ (Fig.~\ref{fig:TA1bis}).

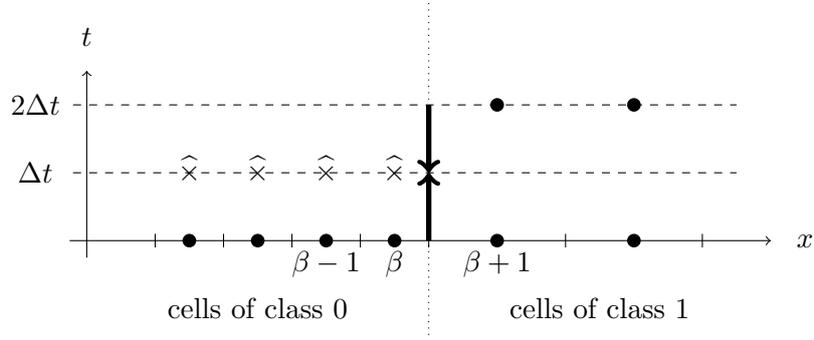
\begin{figure}[!htbp]
\begin{center}
\begin{tikzpicture}[scale=4.5]
    \draw [->,black] (-1.05,0.) -- (1,0.);
    \draw (1.1,0) node {$x$} ;
    \draw [black] (-0.8,-0.02) -- (-0.8, 0.02);
    \fill (canvas cs:x=-0.7cm,y=0cm) circle (0.2mm);
    \draw [black] (-0.6,-0.02) -- (-0.6, 0.02);
    \fill (canvas cs:x=-0.5cm,y=0cm) circle (0.2mm);
    \draw [black] (-0.4,-0.02) -- (-0.4, 0.02);
    \fill (canvas cs:x=-0.3cm,y=0cm) circle (0.2mm);
    \draw [black] (-0.2,-0.02) -- (-0.2, 0.02);
    \fill (canvas cs:x=-0.1cm,y=0cm) circle (0.2mm);
    \draw [black] (-0. ,-0.02) -- (-0. , 0.02);
    \fill (canvas cs:x=0.2cm,y=0cm) circle (0.2mm);
    \draw [black] (0.4 ,-0.02) -- (0.4 , 0.02);
    \fill (canvas cs:x=0.6cm,y=0cm) circle (0.2mm);
    \draw [black] (0.8 ,-0.02) -- (0.8 , 0.02);
    \draw[->,black] (-1,-0.05) -- (-1.0,0.5);
    \draw (-1.0,0.6) node {$t$} ;
    \draw [dashed] (-1.04,0.2) -- (0.9,0.2);
    \draw (-1.15,0.2) node {$\Delta t$} ;
    \draw [dashed] (-1.04,0.4) -- (0.9,0.4);
    \draw (-1.15,0.4) node {$2\Delta t$} ;
    \draw [dotted] (0.0,0.7) -- (0.0,-0.3);
    \draw (-0.5,-0.2) node {$\text{cells of class 0}$} ;
    \draw (0.5,-0.2) node {$\text{cells of class 1}$} ;
    \draw (0.2,-0.07) node {$\beta+1$} ;
    \draw (-0.1,-0.07) node {$\beta$} ;
    \draw (-0.3,-0.07) node {$\beta-1$} ;
    \draw (-0.7,0.21) node {$\widehat{\times}$} ;
    \draw (-0.5,0.21) node {$\widehat{\times}$} ;
    \draw (-0.3,0.21) node {$\widehat{\times}$} ;
    \draw (-0.1,0.21) node {$\widehat{\times}$} ;
    \fill (canvas cs:x=0.2cm,y=0.4cm) circle (0.2mm);
    \fill (canvas cs:x=0.6cm,y=0.4cm) circle (0.2mm);

    \draw[black] [line width=2pt, ->] (0.,0.4) -- (0.,0.2);
    \draw[black] [line width=2pt, ->] (0.,0.0) -- (0.,0.2);
\end{tikzpicture}
\caption{Definition of the interpolated flux at the interface between cell classes\label{fig:TA2}}
\end{center}
\end{figure}

$\bullet$ {\bf Step 3:}\\
For the cells of class rank $0$ sharing a face with a cell of class rank $1$, the update of the solution associated with stage b-0
needs the definition of the flux at the interface $\beta+1/2$,
\begin{equation}
F_{\beta+1/2}(\widehat{W}^{\Delta t})=\frac{1}{2} \big( F_{\beta+1/2}(W^0)+F_{\beta+1/2}(\widehat{W}^{2\Delta t}) \big).
\label{eq:step3_flux}
\end{equation}
This interpolation is represented by mean of arrows in Fig.~\ref{fig:TA2}.

\begin{figure}[h]
\begin{center}
\begin{tikzpicture}[scale=4.5]
    \draw [->,black] (-1.05,0.) -- (1,0.);
    \draw (1.1,0) node {$x$} ;
    \draw [black] (-0.8,-0.02) -- (-0.8, 0.02);
    \fill (canvas cs:x=-0.7cm,y=0cm) circle (0.2mm);
    \draw [black] (-0.6,-0.02) -- (-0.6, 0.02);
    \fill (canvas cs:x=-0.5cm,y=0cm) circle (0.2mm);
    \draw [black] (-0.4,-0.02) -- (-0.4, 0.02);
    \fill (canvas cs:x=-0.3cm,y=0cm) circle (0.2mm);
    \draw [black] (-0.2,-0.02) -- (-0.2, 0.02);
    \fill (canvas cs:x=-0.1cm,y=0cm) circle (0.2mm);
    \draw [black] (-0. ,-0.02) -- (-0. , 0.02);
    \fill (canvas cs:x=0.2cm,y=0cm) circle (0.2mm);
    \draw [black] (0.4 ,-0.02) -- (0.4 , 0.02);
    \fill (canvas cs:x=0.6cm,y=0cm) circle (0.2mm);
    \draw [black] (0.8 ,-0.02) -- (0.8 , 0.02);
    \draw[->,black] (-1,-0.05) -- (-1.0,0.5);
    \draw (-1.0,0.6) node {$t$} ;
    \draw [dashed] (-1.04,0.2) -- (0.9,0.2);
    \draw (-1.15,0.2) node {$\Delta t$} ;
    \draw [dashed] (-1.04,0.4) -- (0.9,0.4);
    \draw (-1.15,0.4) node {$2\Delta t$} ;
    \draw [dotted] (0.0,0.7) -- (0.0,-0.3);
    \draw (-0.5,-0.2) node {$\text{cells of class 0}$} ;
    \draw (0.5,-0.2) node {$\text{cells of class 1}$} ;
    \draw (0.2,-0.07) node {$\beta+1$} ;
    \draw (-0.1,-0.07) node {$\beta$} ;
    \draw (-0.3,-0.07) node {$\beta-1$} ;
    \draw (-0.7,0.21) node {$\widehat{\times}$} ;
    \draw (-0.5,0.21) node {$\widehat{\times}$} ;
    \draw (-0.3,0.21) node {$\widehat{\times}$} ;
    \draw (-0.1,0.21) node {$\widehat{\times}$} ;
    \draw ( 0.2,0.21) node {$\widehat{\times}$} ;
    \fill (canvas cs:x=0.2cm,y=0.4cm) circle (0.2mm);
    \fill (canvas cs:x=0.6cm,y=0.4cm) circle (0.2mm);
    \draw [dashed] [->] (0.8,0.3) -- (0.25,0.2);
    \draw (0.8,0.3) node [right] {$\text{Extrapolated state for MUSCL extrapolation}$} ;
\end{tikzpicture}
\caption{Estimation of the states in class rank 1 for the intermediate time $\Delta t$}
\label{fig:TA4}
\end{center}
\end{figure}
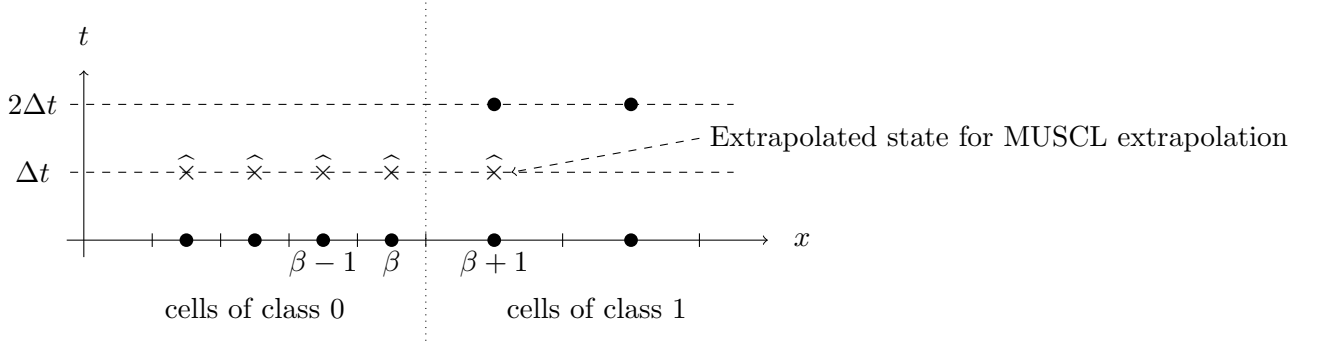

$\bullet$ {\bf Step 4:}\\
The flux is directly given by Step 3 for the interface $\beta+1/2$ but the residual in cells $\beta$, $\beta-1$, etc.
may involve states in the cells of class 1.
The number of cells to treat in class rank 1 depends only on the stencil of the spatial scheme for the cells in class rank 0
(MUSCL formulation for instance). In order to compute the other corrector flux at $t=\Delta t$ for cells of class rank 0
located in the stencil of the spatial schemes (cell $\beta-1$ for instance), it is necessary to predict the states in some cells of class rank $1$ at
$t=\Delta t$. A simple first-order accurate prediction would be
$
\widehat{W}^{\Delta t}_{\beta+1} = W^0_{\beta+1}+\Delta t R(W^0)
$. To keep a second-order time accuracy, the residual at time $\Delta t$ is computed with a parabolic interpolation of the residuals,
\begin{equation}
     R^{\Delta t}_{\beta+1}= \frac{1}{4} R_{\beta+1}(\widehat{W}^{2 \Delta t}) + \frac{3}{4} R_{\beta+1}(W^0),
\end{equation}
which is used to compute the predictor state at time $\Delta t$ using
\begin{equation}
    \widehat{W}^{\Delta t}_{\beta+1} = W^0_{\beta+1} + \Delta t R^{\Delta t}_{\beta+1}.\label{eq:stateStage4}
\end{equation}
The new estimated solution in the cell $\beta+1$ is represented as a predicted state in Fig.~\ref{fig:TA4}.

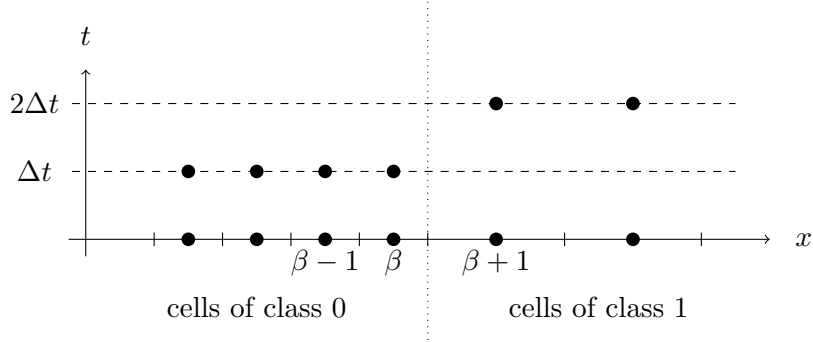
\begin{figure}[h]
\begin{center}
\begin{tikzpicture}[scale=4.5]
    \draw [->,black] (-1.05,0.) -- (1,0.);
    \draw (1.1,0) node {$x$} ;
    \draw [black] (-0.8,-0.02) -- (-0.8, 0.02);
    \fill (canvas cs:x=-0.7cm,y=0cm) circle (0.2mm);
    \draw [black] (-0.6,-0.02) -- (-0.6, 0.02);
    \fill (canvas cs:x=-0.5cm,y=0cm) circle (0.2mm);
    \draw [black] (-0.4,-0.02) -- (-0.4, 0.02);
    \fill (canvas cs:x=-0.3cm,y=0cm) circle (0.2mm);
    \draw [black] (-0.2,-0.02) -- (-0.2, 0.02);
    \fill (canvas cs:x=-0.1cm,y=0cm) circle (0.2mm);
    \draw [black] (-0. ,-0.02) -- (-0. , 0.02);
    \fill (canvas cs:x=0.2cm,y=0cm) circle (0.2mm);
    \draw [black] (0.4 ,-0.02) -- (0.4 , 0.02);
    \fill (canvas cs:x=0.6cm,y=0cm) circle (0.2mm);
    \draw [black] (0.8 ,-0.02) -- (0.8 , 0.02);
    \draw[->,black] (-1,-0.05) -- (-1.0,0.5);
    \draw (-1.0,0.6) node {$t$} ;
    \draw [dashed] (-1.04,0.2) -- (0.9,0.2);
    \draw (-1.15,0.2) node {$\Delta t$} ;
    \draw [dashed] (-1.04,0.4) -- (0.9,0.4);
    \draw (-1.15,0.4) node {$2\Delta t$} ;
    \draw [dotted] (0.0,0.7) -- (0.0,-0.3);
    \draw (-0.5,-0.2) node {$\text{cells of class 0}$} ;
    \draw (0.5,-0.2) node {$\text{cells of class 1}$} ;
    \draw (0.2,-0.07) node {$\beta+1$} ;
    \draw (-0.1,-0.07) node {$\beta$} ;
    \draw (-0.3,-0.07) node {$\beta-1$} ;
    \fill (canvas cs:x=-0.7cm,y=0.2cm) circle (0.2mm);
    \fill (canvas cs:x=-0.5cm,y=0.2cm) circle (0.2mm);
    \fill (canvas cs:x=-0.3cm,y=0.2cm) circle (0.2mm);
    \fill (canvas cs:x=-0.1cm,y=0.2cm) circle (0.2mm);
    \fill (canvas cs:x=0.2cm,y=0.4cm) circle (0.2mm);
    \fill (canvas cs:x=0.6cm,y=0.4cm) circle (0.2mm);

\end{tikzpicture}
\caption{Update of the solution of class 0 at time $\Delta t$. The states at time $2\Delta t$ for class rank 1 are available.}
\label{fig:TA4bis}
\end{center}
\end{figure}

Using the new available information, the cells of class rank 0 can be updated and the stage b-0 is completed. From
now on, the variables $W^{\Delta t}$ are available for any cell of class rank 0 (Fig.~\ref{fig:TA4bis}).
The last step is to apply stages c-0 and d-0.

$\bullet$ {\bf Step 5:}\\
Finally, the cells of class rank $0$ have to be integrated in time from $\Delta t$ to $2\Delta t$ (stages c-0 and d-0).
There are two points to perform the time integration.
First, it is mandatory to keep the interface flux constant for the Heun's stages: the flux computed using Eq.~\eqref{eq:step3_flux} is
imposed to compute the residual in cell $\beta$ for stages c-0 and d-0. Moreover, the high-order spatial interpolation (for instance for interface
$\beta-1/2$) may need the fields in the cells of class rank 1 at time $\Delta t$. In that case, the data computed using Eq.~\eqref{eq:stateStage4}
is kept constant for stages c-0 and d-0.

To conclude, in one time iteration, this temporal adaptive method allows to integrate small cells in time
with the time step of the biggest cells of the domain thanks to a subcycling process.
Brenner~\cite{Brenner_AIAA_1993} showed that for a maximal time step $\Delta t_{max}$ equal to $2^K\Delta t_{min}$,
if most cells are included in the time class of rank $K$, then
the computational cost is divided by $2^K$ in the most favorable case.

\subsection{Conservation Property\label{Heun+TA_conservation}}

As stated in Sec.~\ref{sec:intro}, conservation associated to the flux balance is not always guaranteed,
and this section is devoted to the proof of conservation for the proposed time-adaptive
Heun's scheme.

First, according to the one-dimensional configuration presented in Sec.~\ref{sec:TA_method},
the time integration of cell $\beta+1$ of class rank $1$ may be formulated with predictor and corrector stages
going from $t=0$ to $t=2\Delta t$ as
\begin{equation}
\begin{aligned}
\widehat{W}^{2\Delta t}_{\beta+1}&={W}^{0}_{\beta+1}+\frac{2\Delta t}{\Delta x} \big( F_{\beta+\frac{3}{2}}(W^0)-F_{\beta+\frac{1}{2}}(W^0)\big) \\
{W}^{2\Delta t}_{\beta+1}&={W}^{0}_{\beta+1}+\frac{2\Delta t}{2\Delta x}
\big( F_{\beta+\frac{3}{2}}(\widehat{W}^{2\Delta t})+F_{\beta+\frac{3}{2}}(W^0)-F_{\beta+\frac{1}{2}}(W^0)-F_{\beta+\frac{1}{2}}(\widehat{W}^{2\Delta t})\big). \\
\end{aligned}
\end{equation}
All negative contributions from the interface $\beta+\frac{1}{2}$ must be recovered in the flux balance of the cell $\beta$.

Moreover, focusing on the opposite side of the interface,
the time integration from $t=0$ to $t=2\Delta t$ of the cell $\beta$ of class rank $0$ may be formulated as
\begin{equation}
\begin{aligned}
\widehat{W}^{\Delta t}_{\beta} &= {W}^{0}_{\beta} + \frac{\Delta t}{\Delta x} \big( F_{\beta+\frac{1}{2}}(W^0)-F_{\beta-\frac{1}{2}}(W^0)\big), \\
{W}^{\Delta t}_{\beta} &=         {W}^{0}_{\beta} + \frac{\Delta t}{2\Delta x}
\big(
  F_{\beta+\frac{1}{2}}(\widehat{W}^{\Delta t})
+ F_{\beta+\frac{1}{2}}(W^0)
- F_{\beta-\frac{1}{2}}(W^0)
- F_{\beta-\frac{1}{2}}(\widehat{W}^{\Delta t}) \big)
\end{aligned}
\label{eq:conserv1}
\end{equation}
in a first phase, and
\begin{equation}
    \begin{aligned}
\widehat{W}^{2\Delta t}_{\beta} &= {W}^{\Delta t}_{\beta} + \frac{\Delta t}{\Delta x}
\big(
    F_{\beta+\frac{1}{2}}({W}^{\Delta t})
    - F_{\beta-\frac{1}{2}}(W^{\Delta t})\big), \\
{W}^{2\Delta t}_{\beta} &= {W}^{\Delta t}_{\beta} + \frac{\Delta t}{2\Delta x}
\big(
  F_{\beta+\frac{1}{2}}(\widehat{W}^{\Delta t})
+ F_{\beta+\frac{1}{2}}(\widehat{W}^{2\Delta t})
- F_{\beta-\frac{1}{2}}(W^{\Delta t})
- F_{\beta-\frac{1}{2}}(\widehat{W}^{2\Delta t})\big),
\end{aligned}
\label{eq:conserv2}
\end{equation}
in a second phase.
If ${W}^{\Delta t}_{\beta}$ of Eq.~\eqref{eq:conserv1} is replaced in the second equation of~\eqref{eq:conserv2},
and reminding that Eq.~\eqref{eq:step3_flux} holds, then
\begin{equation}
{W}^{2\Delta t}_{\beta} = {W}^{0}_{\beta} + \frac{\Delta t}{2\Delta x}
\bigg( 2\big(F_{\beta+\frac{1}{2}}(\widehat{W}^{2\Delta t})+F_{\beta+\frac{1}{2}}(W^0)\big)
-F_{\beta-\frac{1}{2}}(W^0)
-F_{\beta-\frac{1}{2}}(\widehat{W}^{\Delta t})
-F_{\beta-\frac{1}{2}}(W^{\Delta t})
-F_{\beta-\frac{1}{2}}(\widehat{W}^{2\Delta t})\bigg).\\
\end{equation}

It appears that the time integration of state $W$ at cells $\beta$ and $\beta+1$
between $t=0$ and $t=2\Delta t$ share exactly the same flux at the interface $\beta+\frac{1}{2}$.

\section{Accuracy and spectral analysis of temporal adaptive method \label{sec:accuracy+spectral}}
\subsection{Time Accuracy\label{Heun+TA_accuracy}}

In this section, a local analysis of the accuracy for the temporal adaptive version of Heun's scheme
is performed near an interface concerned by time synchronization, which means that the interface is shared
by two cells belonging to to adjacent class ranks. For the sake of clarity, the proof is performed in 1D but
the demonstration can be kept in a multi-dimensional framework.

Let's introduce a 1D mesh with a fixed grid size $\Delta x$ with two temporal classes imposed. A generic partial
differential equation is integrated spatially, as for standard finite volume. Indeed, two relations are obtained,
the first one being associated to the exact relation and the second one to the approximated
discrete version:
\begin{equation}
\begin{aligned}
    \Delta x \frac{d \mathcal{W}_{j}}{d t}&= \Delta x \mathcal{R}_j=
    \mathcal{F}_{j+\frac{1}{2}}-\mathcal{F}_{j-\frac{1}{2}}, \\
        \Delta x \frac{d W_{j}}{d t}&= \Delta x R_j=
    F_{j+\frac{1}{2}}-F_{j-\frac{1}{2}},
\end{aligned}
\label{eq:EDP_order}
\end{equation}
with $\mathcal{W}$ the exact state, $\mathcal{R}$ the exact residual and $\mathcal{F}$ the exact flux.
As introduced above, $W$ represents the numerical state, $R$ the numerical residual and $F$ the
numerical flux. In the following, the numerical error on the state for cells of rank $0$ and $1$ between
$t=0$ and $t=2\Delta t$ is defined as:
\begin{equation}
  e({W}^{2\Delta t}) = {\mathcal{W}}^{2 \Delta t} - {W}^{2\Delta t},
\label{eq:error_w}
\end{equation}
This numerical error will be computed at each step of the temporal adaptive method presented in
Sec.~\ref{sec:TA_method}. \GP{From now on, it is assumed that the computation of the flux is $p^{\mbox{\scriptsize th}}$-order accurate in space} and that a first-order finite difference approximates the time derivative, which leads to:
\begin{equation}
\begin{aligned}
F_{j+1/2} &= {\cal{F}}_{j+1/2} + \mathcal{O}(\Delta x^p),\\
\frac{\widehat{W}^{\Delta t} - {W}^{0} }{\Delta t} &= \frac{dW}{dt} = \frac{d \mathcal{W}_{j}}{d t} + \mathcal{O}(\Delta t)
\end{aligned}\label{eq:Basis_localerror}
\end{equation}
\GP{Using Eq.~\eqref{eq:EDP_order}, it is clear that the residual behaves as $\mathcal{O}(\Delta x^{p-1})$ due to 
the relation between the residual and the flux.}
\LM{According to Eq.~\eqref{eq:Basis_localerror},
Eq.~\eqref{eq:EDP_order} can be written as (omitting subscript $j$)}
\begin{equation}
\begin{aligned}
\frac{\widehat{W}^{\Delta t} - {W}^{0} }{\Delta t} &=   \mathcal{R} + \mathcal{O}(\Delta x^{p-1})  + \mathcal{O}(\Delta t), \\
\widehat{W}^{\Delta t} & = {W}^{0} + \Delta t \big( \mathcal{R} + \mathcal{O}(\Delta x^{p-1})  \big) + \mathcal{O}(\Delta t^{2})
= {W}^{0} + \Delta t \mathcal{R} + \mathcal{O}(\Delta t \Delta x^{p-1}, \Delta t^{2}).
\end{aligned}
\label{eq:predictor}
\end{equation}
Equation~\eqref{eq:predictor} is typically what occurs during a predictor stage of the Heun' scheme.

Let us now study the local error performed at cells of class rank $1$.

\subsubsection{Local numerical error of cells of class rank $1$}
According to the numerical error performed at the predictor state from Eq.\eqref{eq:predictor},
and knowing the orders of accuracy of the numerical scheme,
the accuracy of the numerical flux (omitting subscript of interface index) at cells of temporal class rank $1$ is:
\begin{equation}
\begin{aligned}
    F(\widehat{W}^{2\Delta t}) & = F({W}^{0} + 2 \Delta t \mathcal{R} + \mathcal{O}(\Delta t \Delta x^{p-1}, \Delta t^{2})) \\
                               & = \mathcal{F}({W}^{0} + 2 \Delta t \mathcal{R} + \mathcal{O}(\Delta t \Delta x^{p-1}, \Delta t^{2})) + \mathcal{O}(\Delta x^{p}).
\end{aligned}
\label{eq:flux_exp}
\end{equation}
Then according to Eqs.~\eqref{eq:flux_exp} and~\eqref{eq:EDP_order}, it is possible to formulate the residual as
\begin{equation}
\begin{aligned}
    R(\widehat{W}^{2\Delta t}) & = R({W}^{0} + 2 \Delta t \mathcal{R} + \mathcal{O}(\Delta t \Delta x^{p-1}, \Delta t^{2})) \\
                               & = \mathcal{R}({W}^{0} + 2 \Delta t \mathcal{R} + \mathcal{O}(\Delta t \Delta x^{p-1}, \Delta t^{2})) + \mathcal{O}(\Delta x^{p-1})
\end{aligned}
\end{equation}
With the assumption that no numerical error occurred at instant $t=0$, i.e. $W^0=\mathcal{W}^0$ and
$\mathcal{R}=\mathcal{R}(W^0)$, the Taylor expansion of flux (omitting subscript of interface index)
around ${W}^{0}$ gives:
\begin{equation}
\begin{aligned}
F(\widehat{W}^{2\Delta t}) &= \mathcal{F}({W}^{0}) + 2 \Delta t \frac{\partial \mathcal{F}}{\partial \mathcal{W}} \mathcal{R}
+ \mathcal{O}(\Delta t \Delta x^{p-1}, {\Delta t^2}, \Delta x^{p}). \\
 \end{aligned}
 \label{eq:error_flux_2dt}
 \end{equation}
The residual is reformulated thanks to Eqs.~\eqref{eq:error_flux_2dt} and~\eqref{eq:EDP_order} as
\begin{equation}
\begin{aligned}
R(\widehat{W}^{2\Delta t}) &= \mathcal{R}({W}^{0}) + 2 \Delta t \frac{\partial \mathcal{R}}{\partial \mathcal{W}} \mathcal{R}
+ \mathcal{O}(\Delta t \Delta x^{p-2}, \frac{\Delta t^2}{\Delta x}, \Delta x^{p-1}). \\
 \end{aligned}
 \label{eq:error_residual_2dt}
 \end{equation}
Then the state $W$ at cells of class rank $1$ of the corrector step is given by
\begin{equation}
\begin{aligned}
W^{2\Delta t}&={W}^{0}+\frac{2\Delta t}{2} \big( R(W^0)+R(\widehat{W}^{2\Delta t})\big) \\
&= {W}^{0} + \Delta t \big( \mathcal{R} + \mathcal{R} + 2\Delta t \frac{\partial \mathcal R}{\partial \mathcal{W}} \mathcal{R}
+ \mathcal{O}(\Delta t \Delta x^{p-2}, \frac{\Delta t^2}{\Delta x}, \Delta x^{p-1}) \big). \\
\end{aligned}
\label{eq:corre}
\end{equation}
Hence it is possible to formulate the numerical error of $W^{2\Delta t}$ at cells of class rank $1$ (see Eq.~\eqref{eq:error_w})
using the Taylor expansion of ${\mathcal{W}}^{2 \Delta t}$ and Eqs.~\eqref{eq:error_residual_2dt} and~\eqref{eq:corre} as
\begin{equation}
\begin{aligned}
    e(W^{2\Delta t}) &= \mathcal{W}^{0} + 2\Delta t\frac{\partial \mathcal{W}}{\partial t}
    + 2{\Delta t^2}\frac{\partial^2 \mathcal{W}}{\partial t^2}
    + \mathcal{O}(\Delta t^3)
    -W^{2\Delta t}
     = \mathcal{O}(\Delta t^3, \frac{\Delta t^3}{\Delta x}, \Delta t^2 \Delta x^{p-2}, \Delta t \Delta x^{p-1}).
\end{aligned}
\label{eq:error_classes_3}
\end{equation}


\subsubsection{Local numerical error due to time interpolation for cells of class rank $1$}

The flux at interface $\beta-\frac{1}{2}$ needs a special treatment due to the use of a MUSCL formulation:
a special treatment of cells of class rank 1 is performed according to step 4.
Here only cell $\beta+1$ of class rank $1$ is considered, so it is important to evaluate the error generated during the parabolic interpolation.
Thanks to Eq.~\eqref{eq:error_residual_2dt}, the state $\widehat{W}^{\Delta t}_{\beta+1}$ is formulated as:
\begin{equation}
\begin{aligned}
\widehat{W}^{\Delta t}_{\beta+1}&={W}^{0}_{\beta+1}+{\Delta t} \bigg( \frac{3}{4}R(W^0)+\frac{1}{4}R(\widehat{W}^{2\Delta t})\bigg). \\
&={W}^{0}_{\beta+1}+{\Delta t}\bigg( \frac{3}{4}\mathcal{R} +\frac{1}{4}(\mathcal{R}+2\Delta t \mathcal{R} \frac{\partial \mathcal R}{\partial \mathcal{W}}\big) +\mathcal{O}(\frac{\Delta t^2}{\Delta x}, \Delta t \Delta x^{p-2},\Delta x^{p-1})\bigg). \\
\end{aligned}
\end{equation}
Then, the error $e(\widehat{W}^{\Delta t}_{\beta+1})$ according to parabolic interpolation is:
\begin{equation}
\begin{aligned}
    e(\widehat{W}^{\Delta t}_{\beta+1})&=\mathcal{W}^{0}_{\beta+1}+\Delta t\frac{\partial \mathcal{W}}{\partial t}+\frac{\Delta t^2}{2}\frac{\partial^2 \mathcal{W}}{\partial t^2}+\mathcal{O}(\Delta t^3)-\widehat{W}^{\Delta t}_{\beta+1} =\mathcal{O}(\Delta t^3, \frac{\Delta t^3}{\Delta x}, \Delta t^2 \Delta x^{p-2}, \Delta t \Delta x^{p-1}).
\end{aligned}
\label{eq:error_parabolic}
\end{equation}

According to Eq.~\eqref{eq:error_parabolic}, the numerical flux at interface $\beta-\frac{1}{2}$ is expressed as
\begin{equation}
\begin{aligned}
    F_{\beta-\frac{1}{2}}(\widehat{W}^{\Delta t})&=\mathcal{F}_{\beta-\frac{1}{2}}+\Delta t \mathcal{R} \frac{\partial \mathcal{F}_{\beta-\frac{1}{2}}}{\partial \mathcal{W}}+\mathcal{O}(\Delta t^3, \frac{\Delta t^3}{\Delta x}, \Delta t^2 \Delta x^{p-2}, \Delta t \Delta x^{p-1};\Delta x^p). \\
    F_{\beta-\frac{1}{2}}({W}^{0})&= \mathcal{F}_{\beta-\frac{1}{2}}+ \mathcal{O}(\Delta x^p).
\end{aligned}
\label{eq:error_flux_beta-1/2}
\end{equation}

Now let us consider the case of cell $\beta$ of class rank $0$.

\subsubsection{Local numerical error of cells $\beta$ of class rank $0$}

According to the numerical error obtained in Eq.~\eqref{eq:predictor} at the predictor step on $\widehat{W}^{2\Delta t}$
and the Taylor expansion of flux around $W$,
\begin{equation}
\begin{aligned}
    F_{\beta+\frac{1}{2}}(\widehat{W}^{2\Delta t})&=\mathcal{F}_{\beta+\frac{1}{2}}+2\Delta t \mathcal{R} \frac{\partial \mathcal{F}_{\beta+\frac{1}{2}}}{\partial \mathcal{W}}+\mathcal{O}(\Delta t^2, \Delta t \Delta x^{p-1},\Delta x^{p}),\\
    F_{\beta+\frac{1}{2}}({W}^{0})&= \mathcal{F}_{\beta+\frac{1}{2}}+ \mathcal{O}(\Delta x^p),
\end{aligned}
\label{eq:error_flux_beta}
\end{equation}
then the flux $F(\widehat{W}^{\Delta t})_{\beta+\frac{1}{2}}$ is written as

\begin{equation}
\begin{aligned}
    F_{\beta+\frac{1}{2}}(\widehat{W}^{\Delta t})&=\frac{1}{2} \bigg( F_{\beta+\frac{1}{2}}({W}^{0})+F_{\beta+\frac{1}{2}}(\widehat{W}^{2\Delta t})\bigg)\\
    &=\mathcal{F}_{\beta+\frac{1}{2}}+ \Delta t \mathcal{R} \frac{\partial \mathcal{F}_{\beta+\frac{1}{2}}}{\partial \mathcal{W}}+\mathcal{O}(\Delta t^2,\Delta x^{p}, \Delta t \Delta x^{p-1}).
\end{aligned}
\label{eq:error_flux_beta_2}
\end{equation}

Thanks to Eqs.~\eqref{eq:error_flux_beta-1/2}, \eqref{eq:error_flux_beta_2}, and~\eqref{eq:error_flux_beta}, the state $W^{\Delta t}_{\beta}$ is formulated as
\begin{equation}
\begin{aligned}
{W}^{\Delta t}_{\beta}&={W}^{0}_{\beta}+\frac{\Delta t}{2\Delta x} \bigg( F_{\beta+\frac{1}{2}}(W^0)+F_{\beta+\frac{1}{2}}(\widehat{W}^{\Delta t})-F_{\beta-\frac{1}{2}}(\widehat{W}^{\Delta t})-F_{\beta-\frac{1}{2}}({W}^{0})\bigg) \\
&={W}^{0}_{\beta}+\frac{\Delta t}{2}\bigg( 2\mathcal{R}+\Delta t \mathcal{R}\frac{\partial \mathcal{R}}{\partial \mathcal{W}}\\
& \text{ }\text{ }\text{ }\text{ }+\mathcal{O}(\frac{\Delta t^2}{\Delta x}, \frac{\Delta t^3}{\Delta x^2}, \Delta t^2 \Delta x^{p-3}, \Delta t \Delta x^{p-2},\Delta x^{p-1})\bigg).
\end{aligned}
\end{equation}
And the error $e(W^{\Delta t}_{\beta})$ is written as
\begin{equation}
\begin{aligned}
    e(W^{\Delta t}_{\beta})&=\mathcal{W}^{0}_{\beta}+\Delta t\frac{\partial \mathcal{W}}{\partial t}+\frac{\Delta t^2}{2}\frac{\partial^2 \mathcal{W}}{\partial t^2}+\mathcal{O}(\Delta t^3)-W^{\Delta t}_{\beta} \\
    &=\mathcal{O}(\Delta t^3, \frac{\Delta t^3}{\Delta x},\frac{\Delta t^4}{\Delta x^2}, \Delta t^3 \Delta x^{p-3},\Delta t^2 \Delta x^{p-2}, \Delta t \Delta x^{p-1}).
\end{aligned}
\label{eq:error_beta}
\end{equation}
Considering the local error of $W$ at cell $\beta$ until $t=\Delta t$ in Eq.~\eqref{eq:error_beta}, and the error associated to the stencil
of the MUSCL reconstruction technique in Eq.~\eqref{eq:error_parabolic},
the fluxes $F_{\beta-\frac{1}{2}}(W^{\Delta t})$ and $F_{\beta-\frac{1}{2}}(\widehat{W}^{\Delta t})$ are expressed as
\begin{equation}
\left\{
\begin{aligned}
    F_{\beta-\frac{1}{2}}(\widehat{W}^{2\Delta t})&=\mathcal{F}_{\beta-\frac{1}{2}}+2\Delta t \mathcal{R} \frac{\partial \mathcal{F}_{\beta-\frac{1}{2}}}{\partial \mathcal{W}}+\mathcal{O}(\Delta t^3, \frac{\Delta t^3}{\Delta x},\frac{\Delta t^4}{\Delta x^2}, \Delta t^3 \Delta x^{p-3},\Delta t^2 \Delta x^{p-2}, \Delta t \Delta x^{p-1},\Delta x^p),\\
    F_{\beta-\frac{1}{2}}({W}^{\Delta t})&= \mathcal{F}_{\beta-\frac{1}{2}} +\mathcal{O}(\Delta t^3, \frac{\Delta t^3}{\Delta x},\frac{\Delta t^4}{\Delta x^2}, \Delta t^3 \Delta x^{p-3},\Delta t^2 \Delta x^{p-2}, \Delta t \Delta x^{p-1},\Delta x^p)
\end{aligned}
\right.
\label{eq:error_flux_beta-1/2_2}
\end{equation}
Thanks to Eqs.~\eqref{eq:error_flux_beta-1/2_2}, \eqref{eq:error_flux_beta_2}, \eqref{eq:error_flux_beta}, and \eqref{eq:error_beta},
the state $W^{2\Delta t}_{\beta}$ is then formulated as
\begin{equation}
\begin{aligned}
{W}^{2\Delta t}_{\beta}&={W}^{\Delta t}_{\beta}+\frac{\Delta t}{2\Delta x} \bigg( F_{\beta+\frac{1}{2}}(\widehat{W}^{2\Delta t})+F_{\beta+\frac{1}{2}}(\widehat{W}^{\Delta t})-F_{\beta-\frac{1}{2}}(\widehat{W}^{2\Delta t})-F_{\beta-\frac{1}{2}}({W}^{\Delta t})\bigg). \\
&=\mathcal{W}^{\Delta t}_{\beta}+\mathcal{O}(\Delta t^3, \frac{\Delta t^3}{\Delta x},\frac{\Delta t^4}{\Delta x^2}, \Delta t^3 \Delta x^{p-3},\Delta t^2 \Delta x^{p-2}, \Delta t \Delta x^{p-1}) \\
&\text{ }\text{ }\text{ }\text{ }+\frac{\Delta t}{2}\bigg( 2\mathcal{R}+{3}\Delta t \mathcal{R}\frac{\partial \mathcal{F}_{\beta+\frac{1}{2}}/\Delta x}{\partial \mathcal{W}}-2\Delta t \mathcal{R}\frac{\partial \mathcal{F}_{\beta-\frac{1}{2}}/\Delta x}{\partial \mathcal{W}}\\
&\text{ }\text{ }\text{ }\text{ }+\mathcal{O}(\frac{\Delta t^2}{\Delta x},\frac{\Delta t^3}{\Delta x^2},\frac{\Delta t^4}{\Delta x^3}, \Delta t^3 \Delta x^{p-4},\Delta t^2 \Delta x^{p-3},\Delta t \Delta x^{p-2},\Delta x^{p-1})
\bigg)
\end{aligned}
\end{equation}
Then the local error $e({W}^{2\Delta t}_{\beta})$ is
\begin{equation}
\begin{aligned}
    e(W^{2\Delta t}_{\beta})&=\mathcal{W}^{\Delta t}_{\beta}+\Delta t\frac{\partial \mathcal{W}}{\partial t}+\frac{\Delta t^2}{2}\frac{\partial^2 \mathcal{W}}{\partial t^2}+\mathcal{O}(\Delta t^3)-W^{2\Delta t}_{\beta} \\
    &=\mathcal{O}(\Delta t^2, \frac{\Delta t^3}{\Delta x},\frac{\Delta t^4}{\Delta x^2},\frac{\Delta t^5}{\Delta x^3}, \Delta t^4 \Delta x^{p-4},\Delta t^3 \Delta x^{p-3},\Delta t^2 \Delta x^{p-2},\Delta t\Delta x^{p-1})
\end{aligned}
\label{eq:error_beta_2dt}
\end{equation}

\subsubsection{Partial conclusion}
In the case of hyperbolic equations, the CFL condition shows that the time step $\Delta t$ depends linearly on
the grid spacing $\Delta x$. Then the local numerical error performed until
$t=2\Delta t$ on cell $\beta$ of class rank $0$ is $O(\Delta t^2,\Delta x^p)$,
and the local numerical error at cells of class rank $1$ is $O(\Delta t^2,\Delta x^p)$.
It appears that the temporal order of accuracy is kept constant locally.

In the following section, the influence of local numerical errors due to the temporal adaptive method
on the global error is investigated.

\subsection{Numerical assessment of the theoretical behavior \label{Heun+TA_accuracy_sinus}}

A one-dimensional numerical analysis of the global order of accuracy is performed
by solving the advection equation of a sinus wave inside a periodic domain of length $L=1$ ($x \in [0,1]$).
The initial state is
\begin{equation}
\begin{aligned}
y(x,t=0)=sin(2\pi x).
\end{aligned}
\end{equation}
The grid contains $N$ cells, and then the grid size is constant.
Two temporal classes are imposed with $100$ cells in the class of rank $0$
in the middle of the domain ($\frac{N}{2}-50 \leq j \leq \frac{N}{2}+50$).
The computation is performed during $3s$ of physical time with a fixed time step $\Delta t=2.5\,10^{-5}s$, with a $1$-exact spatial scheme.
The accuracy of the temporal adaptive method with Heun's scheme (defined as ``Heun+TA'') is compared with the standard Heun's integrator.
As expected, the computation of the total errors $E_{tot}$ reveals that the temporal adaptive method conserves
the second order space-time accuracy of Heun's scheme, according to Tab.~\ref{tab:order_tab}.

\begin{figure}[!ht]
\begin{center}
\includegraphics [width=10cm]{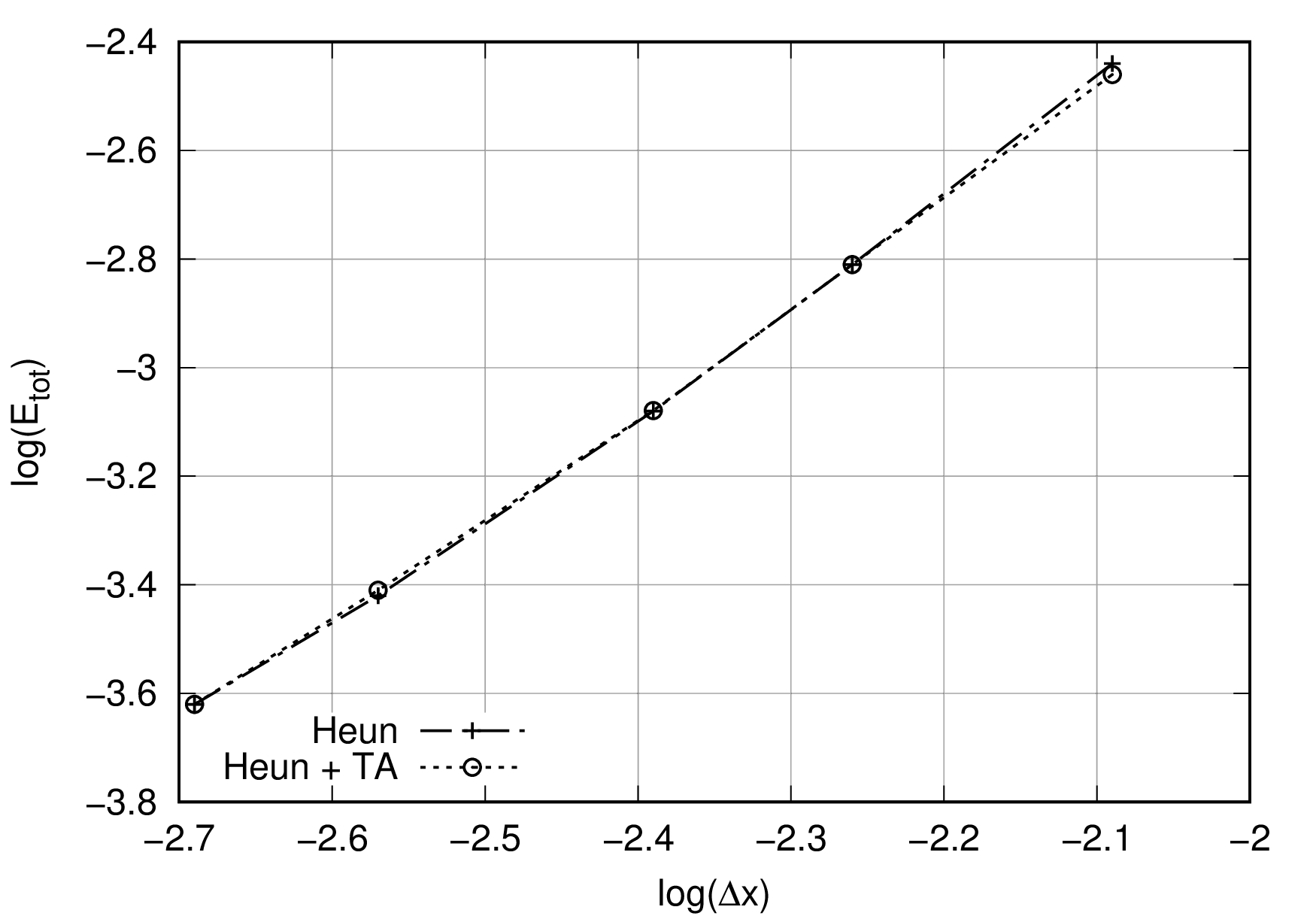}
\caption{Total error performed with second-order spatial schemes
\label{fig:order_accuracy}}
\end{center}
\end{figure}
\begin{table}[!ht]
\begin{center}
\begin{tabular}{|l|c|}
  \hline
  Time integrator & $E_{tot}$ slope \\
  \hline
  Heun & 1.97  \\
  Heun + TA & 1.93  \\
  \hline
\end{tabular}
\caption{Total error slopes \label{tab:order_tab} }
\end{center}
\end{table}

\subsection{Space-Time von Neumann Analysis}
\label{sec:VonNeumannHeun}


The one-dimensional linear advection equation is considered again in this section:
\begin{equation}
\begin{aligned}
\frac{\partial y}{\partial t} +c \frac{\partial y}{\partial x} =0,
\end{aligned}
\label{conv}
\end{equation}
with a harmonic initial condition and with periodic boundary conditions,
\begin{equation}
\left\{
\begin{array}{ll}
y(x, 0) = \exp(ikx) & x \in [0, L] \\
\vspace*{-3mm}\\
y(0, t) = y(L, t) & t \in R^+ \\
\end{array}
\right.
\end{equation}
with $i^2=-1$ and $c>0$. $k$ is the wavenumber of the initial solution. The problem admits an exact solution,
and the comparison between the exact theoretical solution and the numerical approximation gives
information on both dissipation and dispersion of the fully discrete scheme.
The fully discrete relation obtained from
Eq.~\eqref{conv} using a second-order finite volume scheme can be written in the following form:
\begin{equation}
\begin{aligned}
y_j^{n+1}{}=G_j \, y_j^{n},\label{eq:defGj}
\end{aligned}
\end{equation}
where $y_j^n$ represents the harmonic (averaged) solution at discrete time $n$ and in the cell $j$.
The complex coefficient $G_j$ represents the transfer function between two consecutive time solutions.
The coefficient $G_j$ can be expressed as $G_j = |G_j| \, \exp(i \arg(G_j))$.
In the following, $\mu_j = |G_j|$ is the dissipation coefficient, and $\phi_j = arg(G_j)/\hbox{CFL}$ the dispersion coefficient.
Of course, the definition of $G_j$ in the cell $j$ depends on the CFL value,
on the class rank and on the spatial accuracy used for the high order definition of the flux (linear extrapolation for second-order scheme).
For the sake of clarity, the dependency on all the variables
will be omitted.

\begin{figure}[h]
\begin{center}
\begin{tikzpicture}[scale=4.5]
    \draw [->,black] (-1.05,0.) -- (1,0.);
    \draw (1.1,0) node {$x$} ;
    \draw [black] (0.8,-0.02) -- (0.8, 0.02);
    \fill (canvas cs:x=-0.7cm,y=0cm) circle (0.2mm);
    \draw [black] (-0.8,-0.02) -- (-0.8, 0.02);
    \fill (canvas cs:x=0.7cm,y=0cm) circle (0.2mm);
    \draw [black] (0.6,-0.02) -- (0.6, 0.02);
    \fill (canvas cs:x=-0.5cm,y=0cm) circle (0.2mm);
    \draw [black] (-0.6,-0.02) -- (-0.6, 0.02);
    \fill (canvas cs:x=0.5cm,y=0cm) circle (0.2mm);
    \draw [black] (0.4,-0.02) -- (0.4, 0.02);
    \fill (canvas cs:x=0.3cm,y=0cm) circle (0.2mm);
    \draw [black] (0.2,-0.02) -- (0.2, 0.02);
    \fill (canvas cs:x=0.1cm,y=0cm) circle (0.2mm);
    \draw [black] (0. ,-0.02) -- (0. , 0.02);
    \fill (canvas cs:x=-0.1cm,y=0cm) circle (0.2mm);
    \draw [black] (-0.2 ,-0.02) -- (-0.2 , 0.02);
    \fill (canvas cs:x=-0.3cm,y=0cm) circle (0.2mm);
    \draw [black] (-0.4 ,-0.02) -- (-0.4 , 0.02);
    \draw[->,black] (-1,-0.05) -- (-1.0,0.5);
    \draw (-1.0,0.6) node {$t$} ;
    \draw [dashed] (-1.04,0.2) -- (0.9,0.2);
    \draw (-1.15,0.2) node {$\Delta t$} ;
    \draw [dashed] (-1.04,0.4) -- (0.9,0.4);
    \draw (-1.15,0.4) node {$2\Delta t$} ;
    \draw [dotted] (0.4,0.7) -- (0.4,-0.3);
    \draw [dotted] (-0.4,0.7) -- (-0.4,-0.3);
    \draw (-0.,-0.2) node {$\text{cells of class 0}$} ;
    \draw (-0.7,-0.2) node {$\text{cells of class 1}$} ;
    \draw (0.7,-0.2) node {$\text{cells of class 1}$} ;
    \draw (-0.4,0.8) node {$\text{step DOWN}$} ;
    \draw (0.4,0.8) node {$\text{step UP}$} ;
    \draw (0.3,-0.07) node {$\beta$} ;
    \draw (-0.3,-0.07) node {$\gamma$} ;
\end{tikzpicture}
\caption{Scheme of the whole domain configuration}
\label{fig:domainconfing}
\end{center}
\end{figure}
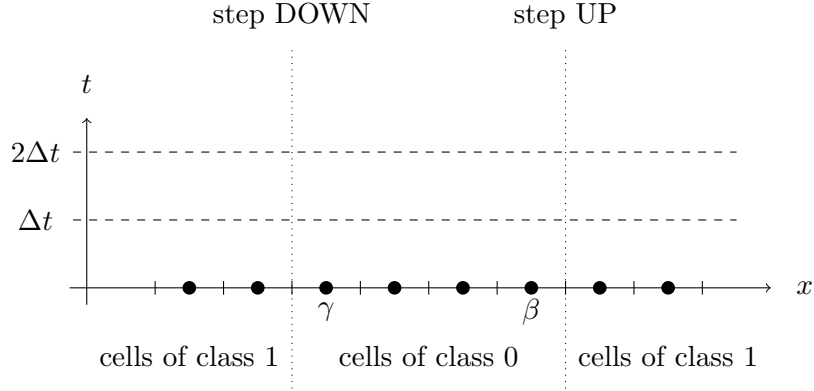

As presented in the previous section, the temporal adaptive method involves sub-cycling of time integrations.
Thus the space-time spectral analysis needs to be performed on several time steps.
In the following, it is performed on two time steps for two classes of cells.
With $c>0$, two configurations of time synchronisation between classes will be studied as
"step DOWN" and "step UP" (Fig.~\ref{fig:domainconfing}).
For this theoretical analysis the cell size will be fixed to $\Delta x$ in the entire numerical domain.
In case of irregular mesh a space-time spectral analysis may be non-trivial~\cite{Vichnevetsky_MCS_23_1981}.
In addition, a numerical analysis of wave packet propagation will be presented in the test case section~\ref{sec:Validation}
in order to study the effect of an irregular grid for our time integration setting.

{\bf Remark:} For a time step $\Delta t$ corresponding to $\mbox{CFL}=0.5$ in cells of class rank $0$,
the time step $2\Delta t$ corresponds to $\mbox{CFL}=1$ in cells of class rank $1$ since $\Delta x$ is constant in the whole domain.
Then the time step $\Delta t$ must be chosen so that $\mbox{CFL} < 0.5$ in order to remain below the stability condition in class rank $1$.

In the following spectral analysis, the domain configuration shown in Fig.~\ref{fig:domainconfing}
is considered with $N=300$ cells.

\subsubsection{Transfer function for Heun's scheme, far from class transition}

The Heun time integration at cell $\beta-10$ (far from class transition) from $t = 0$ to $t = 2\Delta t$ is performed such that
\begin{equation}
\begin{aligned}
y^{\Delta t}_{\beta-10}&=y^{0}_{\beta-10}-{\Delta t} \bigg[ \frac{1}{2}\big( R^0+\widehat{R^0}\big)     \bigg] \\
y^{2\Delta t}_{\beta-10}&=y^{\Delta t}_{\beta-10}-{\Delta t}\bigg[ \frac{1}{2}\big( R^{\Delta t}+\widehat{R}^{\Delta t}\big)     \bigg]
\end{aligned}
\end{equation}
and then the following finite volume formulation is obtained
\begin{equation}
\begin{aligned}
y^{2\Delta t}_{\beta-10}=y^{\Delta t}_{\beta-10}-\frac{\Delta t}{\Delta x} \bigg[ \frac{1}{2}\big( F_{\beta-\frac{19}{2}}(\Delta t,y^{\Delta t})+F_{\beta-\frac{19}{2}}(2\Delta t,\widehat{y}^{\Delta t} ) \big)-\frac{1}{2}\big( F_{\beta-\frac{21}{2}}(\Delta t,y^{\Delta t})+F_{\beta-\frac{21}{2}}(2\Delta t,\widehat{y}^{\Delta t} ) \big)     \bigg]
\end{aligned}
\label{Heun*2}
\end{equation}

\subsubsection{Analysis for the cell $\beta$ near step UP}

The finite volume formulation of the state $y^{2\Delta t}_{\beta}$ at the cell $\beta$ near the step UP is different:
\begin{equation}
\begin{aligned}
y^{2\Delta t}_{\beta}&=y^{\Delta t}_{\beta}-\frac{\Delta t}{\Delta x} \bigg[ \frac{1}{2}\big( F_{\beta+\frac{1}{2}}(\Delta t,y^{\Delta t})+F_{\beta+\frac{1}{2}}(2\Delta t,\widehat{y}^{2\Delta t} ) \big)-\frac{1}{2}\big( F_{\beta-\frac{1}{2}}(\Delta t,y^{\Delta t})+F_{\beta-\frac{1}{2}}(2\Delta t,\widehat{y}^{2\Delta t} ) \big)     \bigg] .
\end{aligned}
\label{stepup_1}
\end{equation}
Considering the MUSCL reconstruction for a simple advection from left to right in one dimension, the linear
extrapolation \GP{of the unknown used} for the flux computation reads:
\begin{equation}
\begin{aligned}
F(t,y )=c~y^L=c~\big( y + \frac{\Delta x}{2}\nabla y \big)
\end{aligned}
\label{MUSCL_1D}
\end{equation}
Then, by substitution in Eq.~\eqref{stepup_1},
\begin{equation}
\begin{aligned}
y^{2\Delta t}_{\beta}=y^{\Delta t}_{\beta}-\frac{c \Delta t}{\Delta x} \bigg[ &\frac{1}{2}\bigg( \frac{1}{2} \big( y_\beta^0 + \frac{\Delta x}{2}(\nabla y)^0_\beta \big)+\frac{1}{2} \big( \widehat{y}_\beta^{2\Delta t} + \frac{\Delta x}{2}(\widehat{\nabla y})^{2\Delta t}_\beta \big) \bigg)+\frac{1}{2}\big( \widehat{y}_\beta^{2\Delta t} + \frac{\Delta x}{2}(\widehat{\nabla y})^{2\Delta t}_\beta \big) \\
&- \frac{1}{2} \big( y_{\beta-1}^{\Delta t} + \frac{\Delta x}{2}(\nabla y)^{\Delta t}_{\beta-1} \big)-\frac{1}{2} \big( \widehat{y}_{\beta-1}^{2\Delta t} + \frac{\Delta x}{2}(\widehat{\nabla y})^{2\Delta t}_{\beta-1} \big)    \bigg]
\end{aligned}
\label{stepup}
\end{equation}
Here the terms $\widehat{y}_{\beta}^{2\Delta t}$ and $(\widehat{\nabla y})^{2\Delta t}_{\beta}$,
may be source of instability because of the violation of CFL condition for these cells.

\subsubsection{Analysis for the cell $\gamma$ near step DOWN}

In the configuration of the step DOWN, the time integrations of the state $y$ at the cell $\gamma$ are formulated as:
\begin{equation}
\begin{aligned}
y^{2\Delta t}_{\gamma}&=y^{\Delta t}_{\gamma}-\frac{\Delta t}{\Delta x} \bigg[ \frac{1}{2}\big( F_{\gamma+\frac{1}{2}}(\Delta t,y^{\Delta t})+F_{\gamma+\frac{1}{2}}(2\Delta t,\widehat{y}^{\Delta t} ) \big)-\frac{1}{2}\big( F_{\gamma-\frac{1}{2}}(\Delta t,y^{\Delta t})+F_{\gamma-\frac{1}{2}}(2\Delta t,\widehat{y}^{2\Delta t} ) \big)     \bigg] \\
y^{2\Delta t}_{\gamma}&=y^{\Delta t}_{\gamma}-\frac{c \Delta t}{\Delta x} \bigg[\frac{1}{2} \big( y_{\gamma}^{\Delta t} + \frac{\Delta x}{2}(\nabla y)^{\Delta t}_{\gamma} \big)+\frac{1}{2} \big( \widehat{y}_{\gamma}^{2\Delta t} + \frac{\Delta x}{2}(\widehat{\nabla y})^{2\Delta t}_{\gamma} \big) \\
& \text{ }\text{ }\text{ }\text{ }\text{ }\text{ }\text{ }\text{ }\text{ }\text{ }\text{ }\text{ }\text{ }\text{ }\text{ }\text{ }\text{ }\text{ }\text{ }-\frac{1}{2}\bigg( \frac{1}{2} \big( y_{\gamma-1}^0 + \frac{\Delta x}{2}(\nabla y)^0_{\gamma-1} \big)+\frac{1}{2} \big( \widehat{y}_{\gamma-1}^{2\Delta t} + \frac{\Delta x}{2}(\widehat{\nabla y})^{2\Delta t}_{\gamma-1} \big) \bigg)-\frac{1}{2}\big( \widehat{y}_{\gamma-1}^{2\Delta t} + \frac{\Delta x}{2}(\widehat{\nabla y})^{2\Delta t}_{\gamma-1} \big)   \bigg]
\end{aligned}
\label{stepdown}
\end{equation}

\subsubsection{Conclusion on the analysis}

To obtain the amplification factor $G$ between $y^{2\Delta t}$ and $y^0$,
all the terms in Eqs.~\eqref{Heun*2}, \eqref{stepup}, and~\eqref{stepdown} need to be expressed as a transfer function from $y^0$:
\begin{equation}
\begin{aligned}
y^{\Delta t}=G_1y^{0} \\
(\widehat{\nabla y})^{2\Delta t}=G_2 y^0 \\
(\nabla y)^0=G_3 y^0 \\
\widehat{y}^{2\Delta t} =G_4 y^0 \\
  ...\\
\end{aligned}
\end{equation}

\begin{figure}[!htbp]\begin{center}
\begin{tabular}{cc}
\includegraphics[width=8cm]{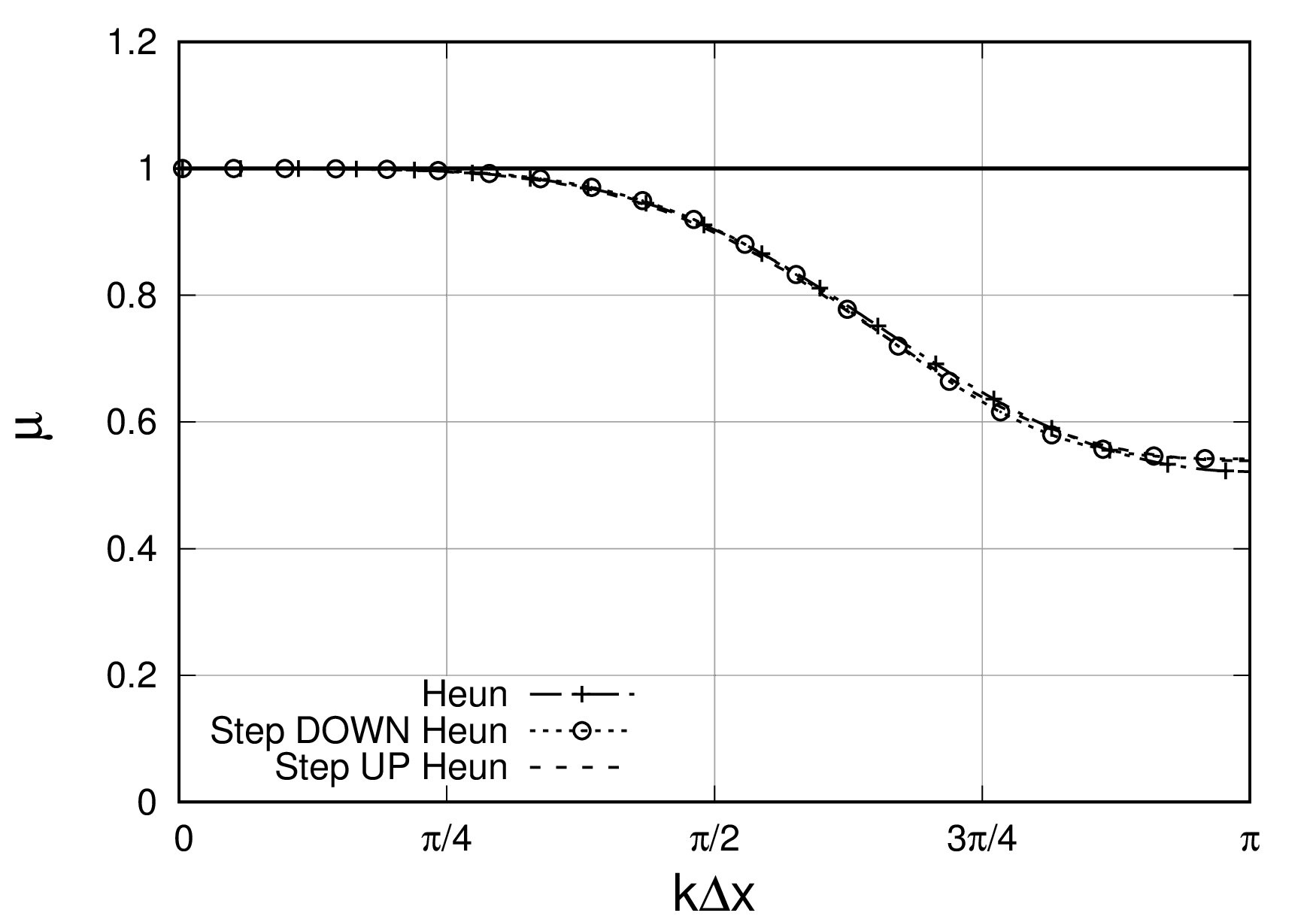} & \includegraphics [width=8cm]{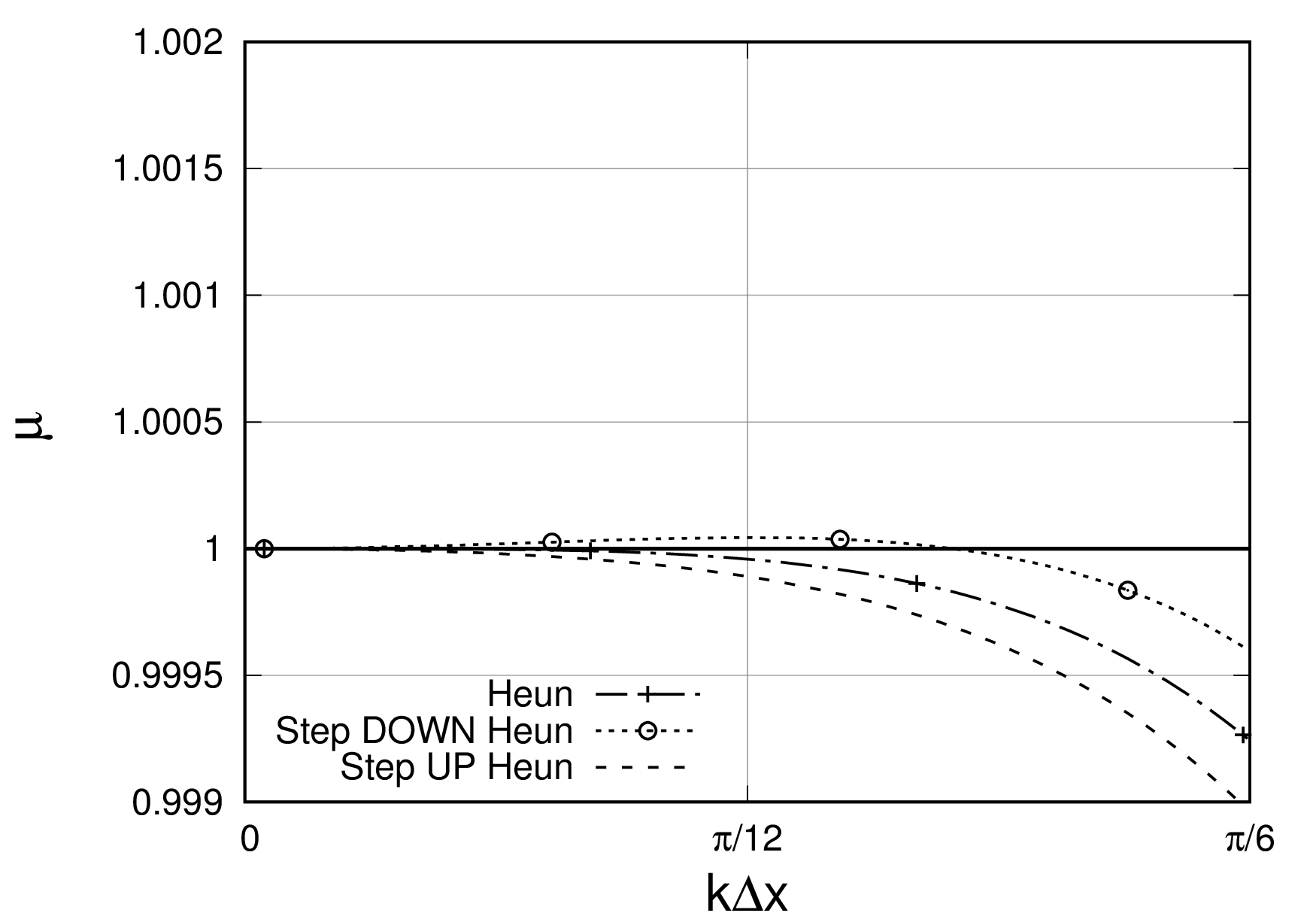}\\
\end{tabular}
\includegraphics [width=8cm]{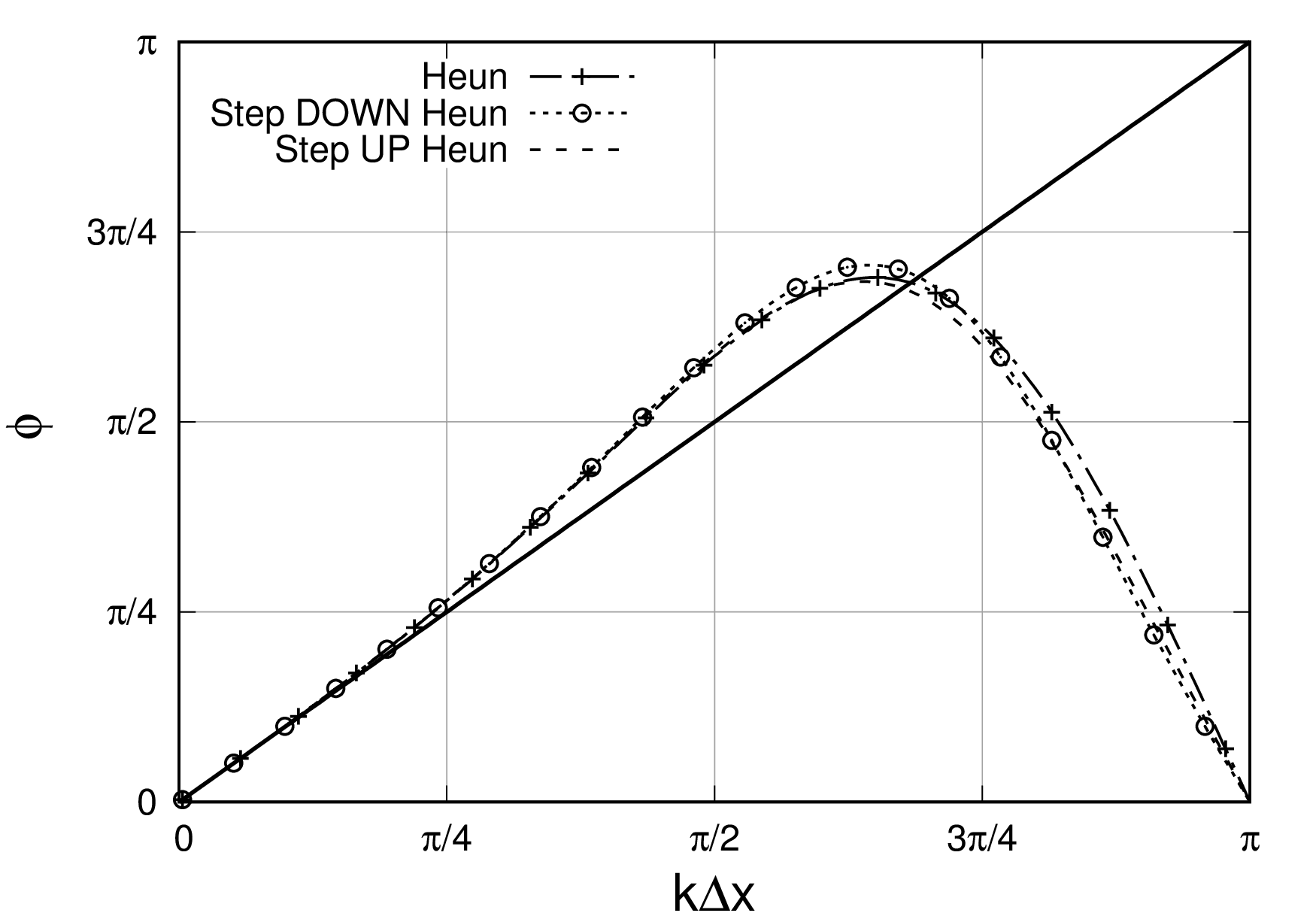}
\caption{Dissipation $\mu$ and dispersion $\phi$ of the Heun scheme with temporal adaptive approach at CFL=0.1
\label{fig:dissiptempHeunFig01_test}}
\end{center}
\end{figure}




\begin{figure}[!htbp]\begin{center}
\begin{tabular}{cc}
\includegraphics[width=8cm]{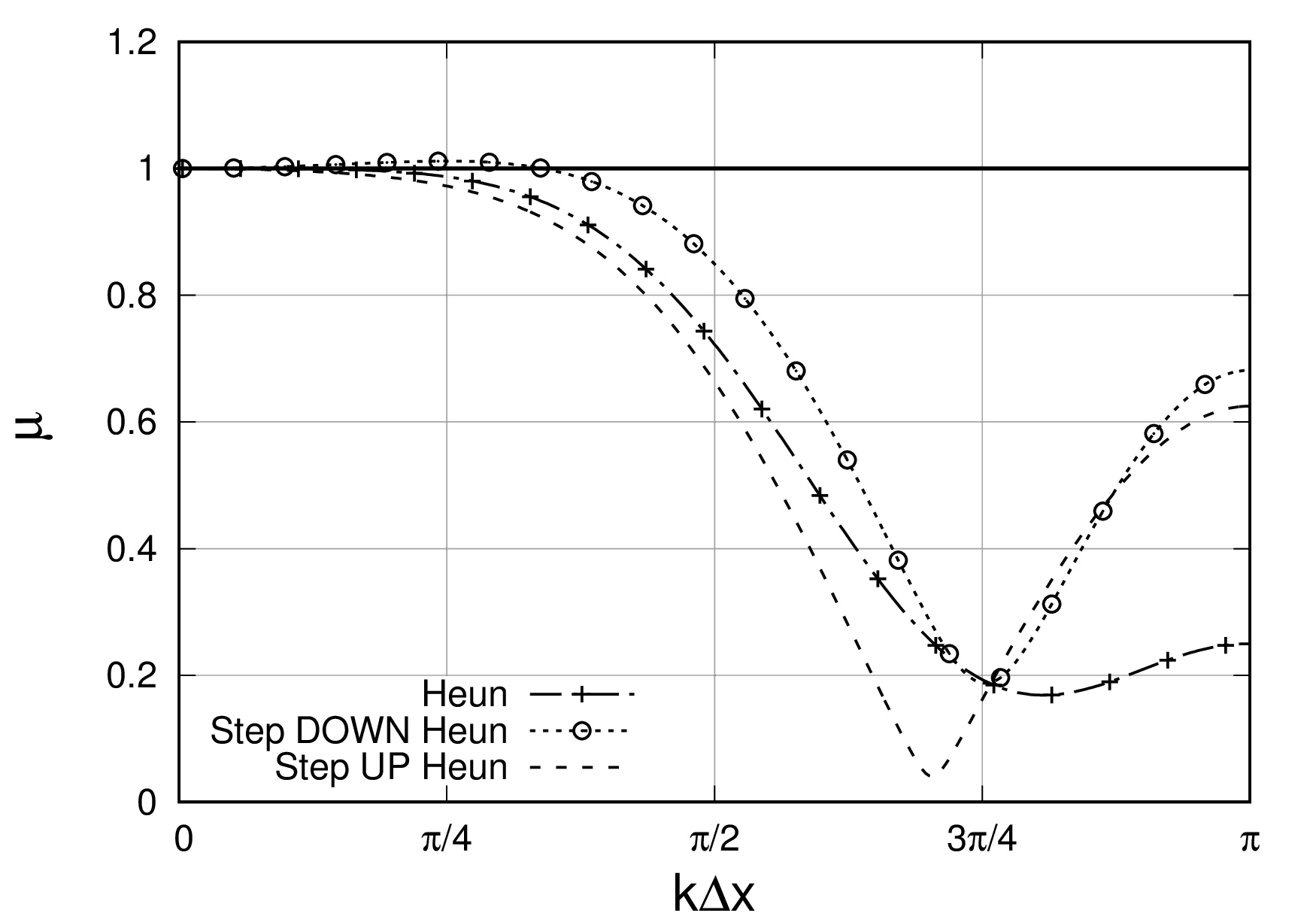} & \includegraphics [width=8cm]{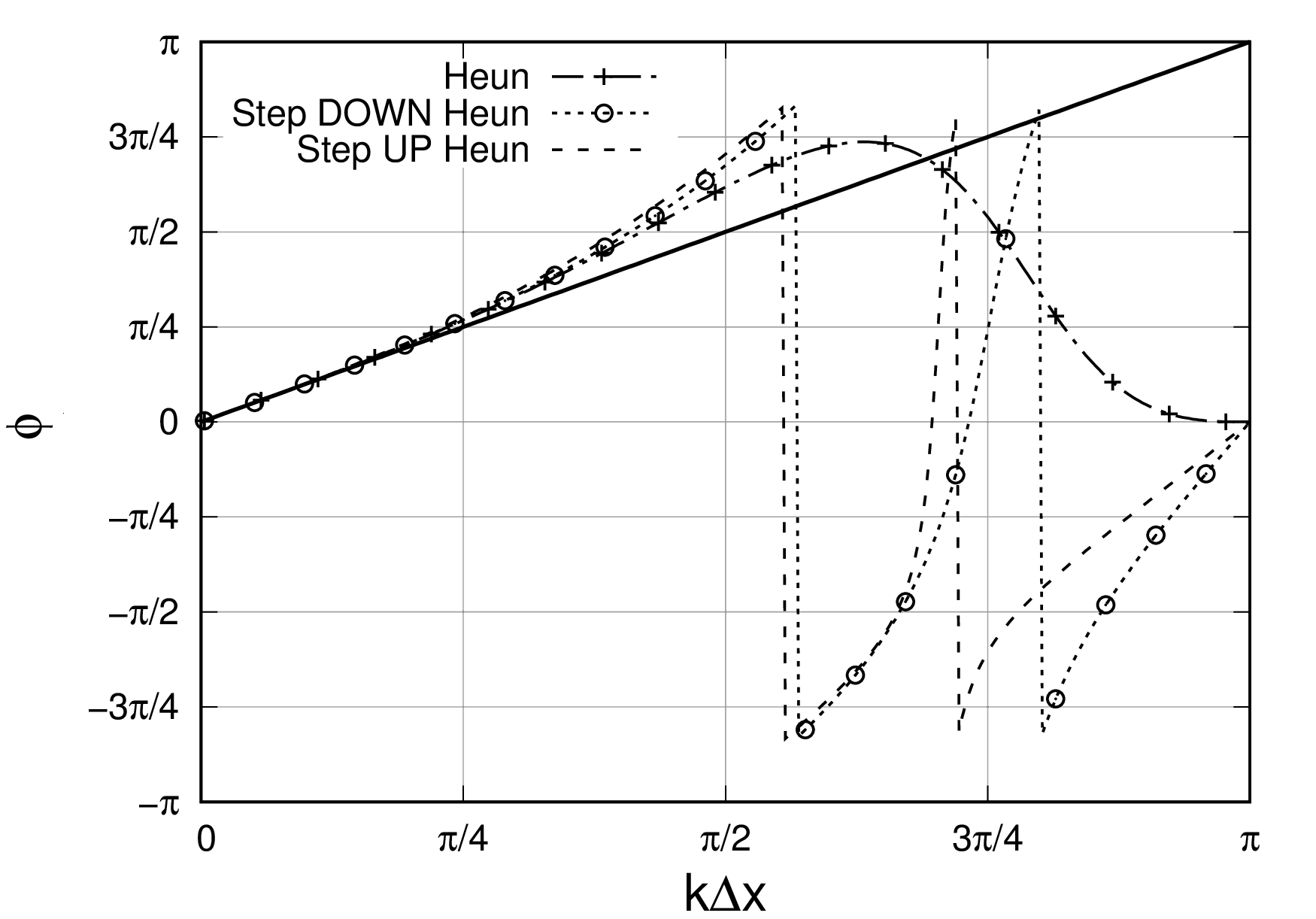}\\
\end{tabular}
\caption{Dissipation $\mu$ and dispersion $\phi$ of the Heun scheme with temporal adaptive approach at CFL=0.3
\label{fig:dissiptempHeunFig03_test}}
\end{center}
\end{figure}



\begin{figure}[!htbp]\begin{center}
\begin{tabular}{cc}
\includegraphics[width=8cm]{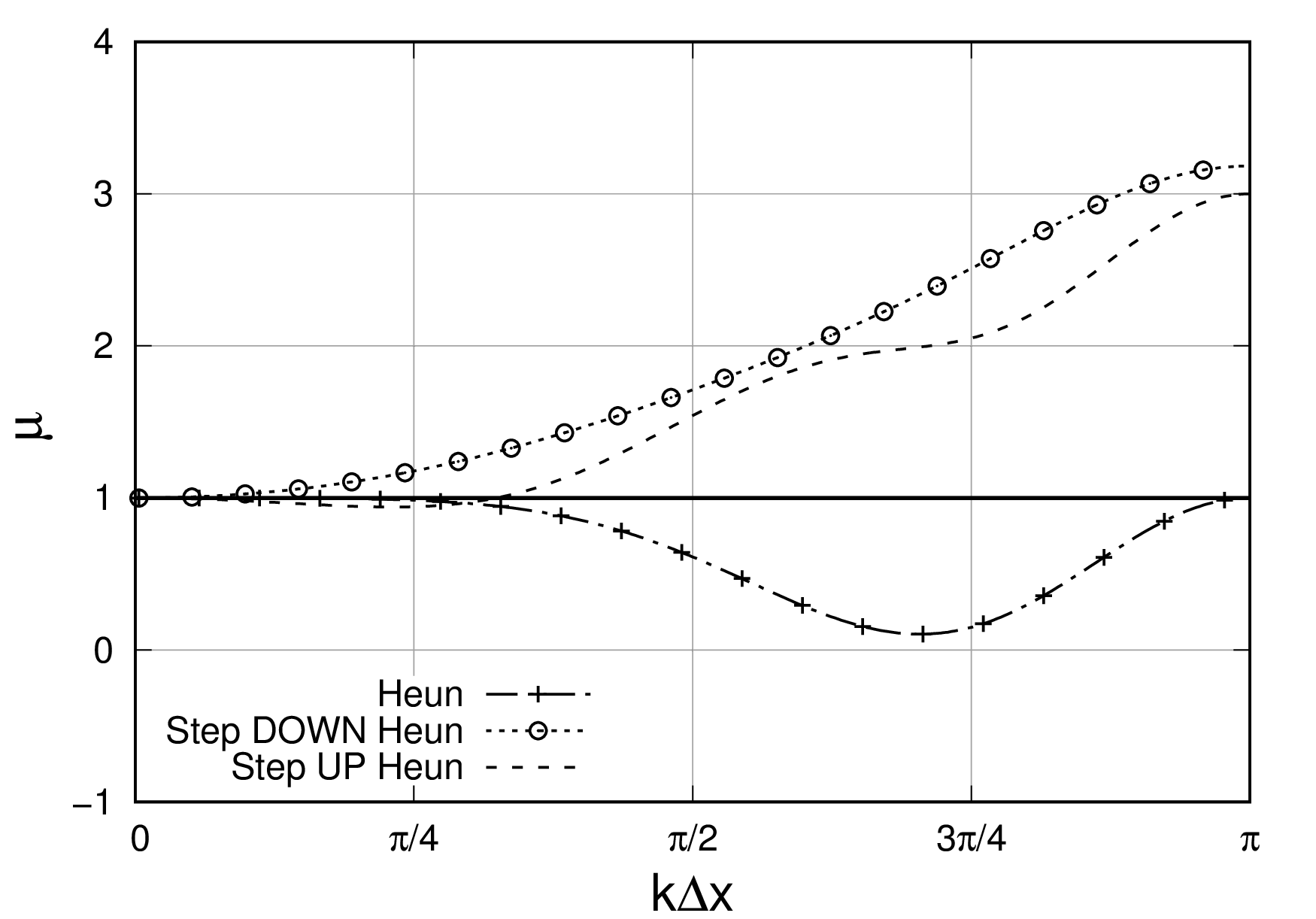} & \includegraphics [width=8cm]{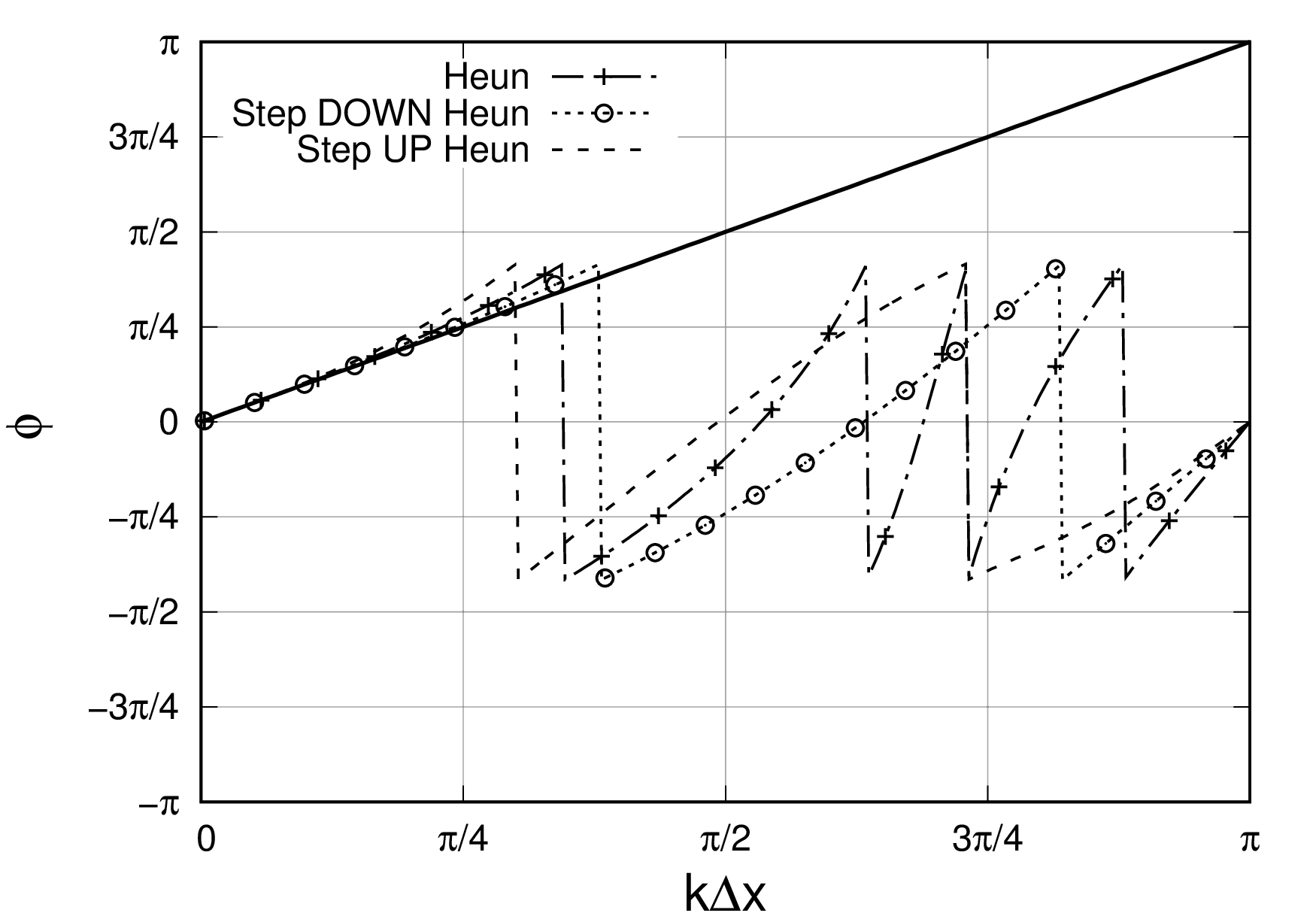}\\
\end{tabular}
\caption{Dissipation $\mu$ and dispersion $\phi$ of the Heun scheme with temporal adaptive approach at CFL=0.6
\label{fig:dissiptempHeunFig06_test}}
\end{center}
\end{figure}



For the step UP configuration, Figs.~\ref{fig:dissiptempHeunFig01_test}
and~\ref{fig:dissiptempHeunFig03_test} show that amplification
is never created. For the step DOWN configuration, it appears that the terms
$\widehat{y}_{\gamma-1}^{2\Delta t}$ and $(\widehat{\nabla y})^{2\Delta t}_{\gamma-1}$ can
create amplification as shown in Figs.~\ref{fig:dissiptempHeunFig01_test}
and~\ref{fig:dissiptempHeunFig03_test}.

At CFL$=0.6$, Fig.~\ref{fig:dissiptempHeunFig06_test} shows that both step UP and DOWN configurations present strong amplification due to the fixed
grid size $\Delta x$ (cf. Remark of Sec.~\ref{sec:VonNeumannHeun}).
It is important to remember that amplification in a certain range of wave numbers does not necessarily imply instability.
Nevertheless, this type of configuration is not allowed to appear in a large part of the computational domain.
\GP{In particular, if several classes are located closely, a wave can be amplified at any step DOWN: CFL below 0.6 is for sure a
good solution.}
%
%
Concerning dispersion behaviour, Figs.~\ref{fig:dissiptempHeunFig01_test} and ~\ref{fig:dissiptempHeunFig06_test} reveal
that the time synchronisation does no affect the numerical speed of
the advected waves for low wave numbers: the dispersion of Heun's scheme is recovered. However,
taking again Heun's scheme as a reference, Fig.~\ref{fig:dissiptempHeunFig03_test} shows that
dispersion is changed for wavenumbers above $\frac{\pi}{2\Delta x}$.

\subsection{Analysis of $q-$waves}\label{sec:qwavesHeun}

For an one-dimensional advection equation at constant speed $c$, any information is transported 
at the velocity $c$ and this result is at the core of the theory for (linear) hyperbolic equations. 
Numerically, numerical parasite waves propagating in the opposite direction of $c$ can be encountered 
and they are referred as $q-$waves. In addition, a wave propagating in the same direction as $c$ is called $p-$wave.
The existence of these $q-$waves was first highlighted by Vichnevetsky~\cite{Vichnevetsky_1982_book}, and then by many 
other authors, such as Poinsot and Veynante for combustion~\cite{Poinsot_2005_book} or Trefethen~\cite{Trefethen_1992_SIAM}. The
existence of $q-$waves is not related to the spectral analysis of a numerical scheme.
Existence of $q-$waves is linked with numerical group velocity according to Sengupta {\it et al.}~\cite{Sengupta_2012_AMC}. In this context, new quantities related to Eq.~\eqref{conv} are introduced according
to the definition proposed in~\cite{Sengupta_2012_AMC}.
The analysis of $q$-waves for the temporal adaptive method is performed with
the time step $2\Delta t$ according to the configuration presented in Sec.~\ref{sec:VonNeumannHeun}.
The numerical phase speed is defined as
 \begin{equation}
\begin{aligned}
c^N=\frac{arg(G)}{2\Delta t \,k},\label{eq:def_cn}
\end{aligned}
\end{equation}
from which is deduced the numerical group velocity,
\begin{equation}
\begin{aligned}
V^{gN}=\frac{d \big(arg(G)\big)}{d k} \frac{c}{\hbox{CFL}\,\Delta x}.\label{eq:def_vgn}
\end{aligned}
\end{equation}
For an advection equation with a positive advection velocity $c>0$ and information going downstream,
$q-$waves propagate upstream with a negative group velocity $V^{gN}$~\cite{Sengupta_2012_AMC}. The
negative group velocity corresponds to a necessary condition to make $q-$waves appear and propagate, but it is not a
sufficient condition. Moreover, the effect of $q-$waves also depends on the amplification factor of the
space-time discretization. Indeed with an excessive dissipation or with filtering, the $q$-waves can
be damped efficiently. In addition,
initial condition, grid resolution and multi-dimensional case have effects on
$q$-waves~\cite{Sengupta_2012_AMC}. Here, the analysis is restricted to the one-dimensional case. 

\begin{figure}[!htbp]\begin{center}
\begin{tabular}{cc}
\includegraphics[width=8cm]{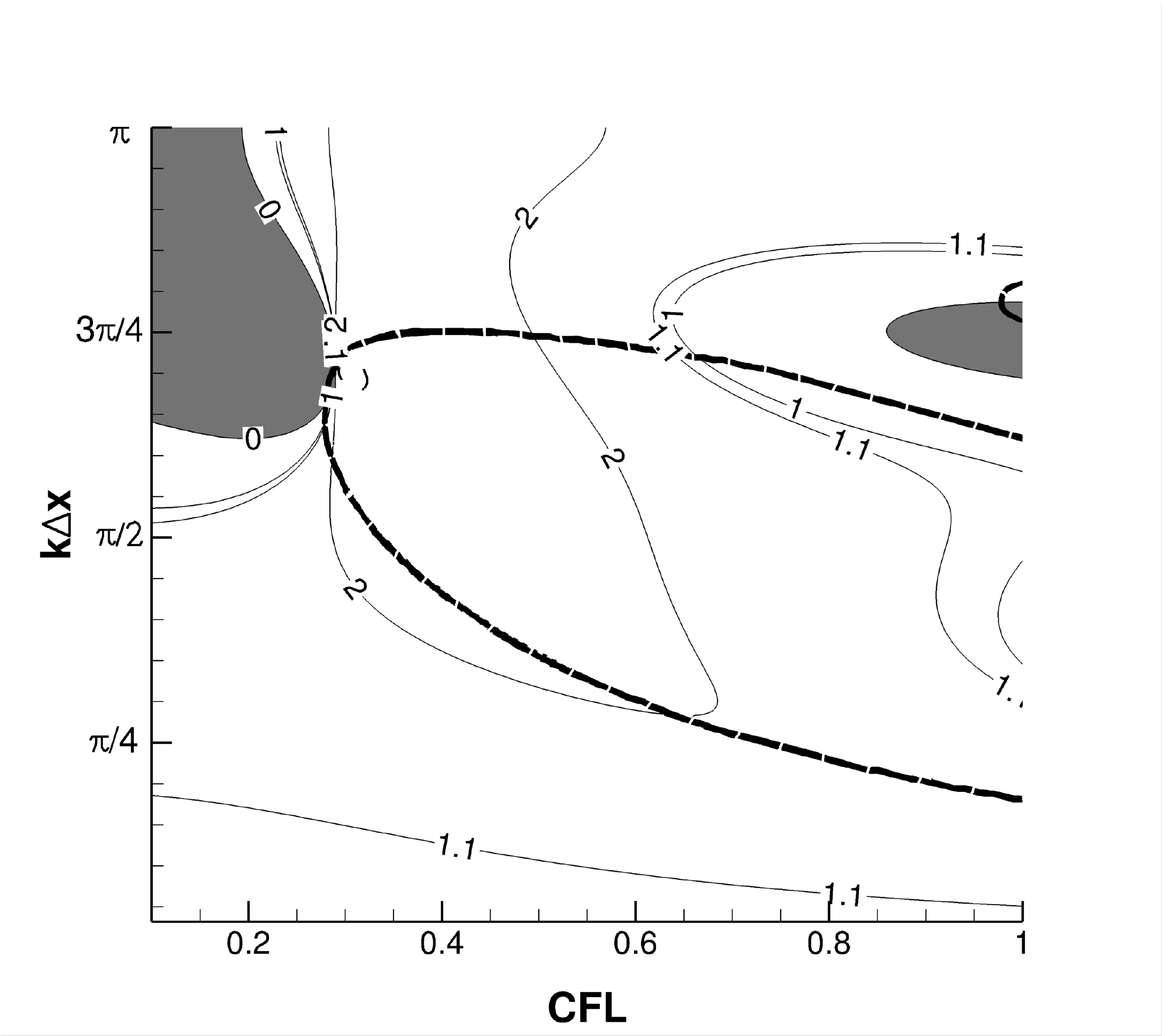} & \includegraphics [width=8cm]{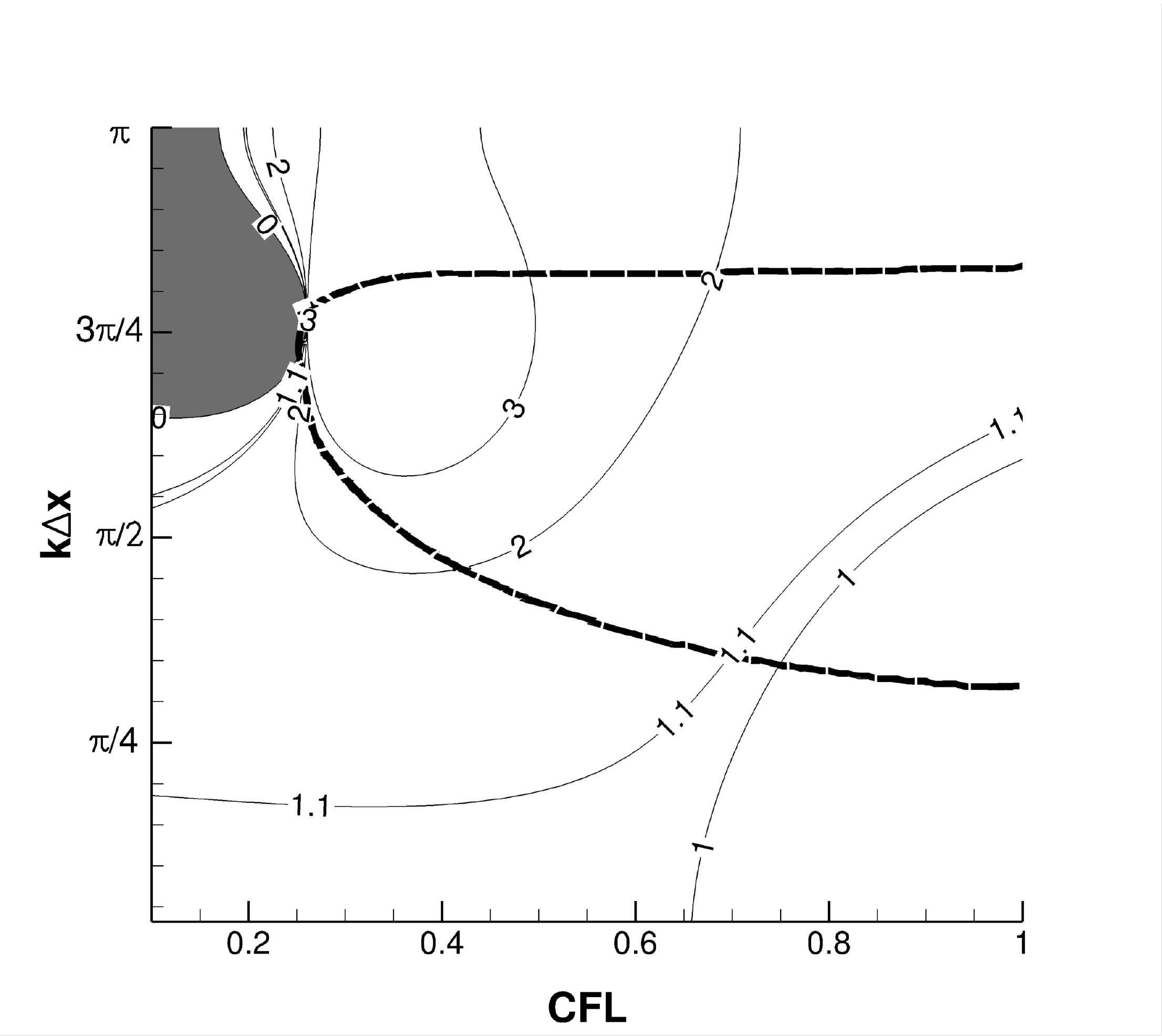}\\
\end{tabular}
\caption{Isocontours of $V^{gN}$ for cell $\beta$ (left - step UP) and $\gamma$ (right - step DOWN)\label{fig:qwavesHeunUP_DOWN}}
\end{center}
\end{figure}

Both configurations (steps UP and DOWN) of temporal adaptation coupled with Heun's time
integration are taken into account, as in Sec.~\ref{sec:VonNeumannHeun}. The numerical phase speed $c^N$ and
the numerical group velocity $V^{gN}$ are computed from $G$ (Eq.~\eqref{eq:defGj}) for cells
$\beta$ and $\gamma$. The analysis of $q-$waves is performed for CFL $\in [0,1]$.
In Figs.~\ref{fig:qwavesHeunUP_DOWN}, dashed curves correspond to $|V^{gN}|/|c|>>1$,
which represents discontinuity of
dispersion $\phi$, and grey zones correspond to negative group velocity, $V^{gN}<0$.

{\bf Remark:}
The CFL domain $[0, 1]$ is larger than the stable CFL domain, and amplification always occurs
for CFL$>0.6$. However, the analysis can still be performed for any value of the CFL number.

\GP{Steps UP and DOWN lead to equivalent areas of negative-$V^{gN}$ group velocity (Fig.~\ref{fig:qwavesHeunUP_DOWN})
and the shape of iso contours are also almost the same for CFL $\in [0,0.8]$. }
%
In the step UP case (cell $\beta$), another negative-$V^{gN}$ group velocity zone is present for CFL$>0.9$.

\begin{figure}[!htbp]\begin{center}
\begin{tabular}{cc}
\includegraphics[width=8cm]{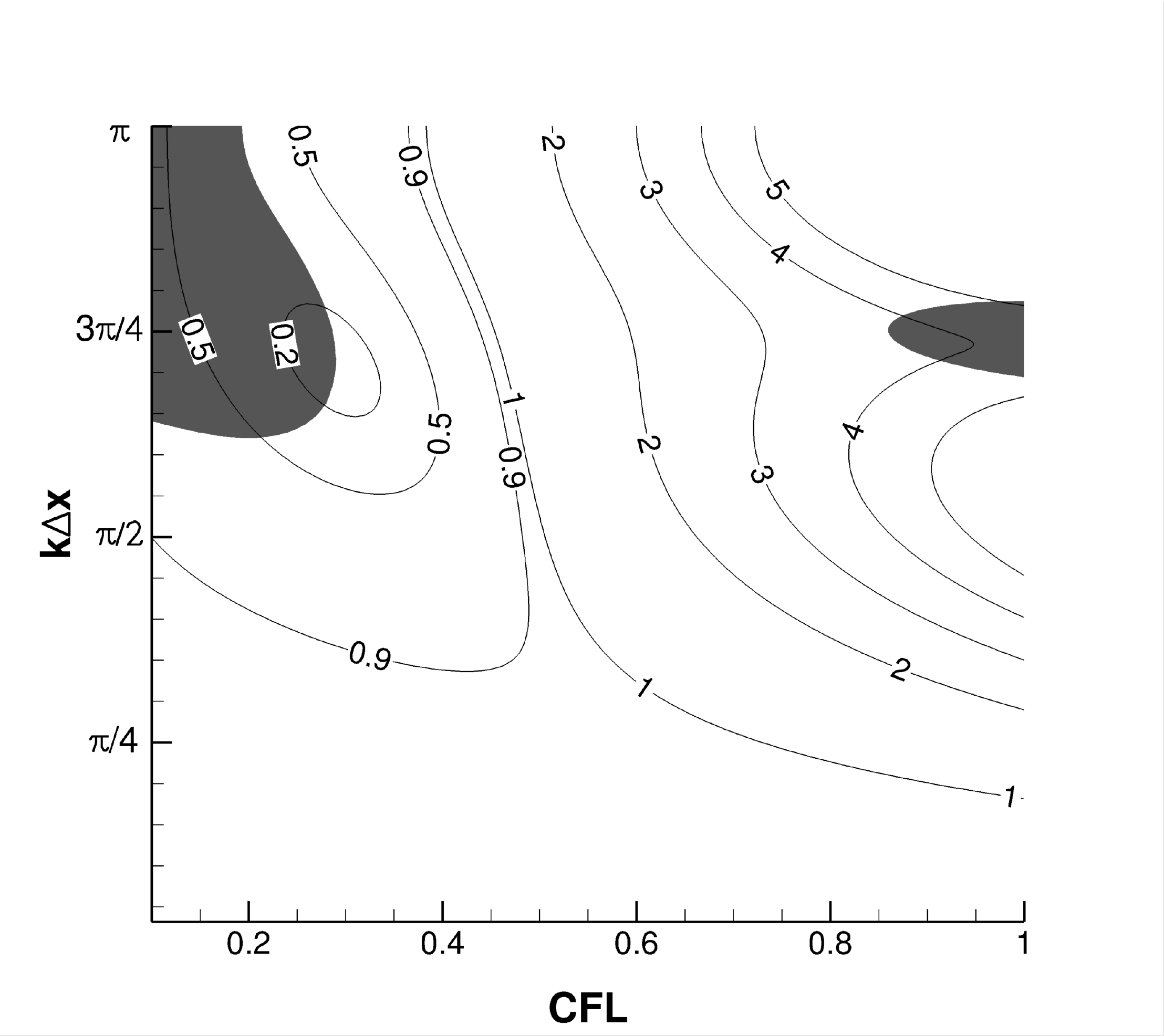} & \includegraphics [width=8cm]{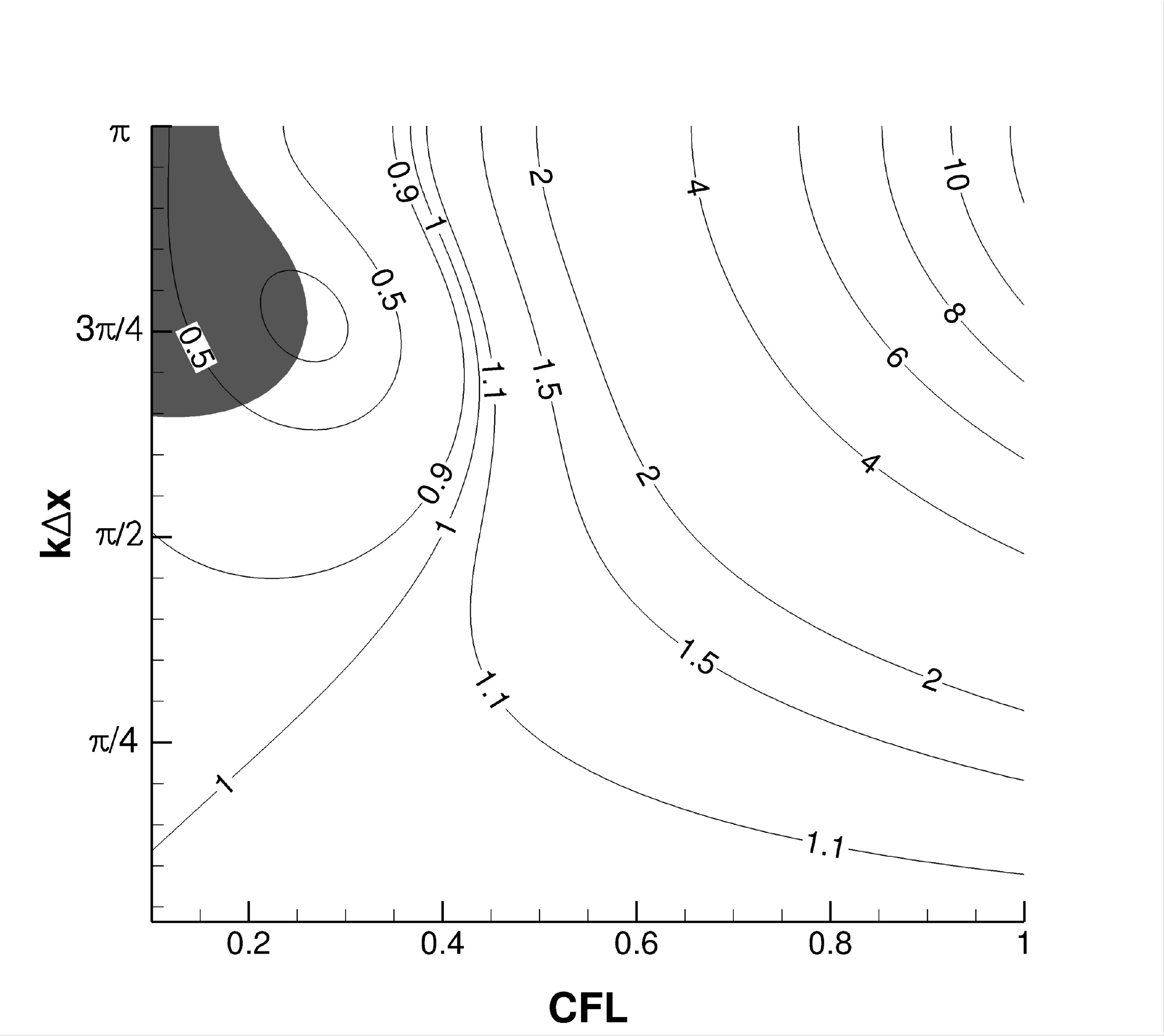}\\
\end{tabular}
\caption{ Isocontours of the dissipation $\mu_{\beta}$ and $\mu_{\gamma}$ with grey zone for negative group velocity\label{fig:qwaves_absG_HeunUP_DOWN}}
\end{center}
\end{figure}

The last question concerns the capability of the temporal adaptive Heun scheme to damp $q-$waves.
To answer it, it is mandatory to couple the observations on negative group velocity with the dissipation property of the time integration.
\GP{As mentioned in the remark of Sec.\ref{sec:VonNeumannHeun}, the numerical scheme is unstable for CFL$>0.5$ and
the negative-$V^{gN}$ zone found for CFL$>0.5$ (Fig.~\ref{fig:qwaves_absG_HeunUP_DOWN}) matches with the dissipation factor
$\mu_{\beta}>3$, leading to amplification. Nevertheless, the zones of negative-$V^{gN}$ for CFL$<0.5$ corresponds to $\mu<0.5$
and the wave is partially dissipated in one iteration.}


\subsection{Partial conclusion}

In this section, the numerical analysis of the proposed time-adaptive version of Heun's scheme was analysed,
focusing on dissipation, dispersion, accuracy and the possibility to encounter $q-$waves.

The next section is dedicated to the very same analysis for our AION scheme that blends
the scheme of Crank-Nicolson with the second-order implicit Runge-Kutta integrator.

\section{Temporal adaptive method with AION scheme \label{sec:TA_AION}}

The standard AION time integration coupled with a 1-exact spatial scheme was shown to be second-order accurate both in
space and time~\cite{Muscat_JCP_XX_2018}. The AION time integrator is built using three underlying schemes applied either in the explicit, implicit or hybrid regions of the mesh.
 The time step is \textit{de facto} limited by the CFL condition of the smallest explicit cells.
For explicit cells, this time integrator was also designed to match with the predictor-corrector formulation
of the standard Heun's scheme. Indeed, it was highlighted in Sec.~\ref{sec:TA_Heun} that the
predictor-corrector formulation is useful for time synchronisation between temporal classes.
Hence, the AION time integration may be coupled with the standard temporal adaptive approach.
In the latter case, the implicit and hybrid cells may belong to the temporal class of highest rank with the
time step limited by the CFL condition of the largest explicit cells. In some cases, it is possible to obtain
configurations for which hybrid cells belonging to the highest rank temporal class share an interface with
explicit cells of lowest rank temporal classes (even though the approach forbids neighbour cells with a
difference in temporal class rank greater than $1$). Thus it is necessary to extend the time adaptive approach to the
hybrid scheme of AION time integrator.

In this section, the goal is to propose a way to couple the AION scheme with time adaptation. Time
adaptation should not be considered for the treatment of implicit cells since they are protected by the
design of the transition parameter. However, it will be mandatory to allow the time synchronisation of temporal
classes in hybrid and/or explicit cells. For $\omega=1$, the AION scheme reverts to Heun's scheme and
the very same time synchronisation can be chosen. The situation differs when synchronization occurs in
the hybrid part of AION scheme, where only one residual is computed (Eq.~\ref{eq:HCS2}).
In the following, the scheme for which time adaptation and synchronization occur in the hybrid part of the
AION scheme will be denoted AION+TA, while Heun+TA will indicate a time adaptation procedure with
synchronization in the explicit cells, as presented in Sec.~\ref{sec:TA_Heun}.

\subsection{Update of the Solution for hybrid time integration\label{sec:TA_method_AION}}

Let us consider again the one-dimensional example introduced in Sec.~\ref{sec:TA_method}. As before,
cells with class rank $1$ will undergo only two stages to reach
$t = 2 \Delta t$ thanks to the hybrid time integration:
\begin{itemize}
    \item[a-1. ] $\displaystyle \widehat{W}^{2 \omega \Delta t} = W^0 + 2\omega \Delta t R(W^0)$
    \item[b-1. ] $\displaystyle W^{2 \Delta t} = W^0 + {2 \Delta t} R^{Hybrid}(\widehat{W}^{2\omega \Delta t})$,
\end{itemize}
but cells with class rank $0$ will undergo four stages:
\begin{itemize}
    \item[a-0. ] $\displaystyle \widehat{W}^{\omega\Delta t} = W^0 + \omega\Delta t R(W^0)$
    \item[b-0. ] $\displaystyle W^{\Delta t} = W^0 + {\Delta t} R^{Hybrid}(\widehat{W}^{\omega\Delta t})$
    \item[c-0. ] $\displaystyle \widehat{W}^{2\omega \Delta t} = W^{\Delta t} + \omega \Delta t R(W^{\Delta t})$
    \item[d-0. ] $\displaystyle W^{2 \Delta t} = W^{\Delta t} + {\Delta t} R^{Hybrid}(\widehat{W}^{2 \omega \Delta t})$.
\end{itemize}
The different steps for the hybrid time integration with adaptation are quite similar with those
presented in Sec.~\ref{sec:TA_method} for the time update using the full explicit Heun's scheme.

\noindent $\bullet$ {\bf Steps 1} and {\bf 2}:\\ 
They areequivalent to those presented for the full explicit scheme in Sec.~\ref{sec:TA_method}\\
$\bullet$ {\bf Step 3:}\\
For the cells of class rank $0$ sharing a face with a cell of class rank $1$, the update of the solution associated with stage b-0
needs the definition of the flux at the interface $\beta+1/2$,
\begin{equation}
F_{\beta+1/2}(\widehat{W}^{\omega\Delta t})=\frac{1}{2} \big( F_{\beta+1/2}(W^0)+F^{Hybrid}_{\beta+1/2}(\widehat{W}^{2\omega\Delta t}) \big).\label{eq:step3_flux_hy}
\end{equation}
$\bullet$ {\bf Step 4:}\\
It is also equivalent to the one presented for the full explicit scheme in Sec.~\ref{sec:TA_method}. The parabolic interpolation of the residual is performed:
\begin{equation}
     R^{\Delta t}_{\beta+1}= \frac{1}{4} R^{Hybrid}_{\beta+1}(\widehat{W}^{2\omega \Delta t}) + \frac{3}{4} R_{\beta+1}(W^0).
\end{equation}
$\bullet$ {\bf Step 5:}\\
The approach to time integrate class rank 0 from $\Delta t$ to $2\Delta t$ is quite equivalent to the full explicit time integration.
Indeed the flux computed in Eq.~\eqref{eq:step3_flux_hy} is used to compute residual in cell $\beta$ for stages c-0 and d-0.
Nevertheless, the following time integration at cell $\beta$ from $\Delta t$ to $2\Delta t$ is chosen	:
\begin{equation}
{W}^{2\Delta t}_{\beta}={W}^{\Delta t}_{\beta}+\frac{\Delta t}{\Delta x} \bigg( \frac{3}{2}F^{Hybrid}_{\beta+\frac{1}{2}}(\widehat{W}^{2 \omega\Delta t})-\frac{1}{2}F_{\beta+\frac{1}{2}}(W^0)-F^{Hybrid}_{\beta-\frac{1}{2}}(\widehat{W}^{2 \omega\Delta t})\bigg).
\end{equation}
This modification with respect to the full explicit Heun's method is important to
ensure space-time conservation of flux from $0$ to $2 \Delta t$, as shown in Sec.~\ref{AION+TA_conservation}.


\section{Analysis of the conservation and the adaptive time accuracy \label{sec:accuracy+spectral_AION}}
\subsection{Conservation Property\label{AION+TA_conservation}}

According to the previous one dimensional configuration presented in Sec.~\ref{sec:TA_method_AION},
the time integration at cell $\beta+1$ of class rank $1$ is formulated as
\begin{equation}
\begin{aligned}
\widehat{W}^{2\omega\Delta t}_{\beta+1}&={W}^{0}_{\beta+1}+\frac{2\omega\Delta t}{\Delta x} \big( F_{\beta+\frac{3}{2}}(W^0)-F_{\beta+\frac{1}{2}}(W^0)\big) \\
{W}^{2\Delta t}_{\beta+1}&={W}^{0}_{\beta+1}+\frac{2\Delta t}{\Delta x} \big( F^{Hybrid}_{\beta+\frac{3}{2}}(\widehat{W}^{2\omega\Delta t})-F^{Hybrid}_{\beta+\frac{1}{2}}(\widehat{W}^{2\omega\Delta t})\big).
\end{aligned}
\end{equation}
As noticed in Sec.~\ref{Heun+TA_conservation}, all negative contributions from the
interface $\beta+\frac{1}{2}$ must be recovered in the flux balance of the cell $\beta$.

In order to demonstrate conservation, it is mandatory to integrate the solution in cell $\beta$ until $t=2\Delta t$, with cell
$\beta$ belonging to class rank $0$. The time integration leads to:
\begin{equation}
\begin{aligned}
\widehat{W}^{\omega \Delta t}_{\beta}&=W^{0}_{\beta}+\frac{\omega\Delta t}{\Delta x} \big( F_{\beta+\frac{1}{2}}(W^0)-F_{\beta-\frac{1}{2}}(W^0)\big),\\
{W}^{\Delta t}_{\beta}&={W}^{0}_{\beta}+\frac{\Delta t}{\Delta x} \bigg( \frac{1}{2}\big(F^{Hybrid}_{\beta+\frac{1}{2}}(\widehat{W}^{2\omega\Delta t})+F_{\beta+\frac{1}{2}}(W^0)\big)-F^{Hybrid}_{\beta-\frac{1}{2}}(\widehat{W}^{\omega\Delta t})\bigg),
\end{aligned}
\label{eq:conserv1_AION}
\end{equation}
in a first phase, and
\begin{equation}
\begin{aligned}
\widehat{W}^{2\omega\Delta t}_{\beta}&={W}^{\Delta t}_{\beta}+ \frac{\omega\Delta t}{\Delta x} \bigg( \frac{1}{2}\big(F^{Hybrid}_{\beta+\frac{1}{2}}(\widehat{W}^{2\omega\Delta t})+F_{\beta+\frac{1}{2}}(W^0)\big)-F_{\beta-\frac{1}{2}}(W^{\Delta t})\bigg), \\
{W}^{2\Delta t}_{\beta}&={W}^{\Delta t}_{\beta}+\frac{\Delta t}{\Delta x} \bigg( \frac{3}{2}F^{Hybrid}_{\beta+\frac{1}{2}}(\widehat{W}^{2\omega\Delta t})-\frac{1}{2}F_{\beta+\frac{1}{2}}(W^0)-F^{Hybrid}_{\beta-\frac{1}{2}}(\widehat{W}^{2\omega\Delta t})\bigg),
\end{aligned}
\label{eq:conserv2_AION}
\end{equation}
in a second phase.

If ${W}^{\Delta t}_{\beta}$ of Eq.~\eqref{eq:conserv1_AION} is replaced in the second equation of Eq.~\eqref{eq:conserv2_AION},
and reminding that Eq.~\eqref{eq:step3_flux_hy} holds, then
\begin{equation}
\begin{aligned}
{W}^{2\Delta t}_{\beta}&={W}^{0}_{\beta}+\frac{\Delta t}{\Delta x} \bigg( 2F^{Hybrid}_{\beta+\frac{1}{2}}(\widehat{W}^{2\omega\Delta t})-F^{Hybrid}_{\beta-\frac{1}{2}}(\widehat{W}^{\omega\Delta t})-F^{Hybrid}_{\beta-\frac{1}{2}}(\widehat{W}^{2\omega\Delta t})\bigg).\\
\end{aligned}
\end{equation}
To conclude, the hybrid time integration in cells $\beta$ and $\beta+1$ from $t=0$ until $t=2\Delta t$ involves the same flux at
the interface $\beta+\frac{1}{2}$ and the procedure is conservative.

In the next section, attention is paid on the accuracy of the proposed time-adaptive AION scheme.

\subsection{Time Accuracy\label{AION+TA_accuracy}}

The goal is to perform the very same analysis as the one performed for the Heun's scheme with time adaptation presented in
Sec.~\ref{Heun+TA_accuracy}. For the conciseness, the full demonstration is not repeated here and only the final form of the error
is provided.

Starting again from flux that are $p^{\mbox{\scriptsize th}}$-order accurate in space and
from a first-order finite difference approximation of the time derivative, error can be computed for cells of class 0 and 1.
Indeed, the numerical error of $W^{2\Delta t}$ at cells of class rank $1$ is:
\begin{equation}
\begin{aligned}
    e(W^{2\Delta t})&=\mathcal{O}(\Delta t^3,\frac{\Delta t^3}{\Delta x},\Delta t^2\Delta x^{p-2},\Delta t\Delta x^{p-1}).
\end{aligned}
\label{eq:error_classes_3_AION}
\end{equation}
and the local error $e({W}^{2\Delta t})$ for cells of class 0 is:
\begin{equation}
\begin{aligned}
    e(W^{2\Delta t})&=\mathcal{O}({\Delta t^2},\frac{\Delta t^3}{\Delta x},\frac{\Delta t^4}{\Delta x^2},\frac{\Delta t^5}{\Delta x^3},\Delta t\Delta x^{p-1},\Delta t^2\Delta x^{p-2},\Delta t^3 \Delta x^{p-3},\Delta t^4 \Delta x^{p-4}).
\end{aligned}
\label{eq:error_beta_2dt_AION}
\end{equation}
According to the CFL condition that links $\Delta t$ and $\Delta x$ for a hyperbolic equation ($\Delta x \simeq \Delta t$),
the local numerical error is $\mathcal{O}(\Delta t^2,\Delta x^p)$ for both class ranks. In the next section, our goal is to
recover this theoretical behaviour with numerical simulations.

\subsection{Numerical assessment of the theoretical behaviour \label{AION+TA_accuracy_sinus} }

The set of equation and the mesh are the same as introduced in Sec.~\ref{Heun+TA_accuracy_sinus}. 
In this context, we remind that $y_k^t$ represents the averaged unknown in the cell $k$ at time $t$. 
Here, the main change concerns the value of $\omega$ over the computational domain. 
The parameter $\omega$ is manufactured according to:
\begin{equation}
\left\{
\begin{array}{ll}
  \displaystyle \omega_j=\alpha \, \omega_{j-1} &  \hbox{ for } \displaystyle\frac{N}{2}-50 \leq j \leq \frac{N}{2}  \\
  \vspace*{-3mm} &\\
  \displaystyle \omega_j=\frac{1}{\alpha} \, \omega_{j-1} & \hbox{ for } \displaystyle\frac{N}{2}+1 \leq j \leq \frac{N}{2}+50  \\
  \vspace*{-3mm} &\\
  \displaystyle \omega_j = 1 & \hbox{elsewhere,}
\end{array}
\right.
\end{equation}
with $\alpha=0.90$, $N$ the number of cells and $j$ the cell index. The hybrid time-adaptive integrator is applied if
$0.6<\omega<1$. The value
$\omega=0.72$ is chosen arbitrarily for switching between cells of class rank 0 and those of class rank 1 for
the time-adaptive AION scheme. For completeness, errors are also provided for Heun's scheme, time-adaptive Heun'
scheme, AION scheme and time-adaptive AION scheme in Tab.~\ref{tab:order_tab_AION}. As before, the time
adaptive versions of the schemes are referred as +TA. It is numerically demonstrated that the temporal
adaptive methods recover the expected orders of accuracy.

\begin{figure}[!ht]
\begin{center}
\includegraphics[width=10cm]{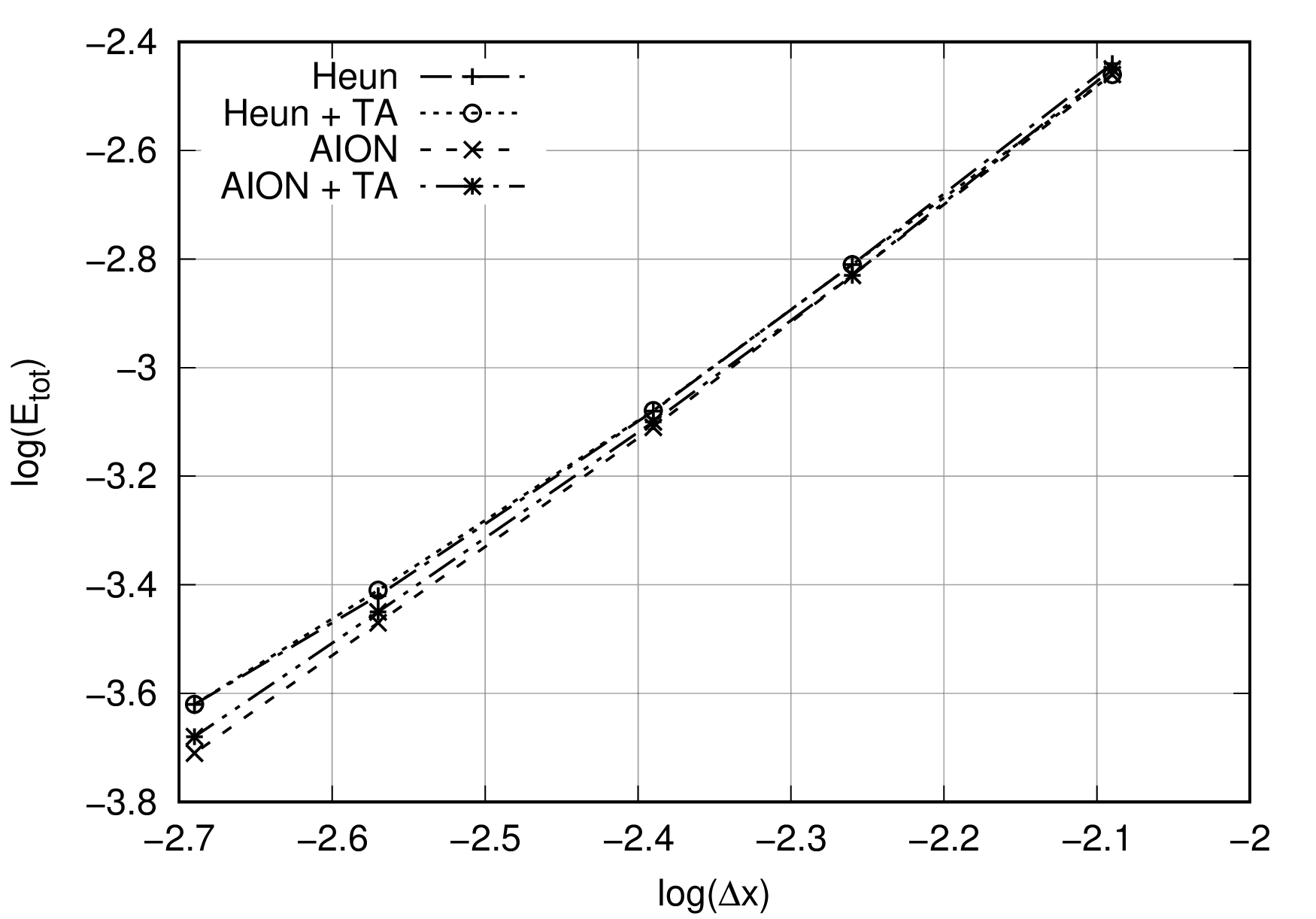}
\caption{Total error performed with second order spatial scheme
\label{fig:order_accuracy_AION}}
\end{center}
\end{figure}
\begin{table}[!ht]
\begin{center}
\begin{tabular}{|l|c|}
  \hline
  Time integrator & $E_{tot}$ slope \\
  \hline
  Heun & 1.97  \\
  Heun + TA & 1.93  \\
  AION & 2.08  \\
  AION + TA & 2.04  \\
  \hline
\end{tabular}
\caption{Total error slopes \label{tab:order_tab_AION} }
\end{center}
\end{table}

The analysis presented in this section gives an information on the local and global accuracy of the scheme, but does not give the two essential behaviours for unsteady simulations: dissipation and dispersion.
Section~\ref{sec:VonNeumannAION} is
dedicated to the Von Neumann analysis of the time adaptive AION scheme.

\subsection{Space-Time von Neumann Analysis}
\label{sec:VonNeumannAION}

The procedure to perform the von Neumann analysis of the time-adaptive AION scheme follows exactly the same steps as
the one for the time-adaptive Heun's scheme presented in Sec.~\ref{sec:VonNeumannHeun}. For conciseness, all the details
are not provided but the main steps to derive mathematical expressions are summarized.

First, the computational domain introduced in Fig.~\ref{fig:domainconfing} is considered again, and $\omega$ is set manually
according to:
\begin{equation}
\left\{
\begin{array}{ll}
  \displaystyle \omega_j=\alpha \, \omega_{j-1} &  \hbox{ for } \displaystyle\frac{N}{2}-50 \leq j \leq \frac{N}{2}  \\
  \vspace*{-3mm} &\\
  \displaystyle \omega_j=\frac{1}{\alpha} \, \omega_{j-1} & \hbox{ for } \displaystyle\frac{N}{2}+1 \leq j \leq \frac{N}{2}+50  \\
  \vspace*{-3mm} &\\
  \displaystyle \omega_j = 1 & \hbox{elsewhere.}
\end{array}
\right.
\end{equation}
with $\alpha=0.90$, $N$ the number of cells and $j$ the cell index. The hybrid time-adaptive integrator is applied if
$0.6<\omega<1$.

\subsubsection{Step UP}

In a second step, expressions for the time integration of the different cells are theoretically recovered.
Indeed, it leads to the following expressions for cell $\beta$ (step UP) until $t=2\Delta t$:
\begin{equation}
\begin{aligned}
y^{\Delta t}_{\beta}&=y^{0}_{\beta}-\frac{\Delta t}{\Delta x} \bigg[ \frac{1}{2}\big( F_{\beta+\frac{1}{2}}(\Delta t,y^0)+F^{Hybrid}_{\beta+\frac{1}{2}}(2\Delta t,\widehat{y}^{2\omega\Delta t} ) \big)- F^{Hybrid}_{\beta-\frac{1}{2}}(\Delta t,\widehat{y}^{\omega\Delta t})   \bigg] \\
y^{2\Delta t}_{\beta}&=y^{\Delta t}_{\beta}-\frac{\Delta t}{\Delta x} \bigg[ \frac{3}{2} F^{Hybrid}_{\beta+\frac{1}{2}}(2\Delta t,\widehat{y}^{2\omega\Delta t} )-\frac{1}{2}F_{\beta+\frac{1}{2}}(\Delta t,y^0) - F^{Hybrid}_{\beta-\frac{1}{2}}(2\Delta t,\widehat{y}^{2\omega\Delta t})   \bigg]
\end{aligned}
\end{equation}
Thanks to the hybrid reconstruction for one dimensional advection with a positive advection velocity $c>0$, it comes:
\begin{equation}
\begin{aligned}
F^{Hybrid}(2\Delta t,y^{2\Delta t})={c}\bigg\{&\omega \big( \frac{y^0+\widehat{y}^{2\omega\Delta t}}{2}\big)+\frac{1}{2}\frac{\Delta x}{2}(\nabla y)^0+(\omega-\frac{1}{2})\frac{\Delta x}{2}(\widehat{\nabla y})^{2\omega\Delta t} \\
&+(1-\omega)\big\{ y^{2\Delta t}_\beta+\frac{\Delta x}{2}(\nabla y)^{2\Delta t}_\beta -\frac{1-\omega}{2}(\Delta_t y)^{2\Delta t} \big\}\bigg\}\\
\end{aligned}
\label{eq:flux_hybrid}
\end{equation}
and the following finite volume formulation in case of hybrid time integration is obtained:
\begin{equation}
\begin{aligned}
y^{2\Delta t}_{\beta}=y^{\Delta t}_{\beta}-\frac{c\Delta t}{\Delta x} &\bigg[ \frac{3}{2}\bigg\{\omega_\beta \big( \frac{y^0_\beta+\widehat{y}_\beta^{2\omega\Delta t}}{2}\big)+\frac{1}{2}\frac{\Delta x}{2}(\nabla y)_\beta^0+(\omega_\beta-\frac{1}{2})\frac{\Delta x}{2}(\widehat{\nabla y})^{2\omega\Delta t}_\beta \\
&+(1-\omega_\beta)\big\{ y^{\Delta t}_\beta+\frac{\Delta x}{2}(\nabla y)^{\Delta t}_\beta -\frac{1-\omega_\beta}{2}(\Delta_t y)^{\Delta t}_\beta \big\}\bigg\}\\
&-\frac{1}{2}\bigg\{y^0_\beta+\frac{\Delta x}{2}(\nabla y)_\beta^0 \bigg\}\\
&-\bigg\{\omega_{\beta-1} \big( \frac{y^0_{\beta-1}+\widehat{y}_{\beta-1}^{2\omega\Delta t}}{2}\big)+\frac{1}{2}\frac{\Delta x}{2}(\nabla y)_{\beta-1}^0+(\omega_{\beta-1}-\frac{1}{2})\frac{\Delta x}{2}(\widehat{\nabla y})^{2\omega\Delta t}_{\beta-1} \\
&\text{ }\text{ }\text{ }\text{ }\text{ }+(1-\omega_{\beta-1})\big\{ y^{2\Delta t}_{\beta-1}+\frac{\Delta x}{2}(\nabla y)^{2\Delta t}_{\beta-1} -\frac{1-\omega_{\beta-1}}{2}(\Delta_t y)^{2\Delta t}_{\beta-1} \big\}\bigg\}\bigg]
\end{aligned}
\label{stepUP_AION}
\end{equation}
Here again, the terms $\widehat{y}_{\beta}^{2\omega\Delta t}$ and $(\widehat{\nabla y})^{2\omega\Delta t}_{\beta}$ could
be source of instability, as for the time-adaptive Heun's scheme.

\subsubsection{Step DOWN}

For the cell $\gamma$ (step DOWN), the hybrid time integration can be formulated as:
\begin{equation}
\begin{aligned}
y^{2\Delta t}_{\gamma}=y^{\Delta t}_{\gamma}-\frac{c\Delta t}{\Delta x_{\gamma}} &\bigg[\bigg\{\omega_{\gamma} \big( \frac{y^0_{\gamma}+\widehat{y}_{\gamma}^{2\omega\Delta t}}{2}\big)+\frac{1}{2}\frac{\Delta x}{2}(\nabla y)_{\gamma}^0+(\omega_{\gamma}-\frac{1}{2})\frac{\Delta x}{2}(\widehat{\nabla y})^{2\omega\Delta t}_{\gamma} \\
&\text{ }\text{ }\text{ }\text{ }\text{ }+(1-\omega_{\gamma})\big\{ y^{2\Delta t}_{\gamma}+\frac{\Delta x}{2}(\nabla y)^{2\Delta t}_{\gamma} -\frac{1-\omega_{\gamma}}{2}(\Delta_t y)^{2\Delta t}_{\gamma} \big\}\bigg\}\\
&-\frac{3}{2}\bigg\{ \omega_{\gamma-1} \big( \frac{y^0_{\gamma-1}+\widehat{y}_{\gamma-1}^{2\omega\Delta t}}{2}\big)+\frac{1}{2}\frac{\Delta x}{2}(\nabla y)_{\gamma-1}^0+(\omega_{\gamma-1}-\frac{1}{2})\frac{\Delta x}{2}(\widehat{\nabla y})^{2\omega\Delta t}_{\gamma-1} \\
& \text{ }\text{ }\text{ }\text{ }\text{ }\text{ }\text{ }+(1-\omega_{\gamma-1})\big\{ y^{\Delta t}_{\gamma-1}+\frac{\Delta x}{2}(\nabla y)^{\Delta t}_{\gamma-1} -\frac{1-\omega_{\gamma-1}}{2}(\Delta_t y)^{\Delta t}_{\gamma-1} \big\}\bigg\}\\
&+\frac{1}{2}\bigg\{y^0_{\gamma-1}+\frac{\Delta x}{2}(\nabla y)_{\gamma-1}^0\bigg\}\bigg]
\end{aligned}
\label{stepDOWN_AION}
\end{equation}

\subsubsection{Mathematical expressions}

In order to obtain the amplification factor $G$ between $y^{2\Delta t}$ and $y^0$, all terms in Eqs.~\eqref{stepUP_AION}~and~
\eqref{stepDOWN_AION} must first be expressed according to state $y^{\Delta t}$:
\begin{equation}
\begin{aligned}
y^{0}=\frac{1}{G_0}y^{\Delta t}\\
\widehat{y}^{2\omega\Delta t}=G_1y^0=\frac{G_1}{G_0}y^{\Delta t}\\
(\nabla y)^0=\frac{G_3}{G_0}y^{\Delta t}\\
  ...\\
\end{aligned}
\end{equation}
because of the presence of term $y^{2\Delta t}$ in the right-hand sides of Eqs.~\eqref{stepUP_AION} and~\eqref{stepDOWN_AION}.
The final expression is finally derived from $y^{\Delta t}$:
\begin{equation}
\begin{aligned}
y^{2\Delta t}=Gy^{\Delta t}=GG_0 y^0
\end{aligned}
\end{equation}

\subsubsection{Partial conclusion}

\begin{figure}[!htbp]\begin{center}
\begin{tabular}{cc}
\includegraphics[width=8cm]{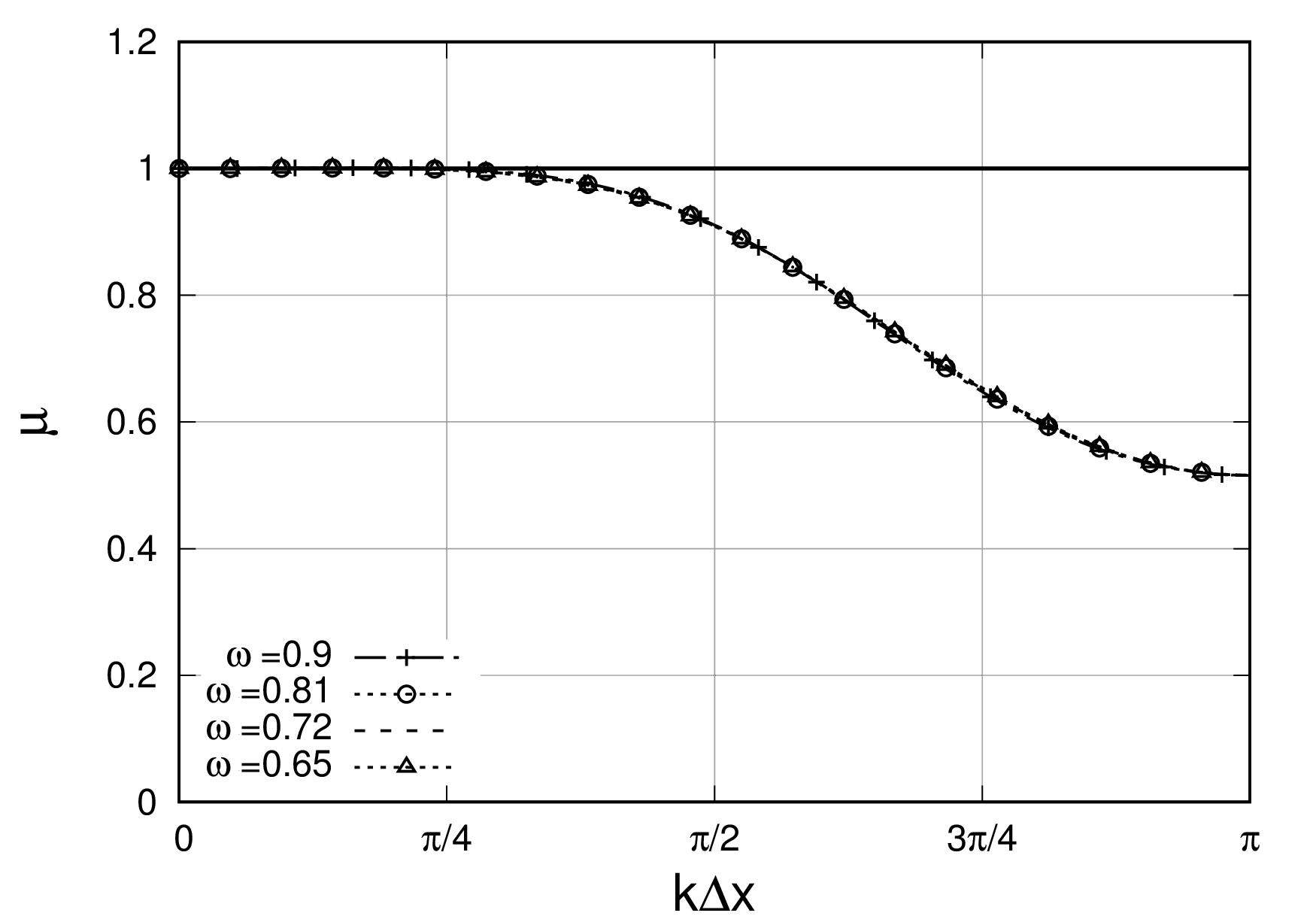} & \includegraphics [width=8cm]{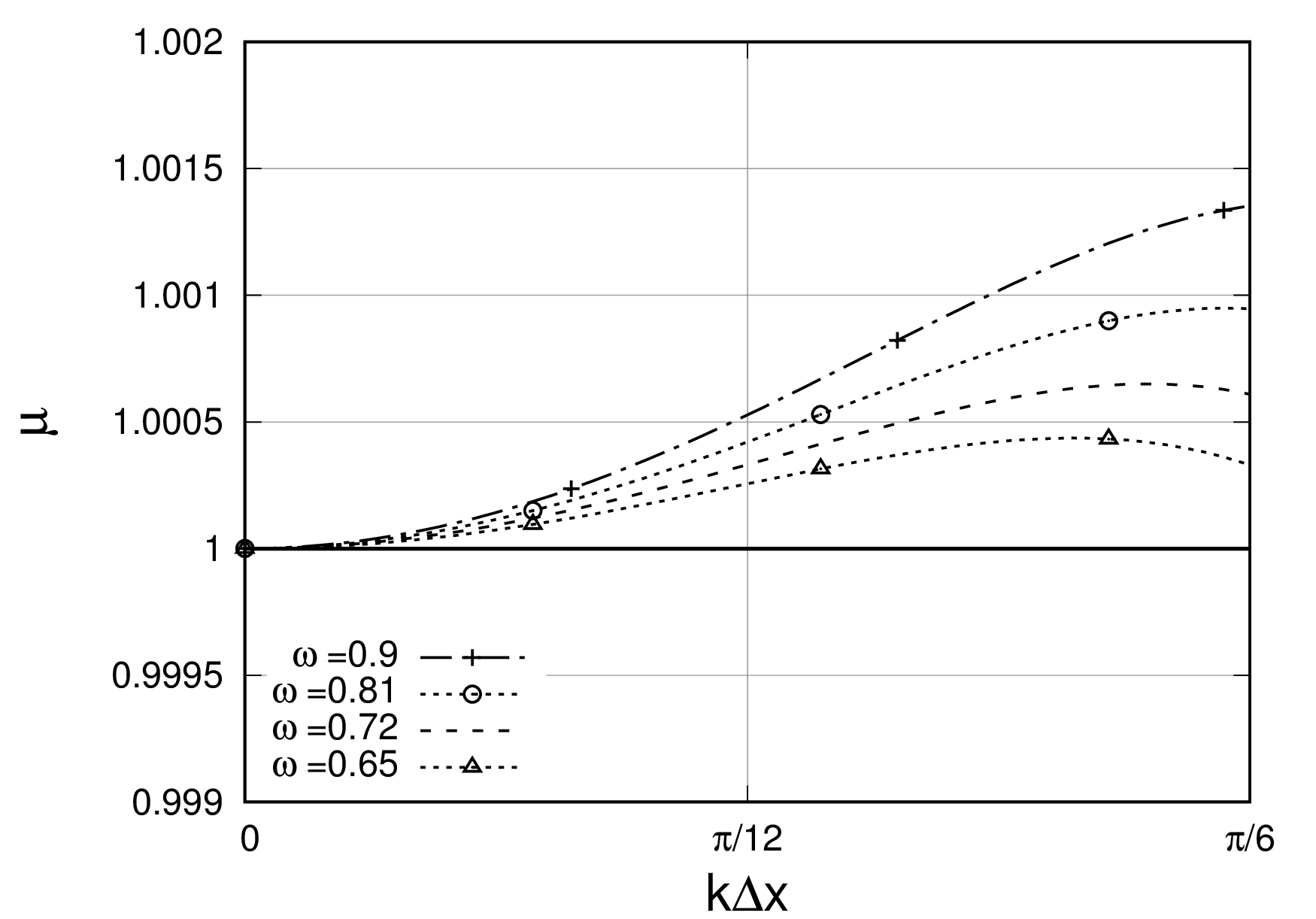}\\
\end{tabular}
\caption{Step DOWN configuration: spectral behaviour of the hybrid time integration for several values of $\omega$
(at CFL$=0.1$)}\label{fig:abs_temp_AION_DOWN_01}
\end{center}
\end{figure}

\begin{figure}[!htbp]\begin{center}
\begin{tabular}{cc}
\includegraphics[width=8cm]{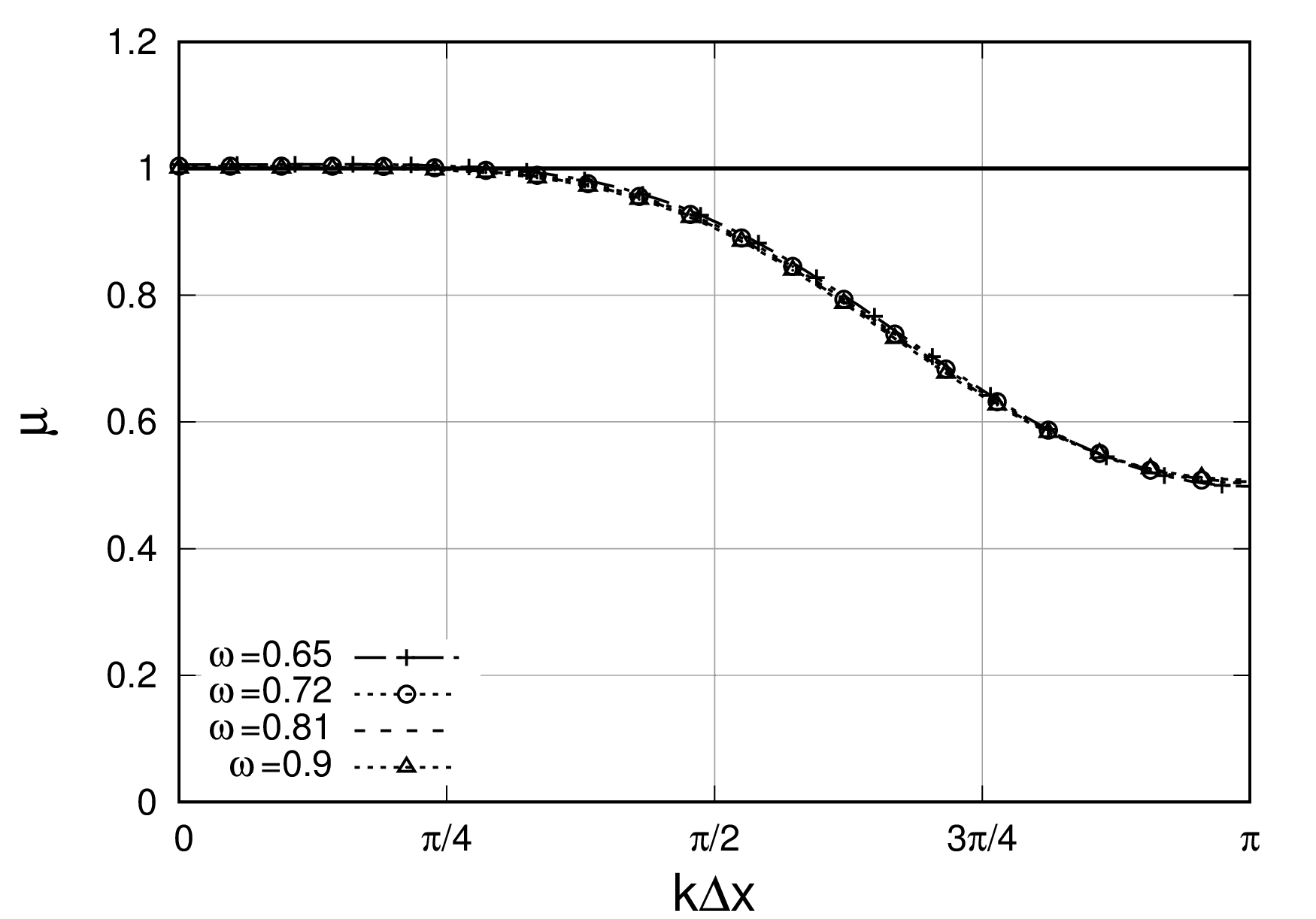} & \includegraphics [width=8cm]{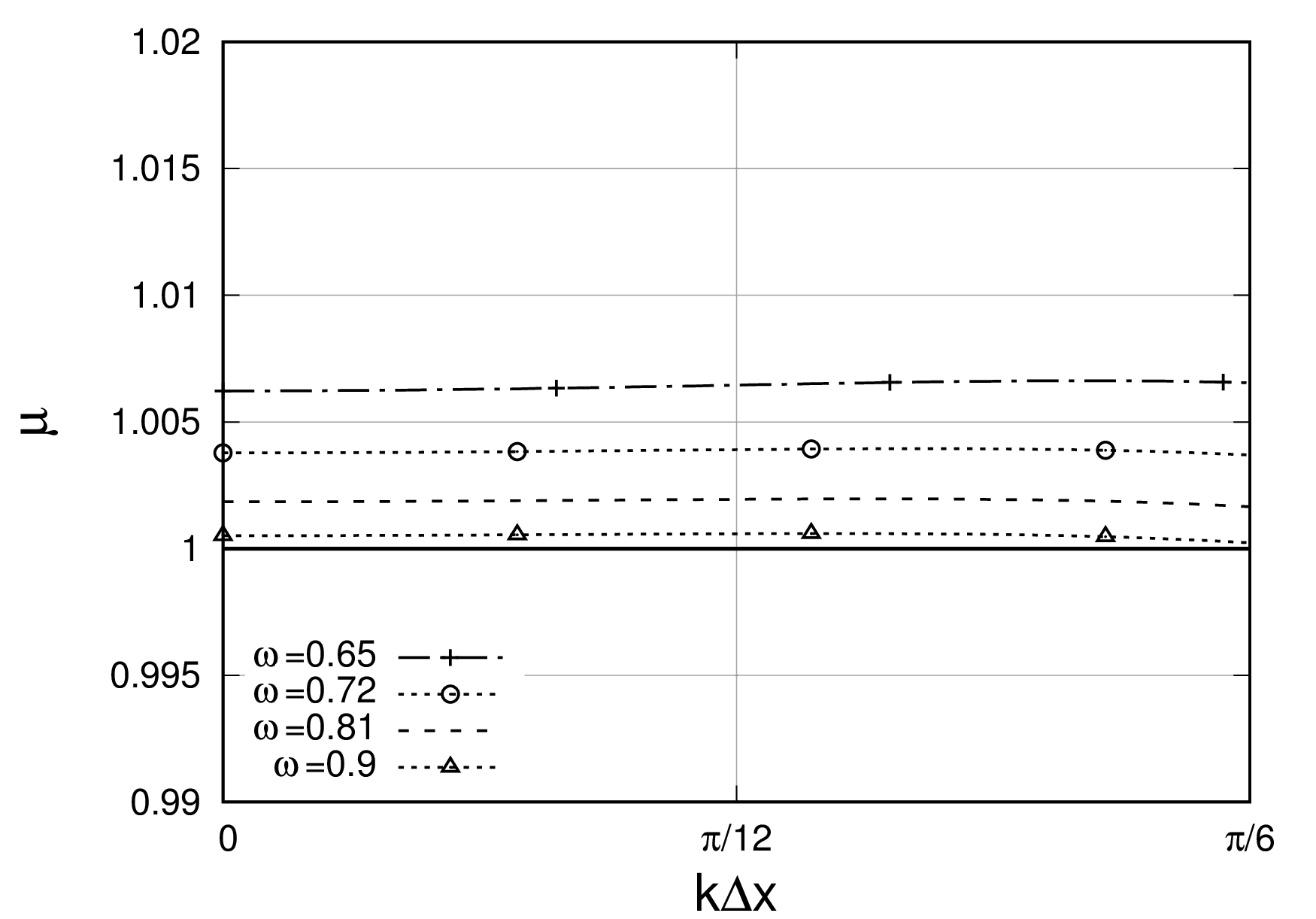}\\
\end{tabular}
\caption{Step UP configuration: spectral behaviour of hybrid time integration for several values of $\omega$ (CFL$=0.1$)}\label{fig:abs_temp_AION_UP_01}
\end{center}
\end{figure}

\begin{figure}[!htbp]\begin{center}
\begin{tabular}{cc}
\includegraphics[width=8cm]{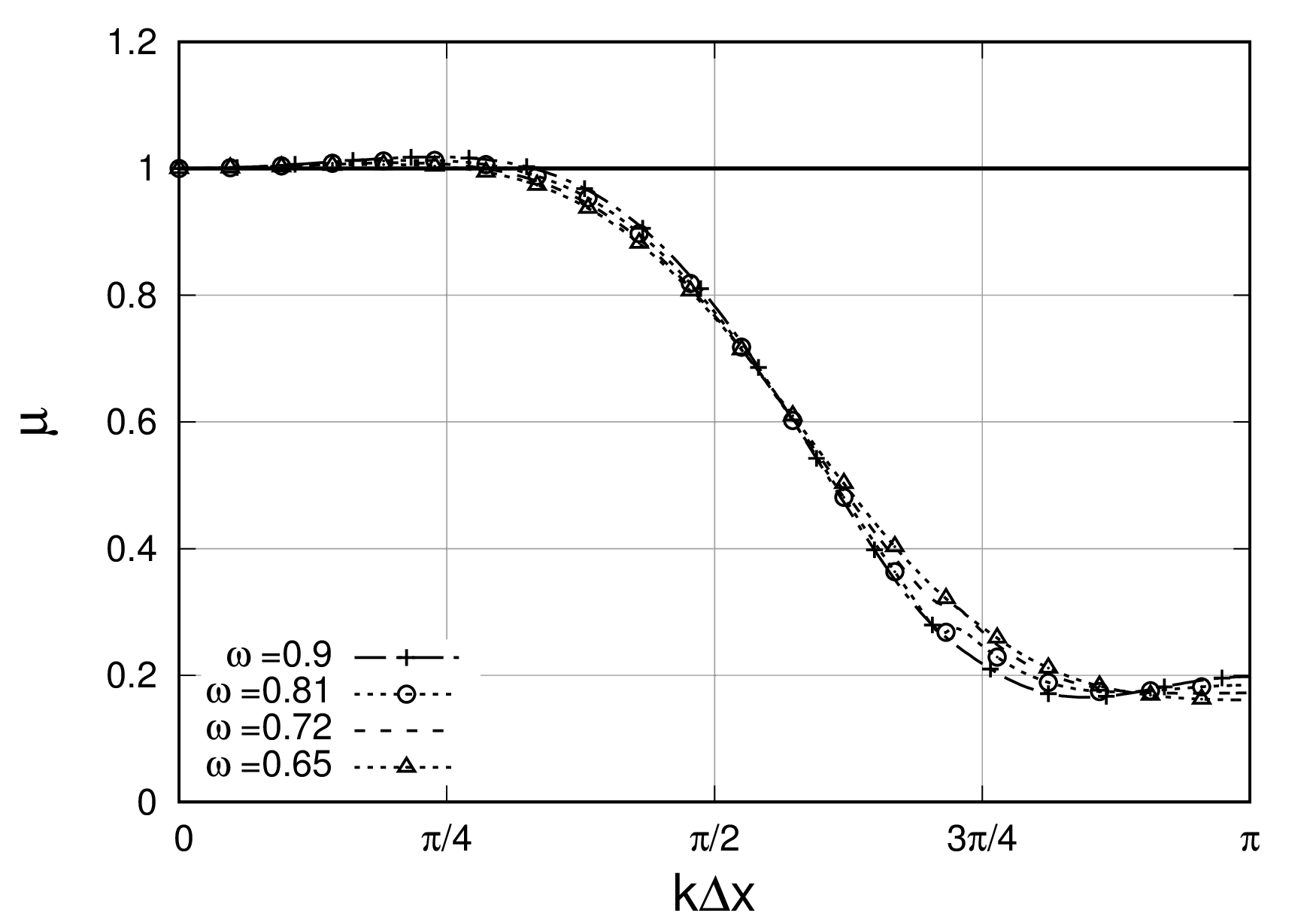} & \includegraphics [width=8cm]{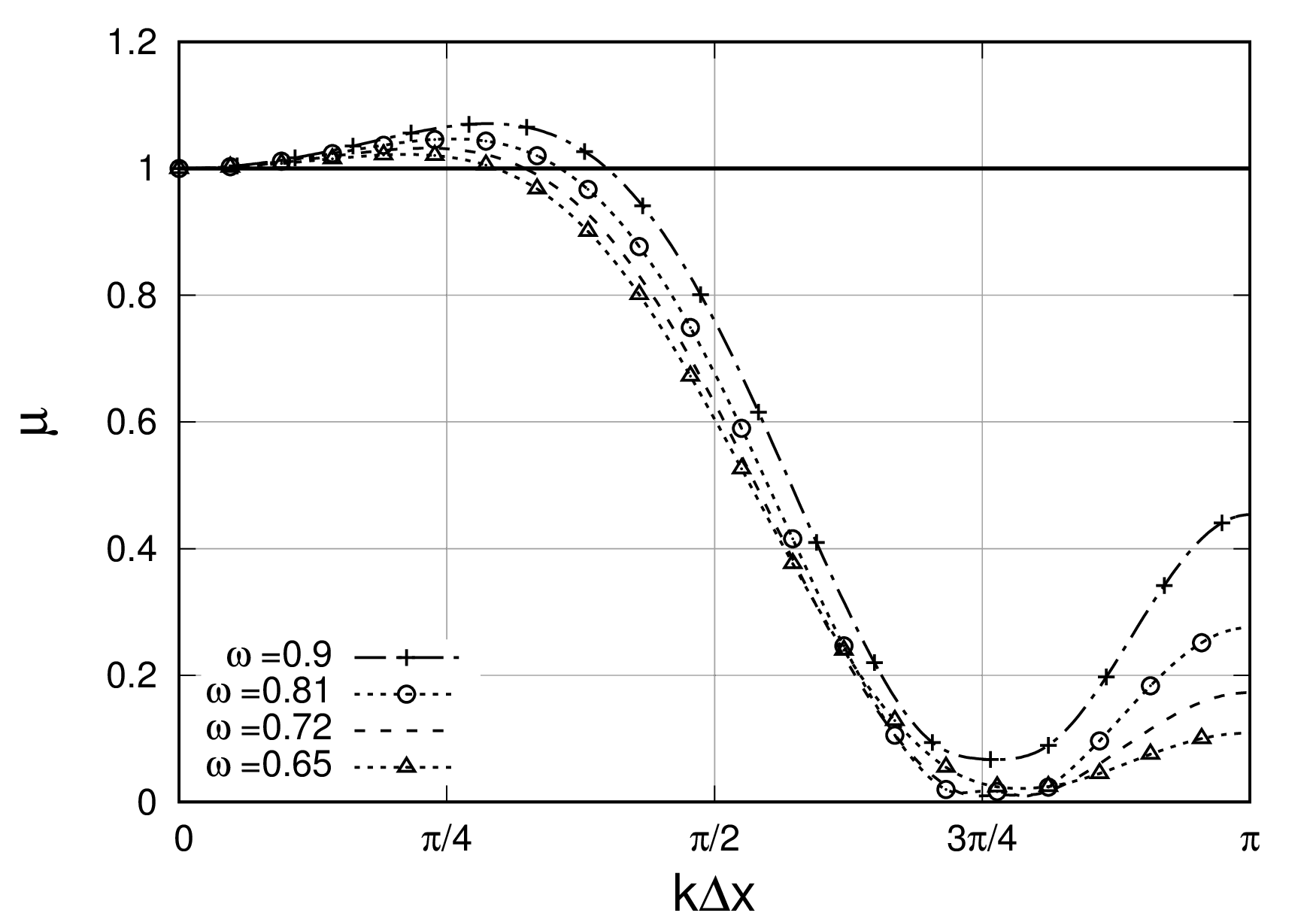}\\
\end{tabular}
\caption{Step DOWN configuration: spectral behaviour of hybrid time integration for several values of $\omega$ with
CFL$=0.3$ (left) and CFL$=0.6$ (right)}\label{fig:abs_temp_AION_DOWN_0306}
\end{center}
\end{figure}

\begin{figure}[!htbp]\begin{center}
\begin{tabular}{cc}
\includegraphics[width=8cm]{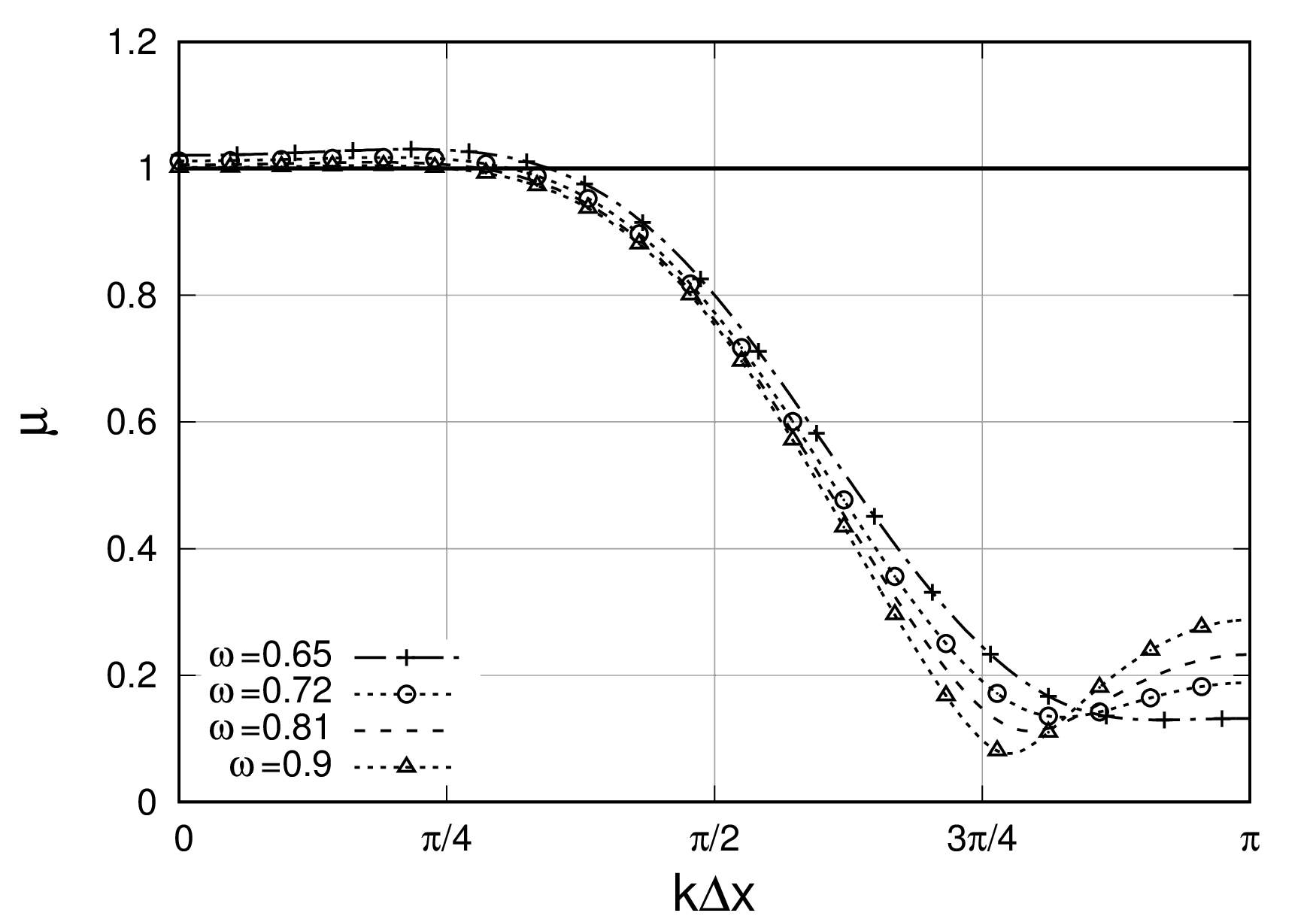} & \includegraphics [width=8cm]{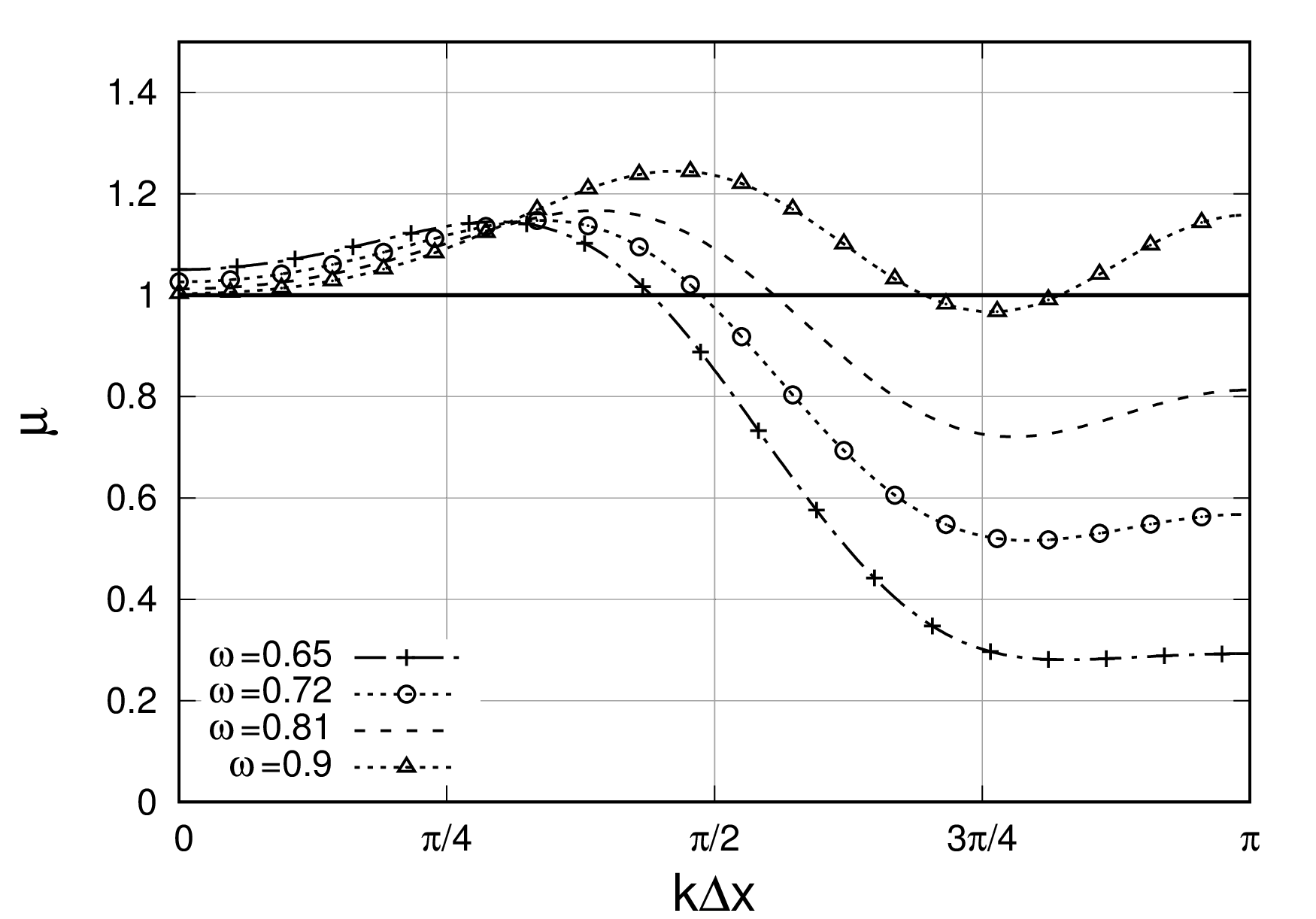}\\
\end{tabular}
\caption{Step UP configuration: spectral behaviour of hybrid time integration for several value of $\omega$  with
CFL$=0.3$ (left) and CFL$=0.6$ (right)}\label{fig:abs_temp_AION_UP_0306}
\end{center}
\end{figure}

The analysis is performed for several values of the parameter $\omega_j$ of the hybrid scheme in cell index $j$
since $\omega_j$ represents the blending of two schemes. In the following, four values are considered: 0.9, 0.81, 0.72 and 0.65.
Figs.~\ref{fig:abs_temp_AION_DOWN_01} and \ref{fig:abs_temp_AION_UP_01} show the dissipation term as a function of $k \Delta x$
for the step DOWN and UP configurations respectively at CFL=0.1.
A focus for the low normalised wave numbers ($0<k\Delta x < \pi/6$) shows that
amplification occurs for both configurations, something which was not found for the Heun+TA scheme. In addition, amplification
becomes greater when the CFL number is increased to 0.3 or 0.6, as shown in Fig.~\ref{fig:abs_temp_AION_DOWN_0306} for the
step DOWN configuration and in Fig.~\ref{fig:abs_temp_AION_UP_0306} for the step UP configuration.

\begin{figure}[!htbp]
\begin{center}
\begin{tabular}{cc}
\includegraphics[width=8cm]{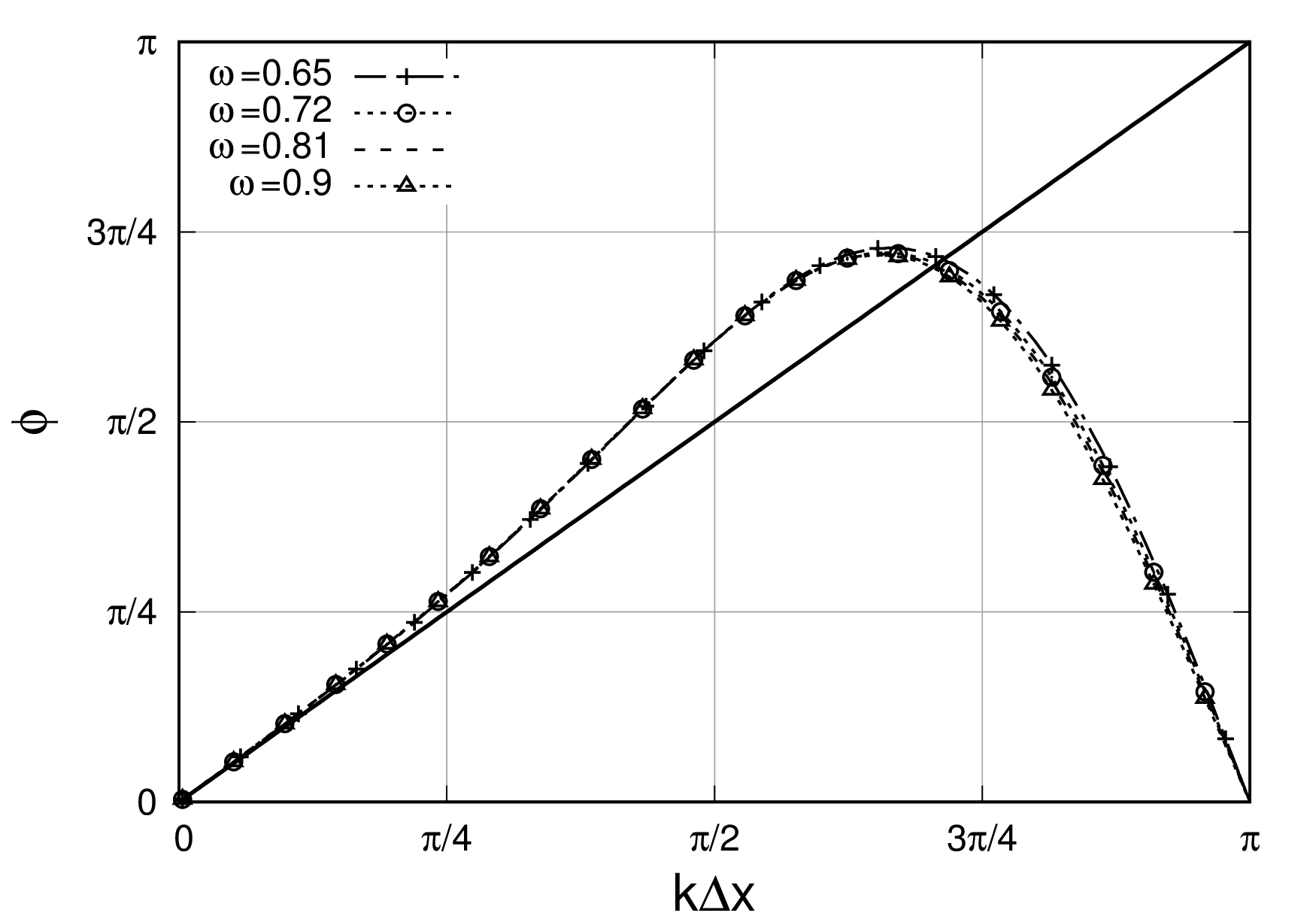} & \includegraphics [width=8cm]{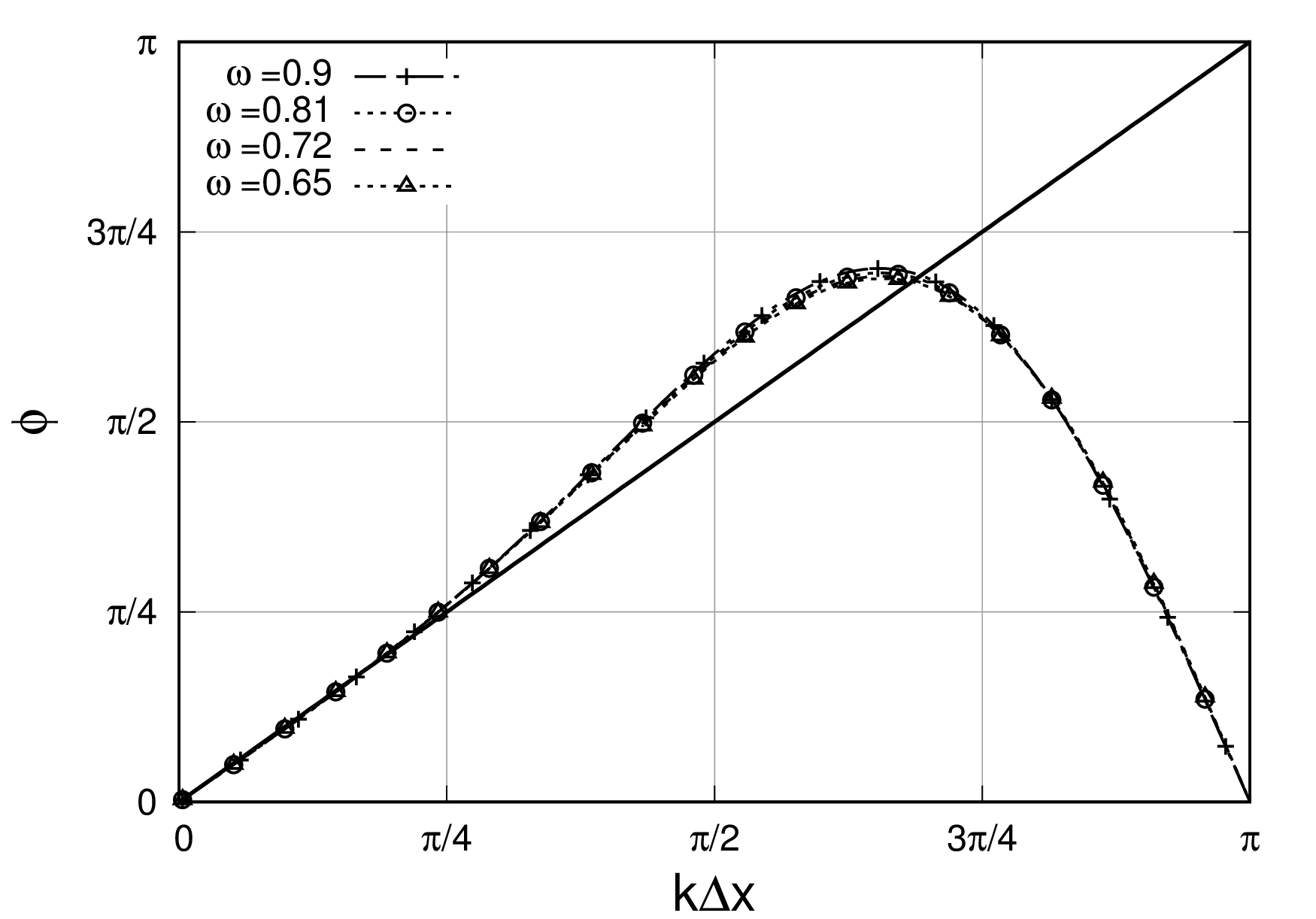}\\
\end{tabular}
\caption{Dissipation of the hybrid time integration for several values of $\omega$ for the Step UP (left) and DOWN (right) configurations at CFL=0.1}\label{fig:phase_temp_AION_UP_DOWN_01}
\end{center}
\end{figure}

\begin{figure}[!htbp]
\begin{center}
\begin{tabular}{cc}
\includegraphics[width=8cm]{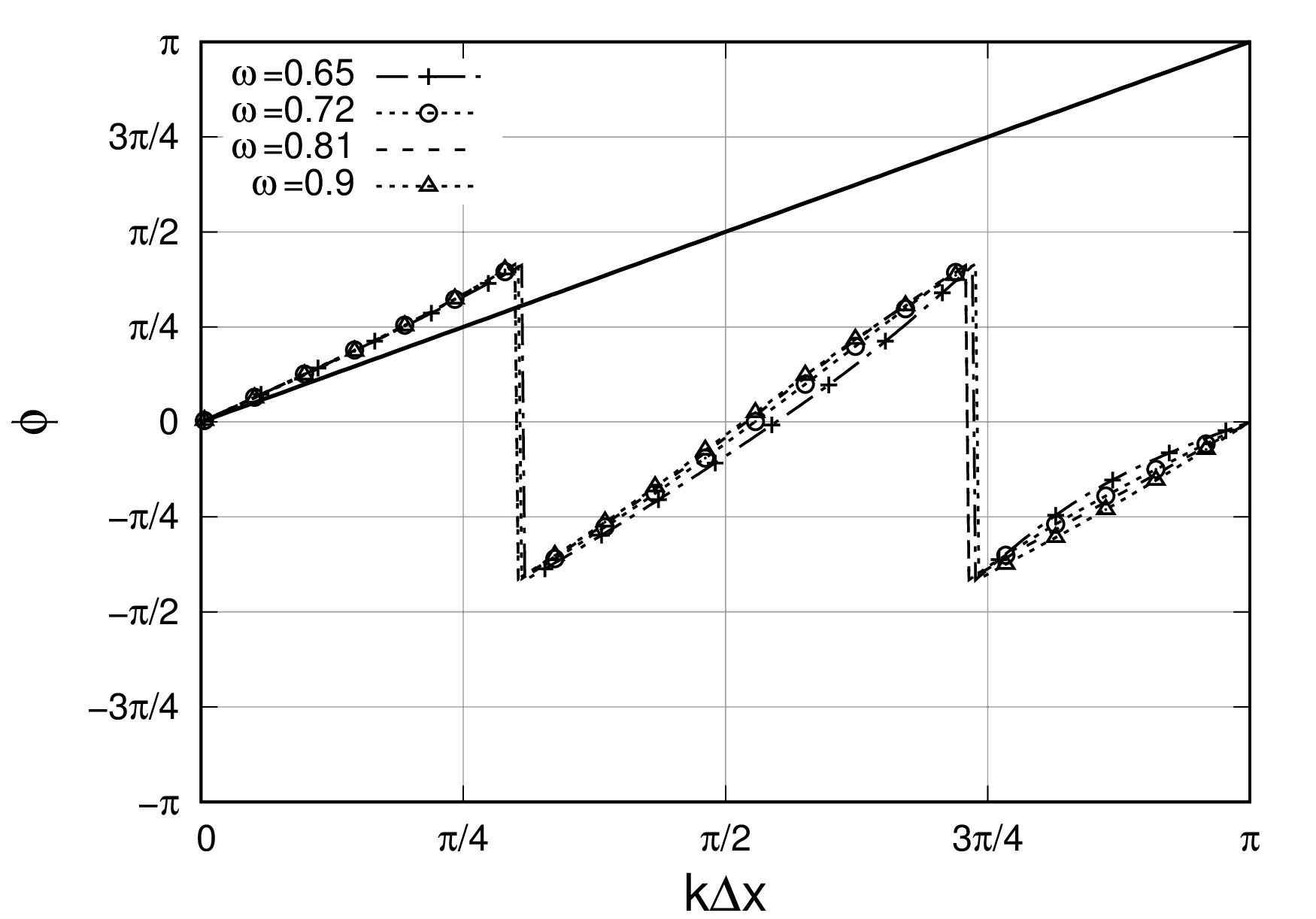} & \includegraphics [width=8cm]{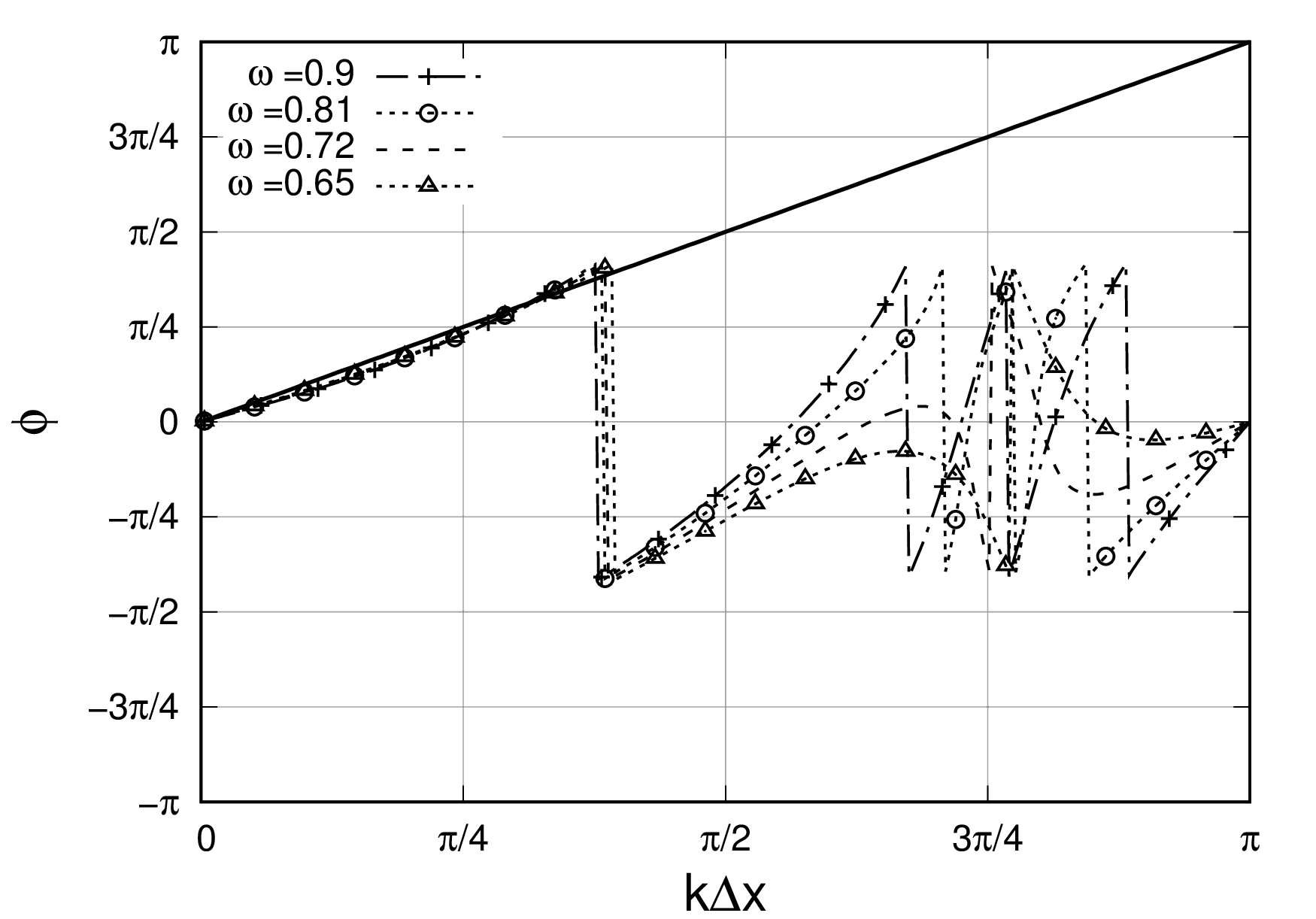}\\
\end{tabular}
\caption{Dissipation of the hybrid time integration for several values of $\omega$ for the Step UP (left) and DOWN (right) configurations at CFL=0.6}\label{fig:phase_temp_AION_UP_DOWN_06}
\end{center}
\end{figure}

\begin{figure}[!htbp]\begin{center}
\begin{tabular}{cc}
\includegraphics[width=8cm]{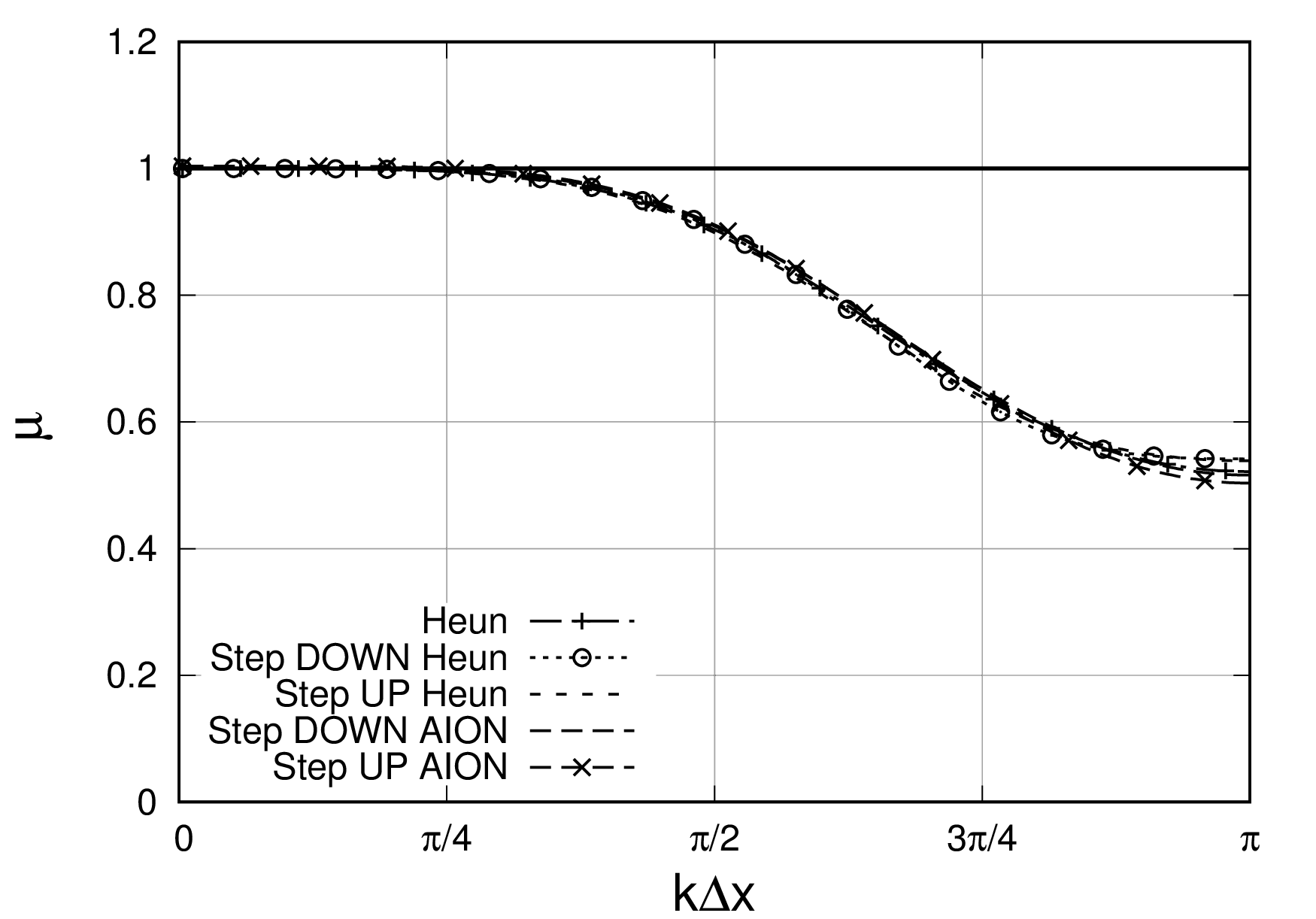} & \includegraphics [width=8cm]{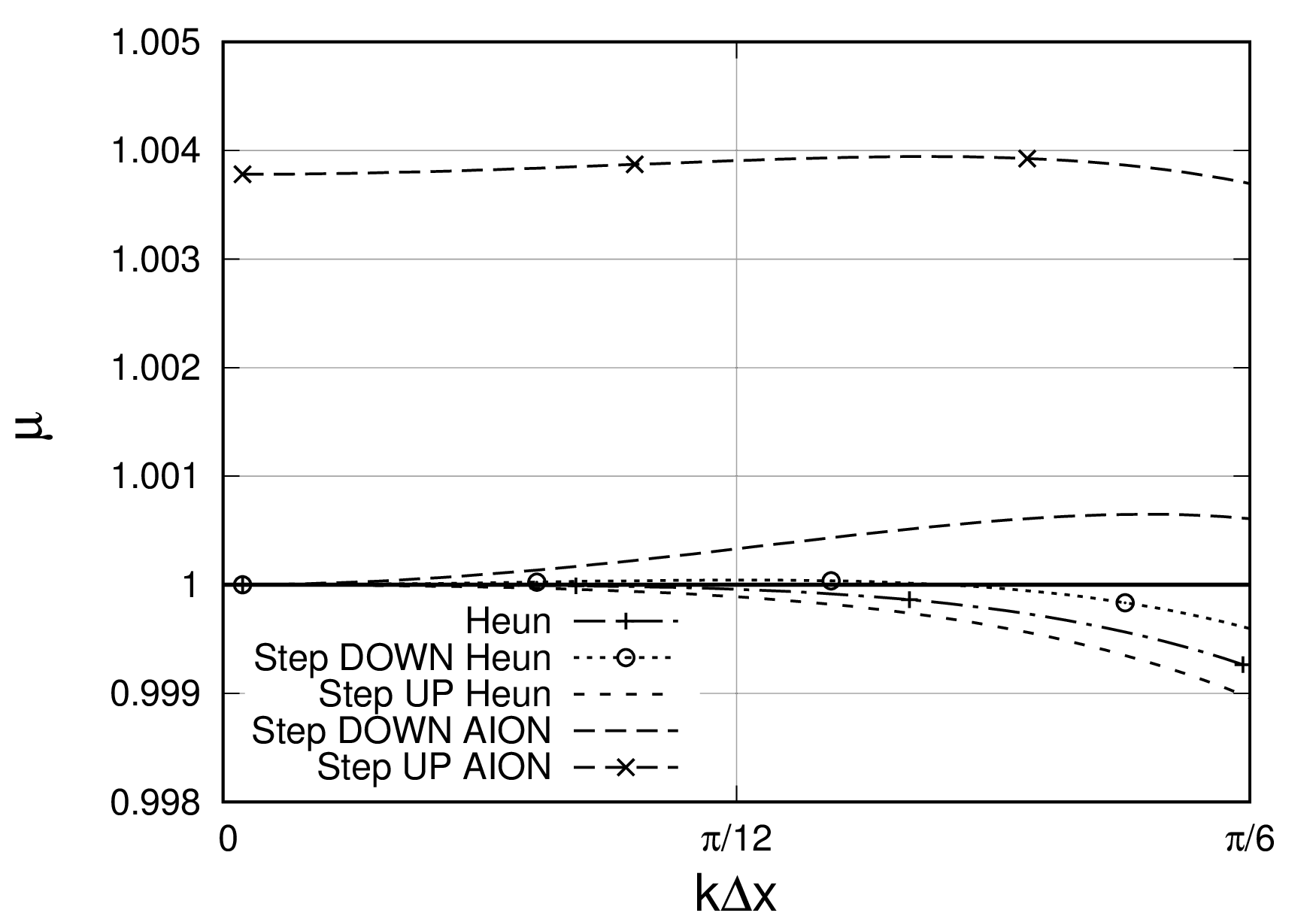}\\
\end{tabular}
\includegraphics [width=8cm]{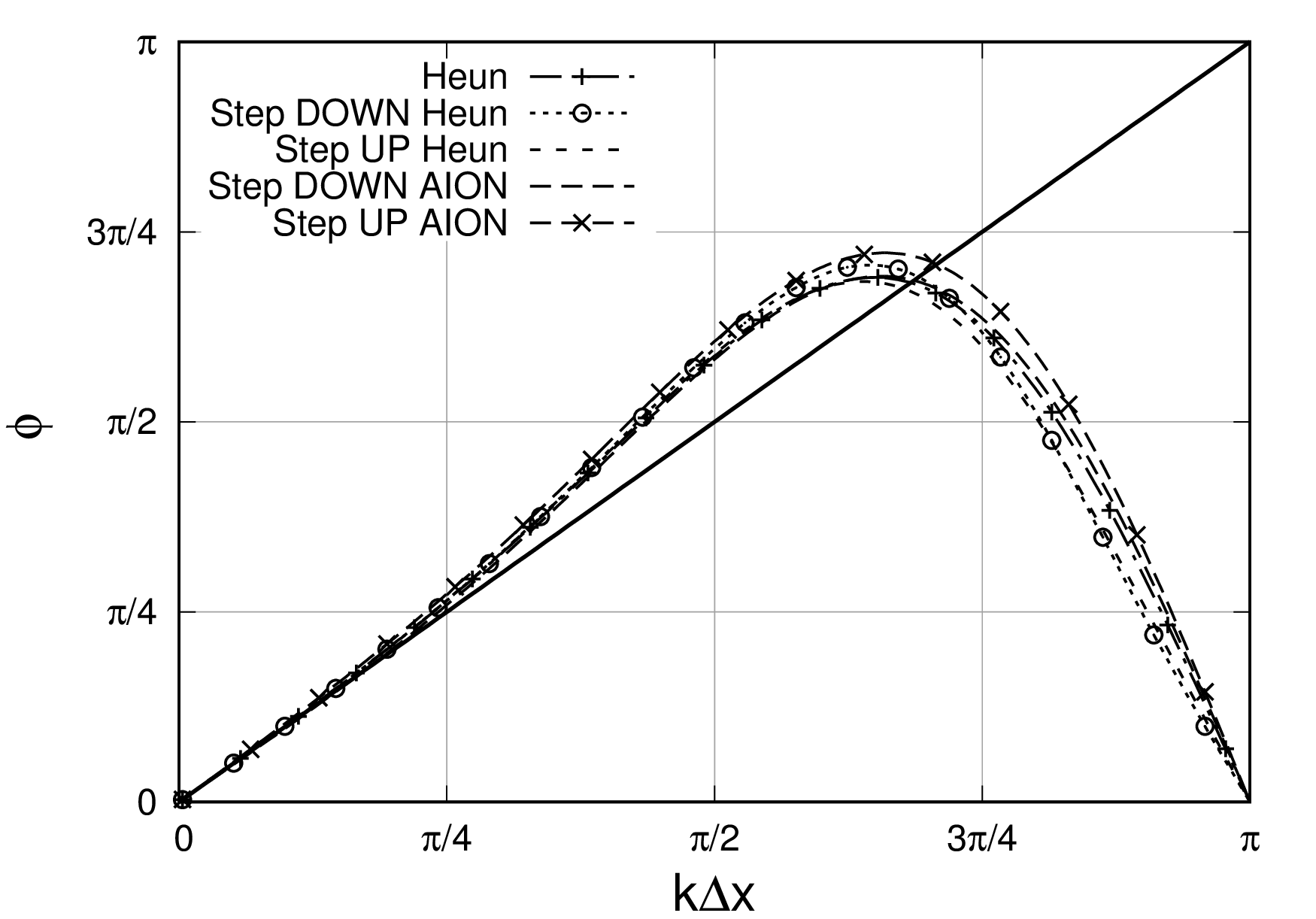}
\caption{ Dissipation $\mu$ and dispersion $\phi$ behaviour at step configurations for AION+TA and Heun+TA integration with CFL=0.1 \label{fig:dissiptempAIONFig01_test}}
\end{center}
\end{figure}

\begin{figure}[!htbp]\begin{center}
\begin{tabular}{cc}
\includegraphics[width=8cm]{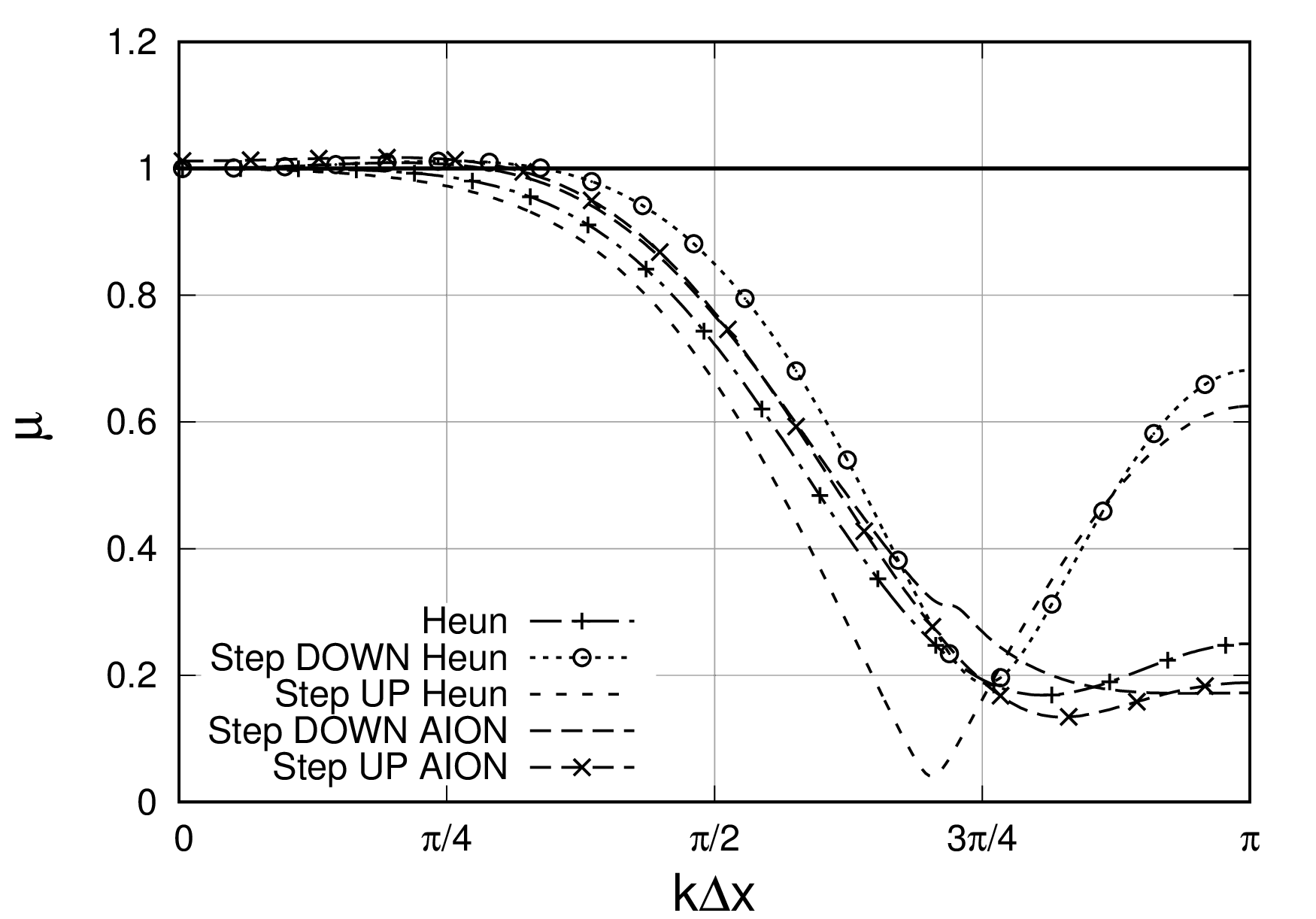} & \includegraphics [width=8cm]{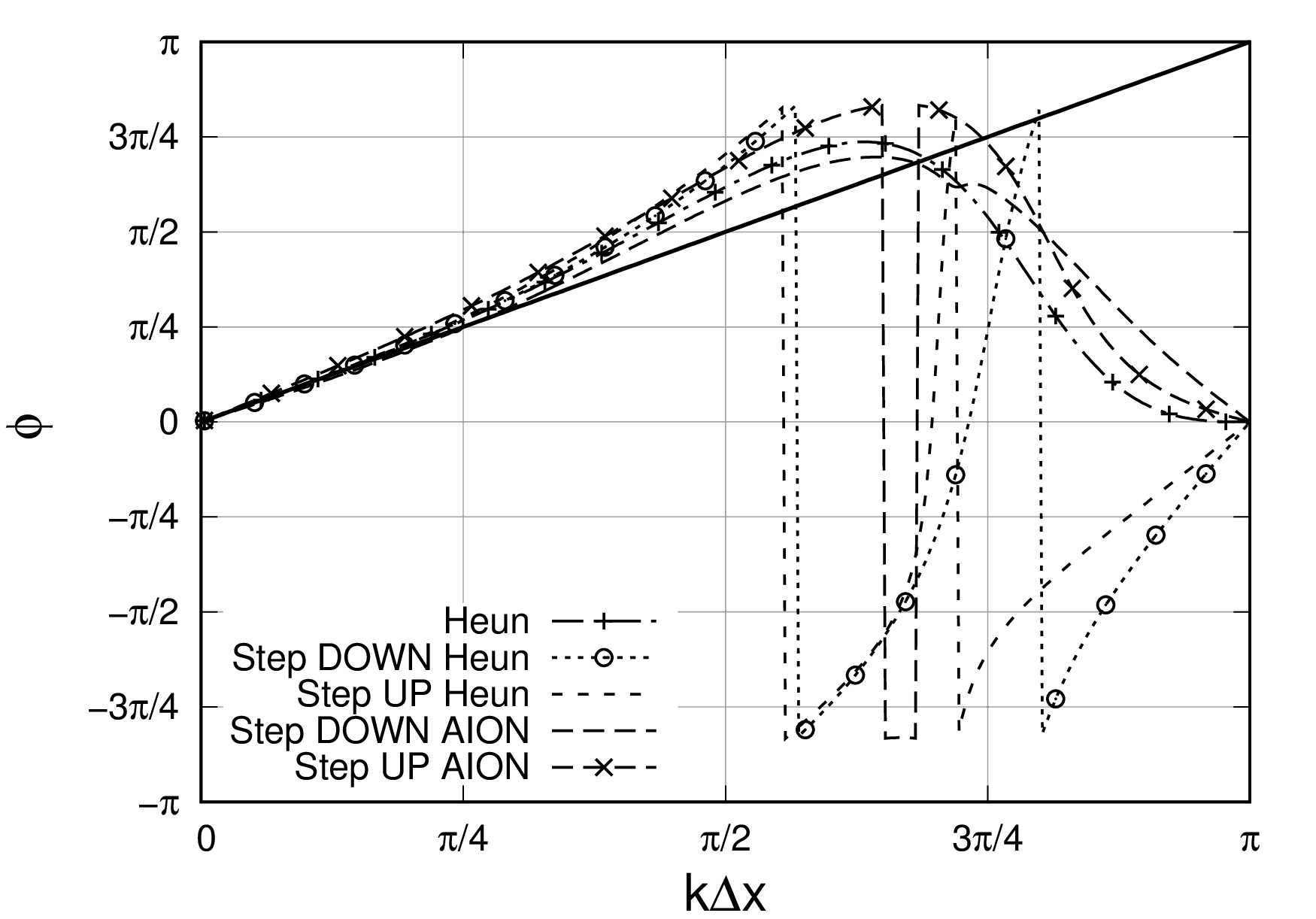}\\
\end{tabular}
\caption{ Dissipation $\mu$ and dispersion $\phi$ behaviour at step configurations for AION+TA and Heun+TA integration with CFL=0.3 \label{fig:dissiptempAIONFig03_test}}
\end{center}
\end{figure}

\begin{figure}[!htbp]\begin{center}
\begin{tabular}{cc}
\includegraphics[width=8cm]{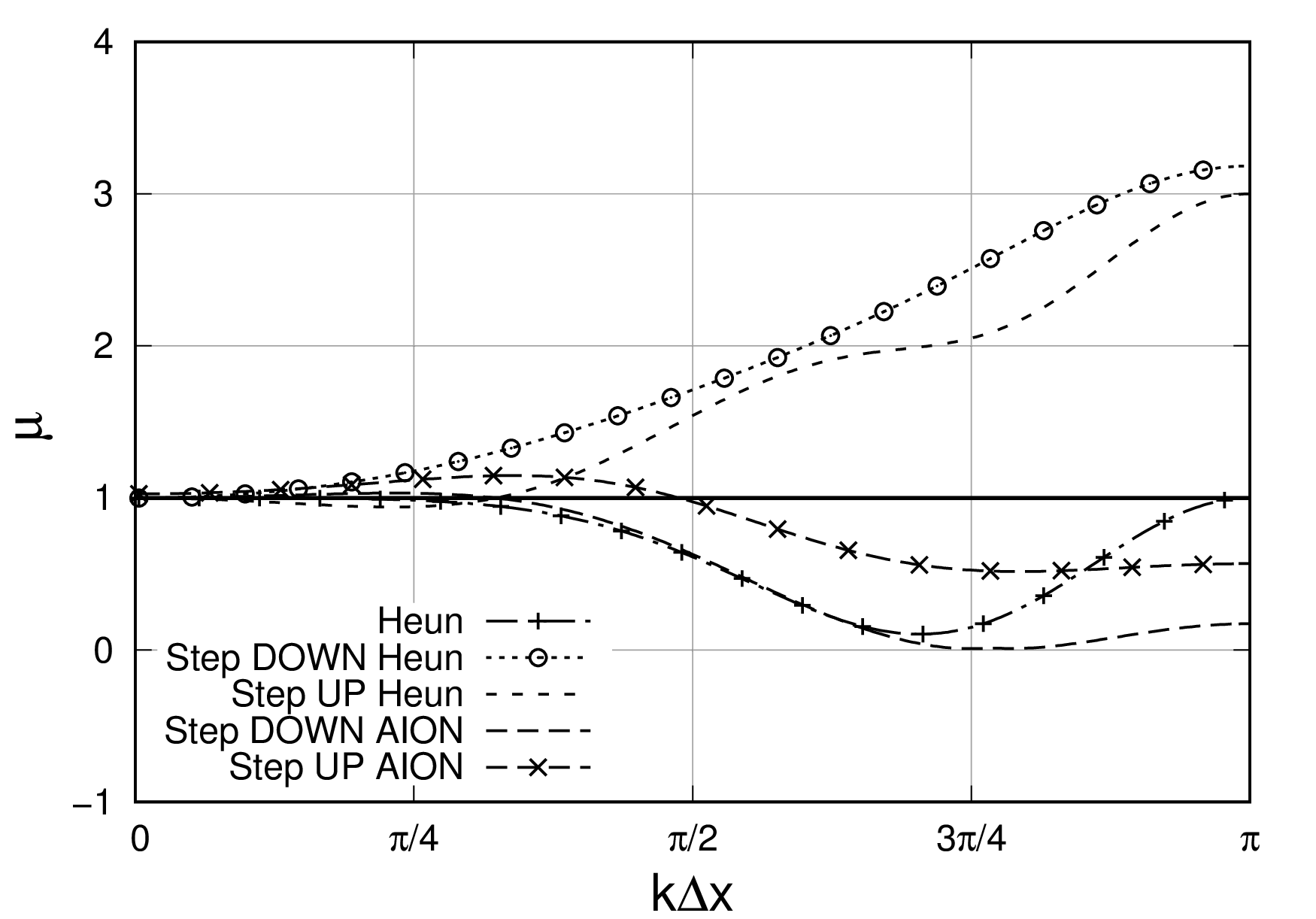} & \includegraphics [width=8cm]{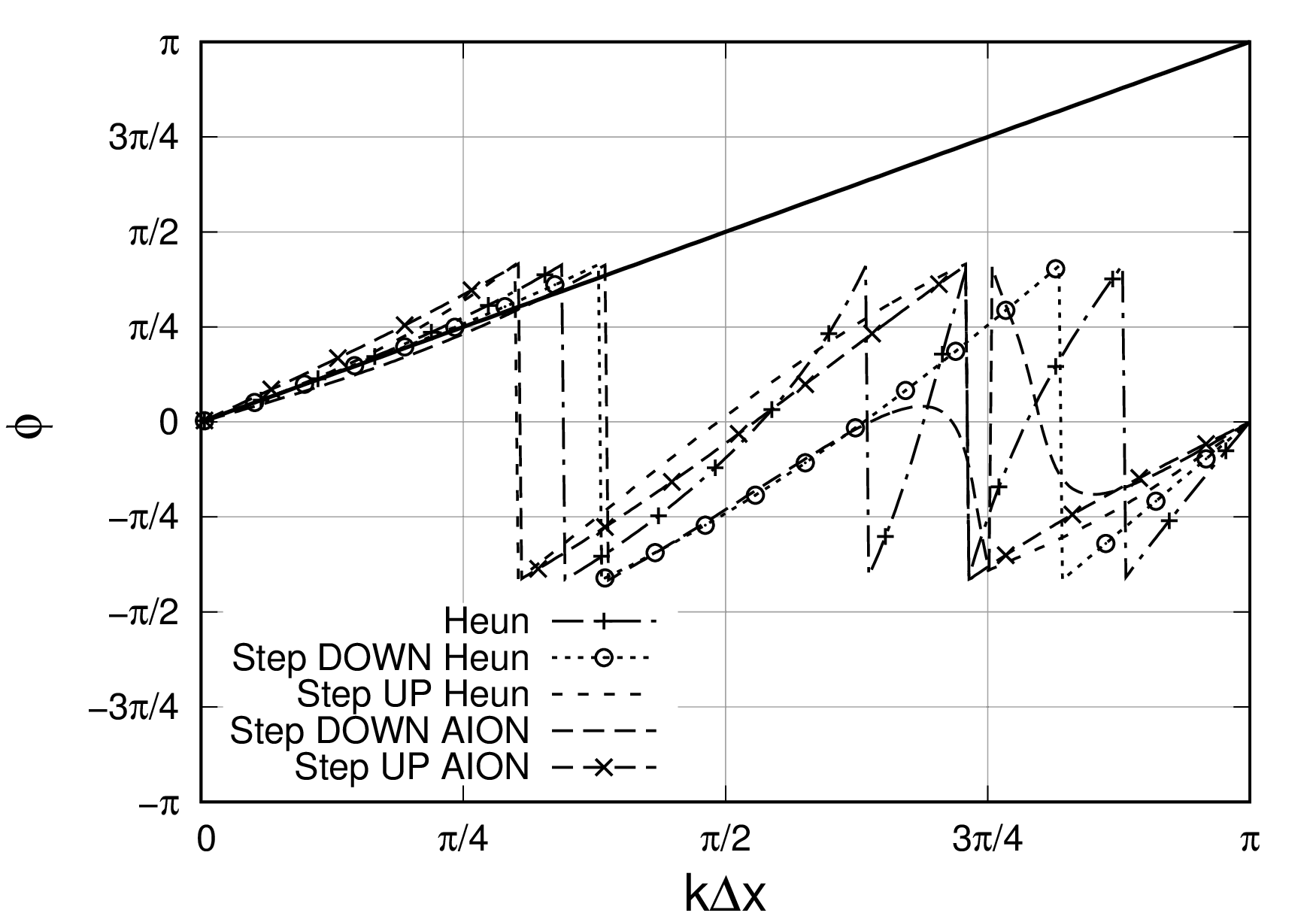}\\
\end{tabular}
\caption{ Dissipation $\mu$ and dispersion $\phi$ behaviour at step configurations for AION+TA and Heun+TA time integration with CFL=0.6 \label{fig:dissiptempAIONFig06_test}}
\end{center}
\end{figure}

Fig.~\ref{fig:phase_temp_AION_UP_DOWN_01} shows that the value of $\omega$ has a minor influence in the dispersion behaviour
at CFL=0.1. The same result is obtained at CFL=0.6 in Fig.~\ref{fig:phase_temp_AION_UP_DOWN_06}, but the shape of the analysis is more complex, due to the discontinuity in phases.
This discontinuity in phase was encountered for the AION scheme in \cite{Muscat_JCP_XX_2018} and is a consequence
of the specific shape of the transfer function between the parameter $k\,\Delta x$ and the dispersion $\Phi$.
Actually, the spectral behaviour for AION+TA and Heun+TA schemes is provided in Figs.~\ref{fig:dissiptempAIONFig01_test}-\ref{fig:dissiptempAIONFig06_test} for two values of the CFL number and for the specific choice $\omega=0.72$.
Moreover, the stability limitation of CFL$\leq 0.5$ with Heun time integration is overcome thanks to the hybrid time
integration: Fig.~\ref{fig:dissiptempAIONFig06_test} shows an amplification of the Heun's scheme for any wavenumber
while amplification only occurs in a short range of wavenumbers for the AION scheme.

\subsection{Analysis of $q-$waves} \label{sec:qwavesHybIRK2}

\begin{figure}[!htbp]\begin{center}
\begin{tabular}{cc}
\includegraphics[width=8cm]{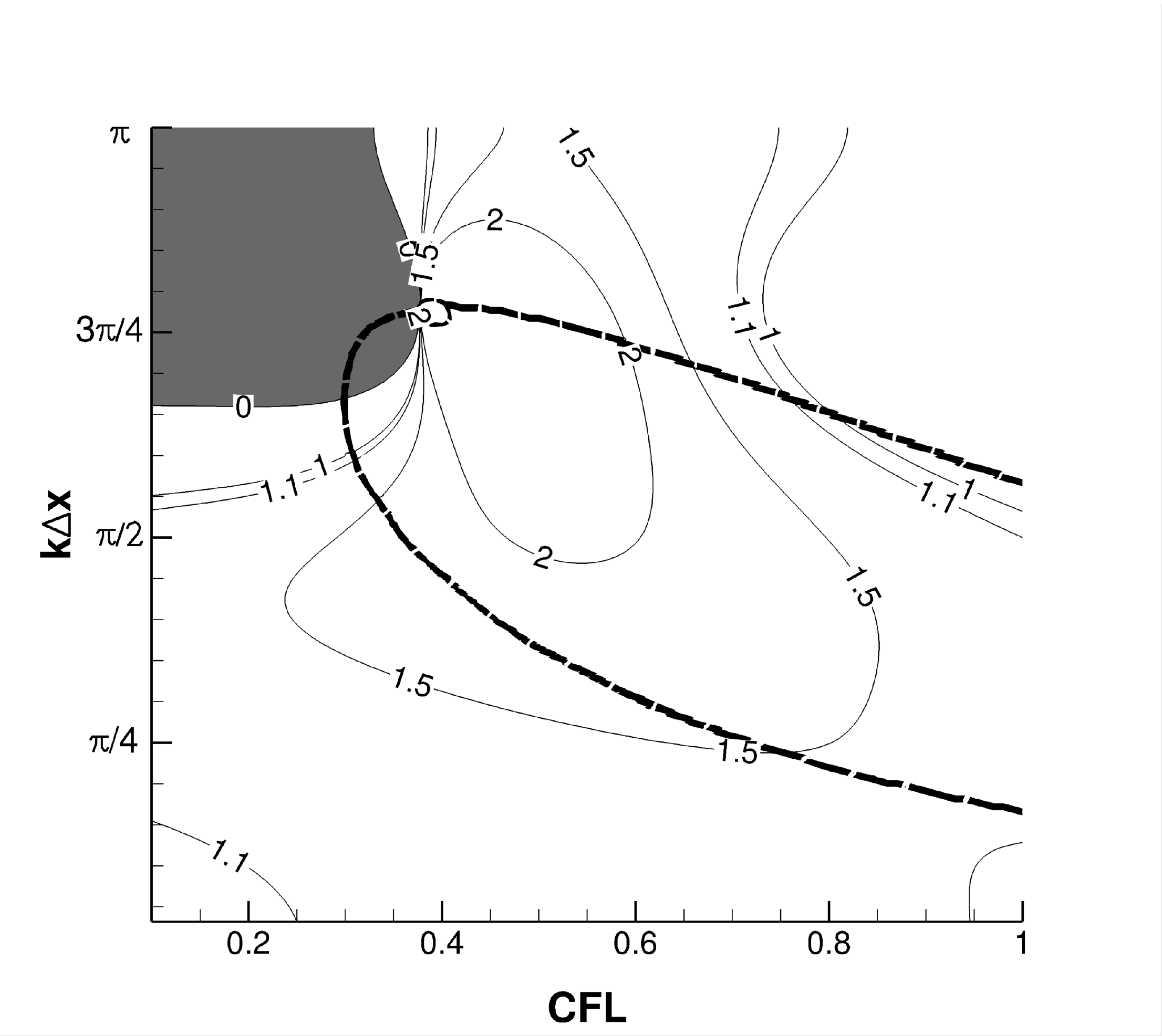} & \includegraphics [width=8cm]{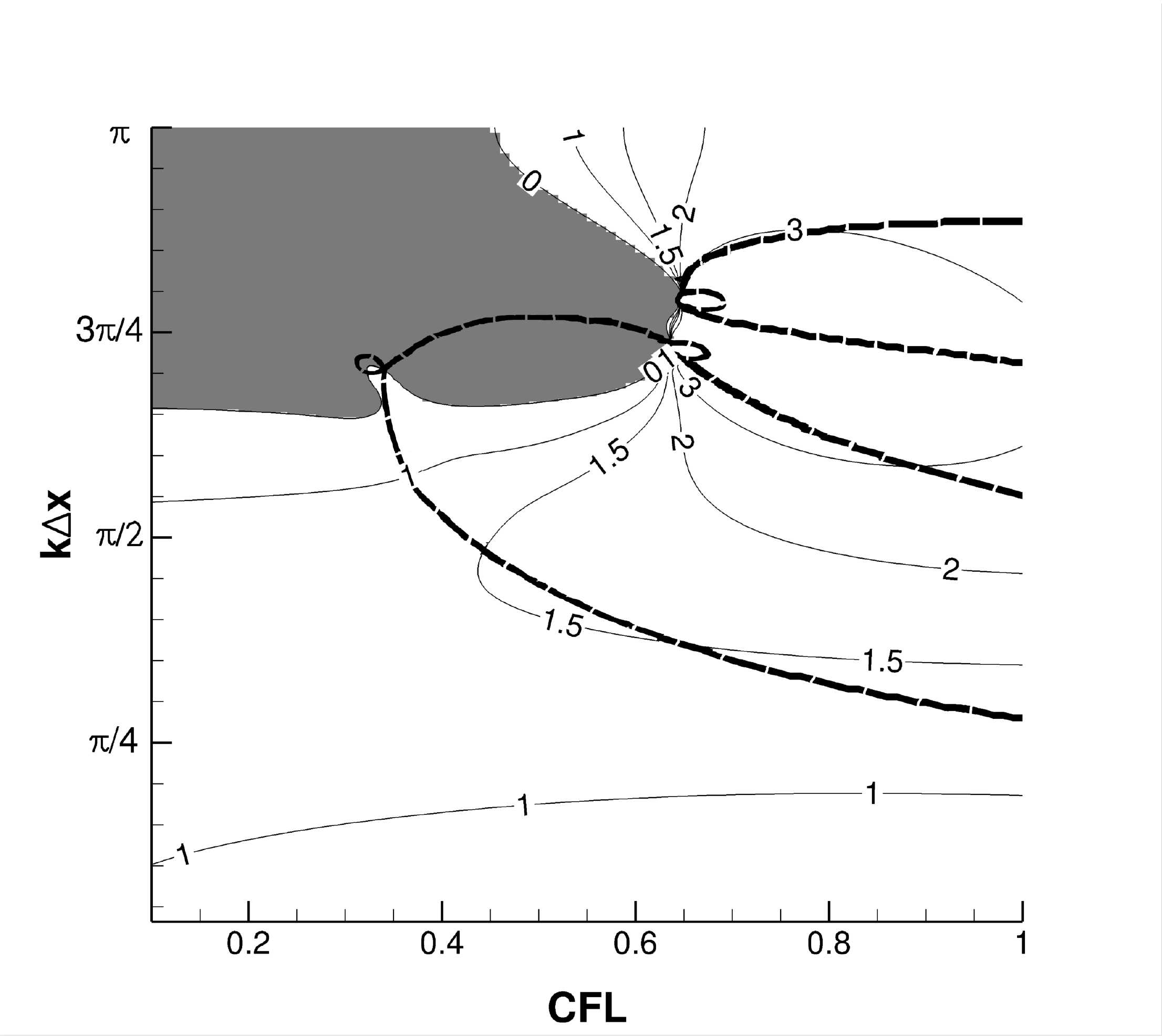}\\
\end{tabular}
\caption{ Isocontours of $V^{gN}$ for step UP and step DOWN with AION+TA time integration scheme\label{fig:qwavesAIONUP_DOWN}}
\end{center}
\end{figure}

The analysis presented in Sec.~\ref{sec:qwavesHeun} is now applied to AION+TA configuration, and
the effect of hybrid time integration
on the occurrence of $q-$waves is explained for configurations UP and DOWN. The analysis follows again the definition of the group velocity and
areas of negative group velocity are looked for.
The grey zone in Fig.~\ref{fig:qwavesAIONUP_DOWN} shows the area of negative-$V^{gN}$ group velocity for the AION+TA configuration and steps UP and DOWN. It can be concluded that negative group velocity waves appear essentially for
large wavenumbers and the area is larger for the step DOWN configuration than for the step UP and the associated CFL values differ.
Indeed, in case of hybrid time integration of Step DOWN, the CFL limit of the negative-$V^{gN}$ group velocity zone is CFL$=0.65$ whereas the CFL limit is $0.3$ in Heun part (Fig.~\ref{fig:qwaves_absG_HeunUP_DOWN}). The last question concerns the damping
of these $q-$waves using the dissipation of the scheme.

\begin{figure}[!htbp]\begin{center}
\begin{tabular}{cc}
\includegraphics[width=8cm]{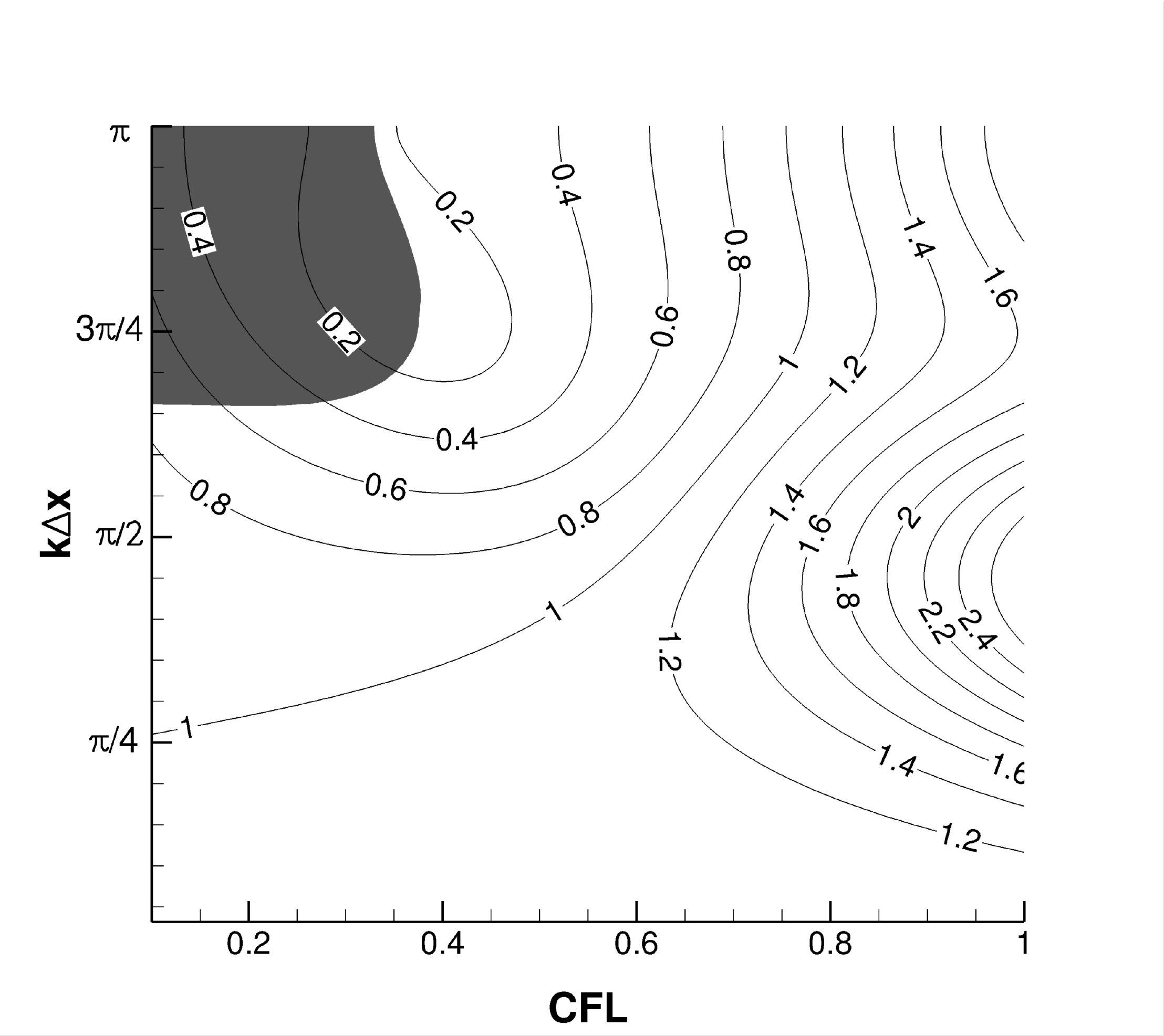} & \includegraphics [width=8cm]{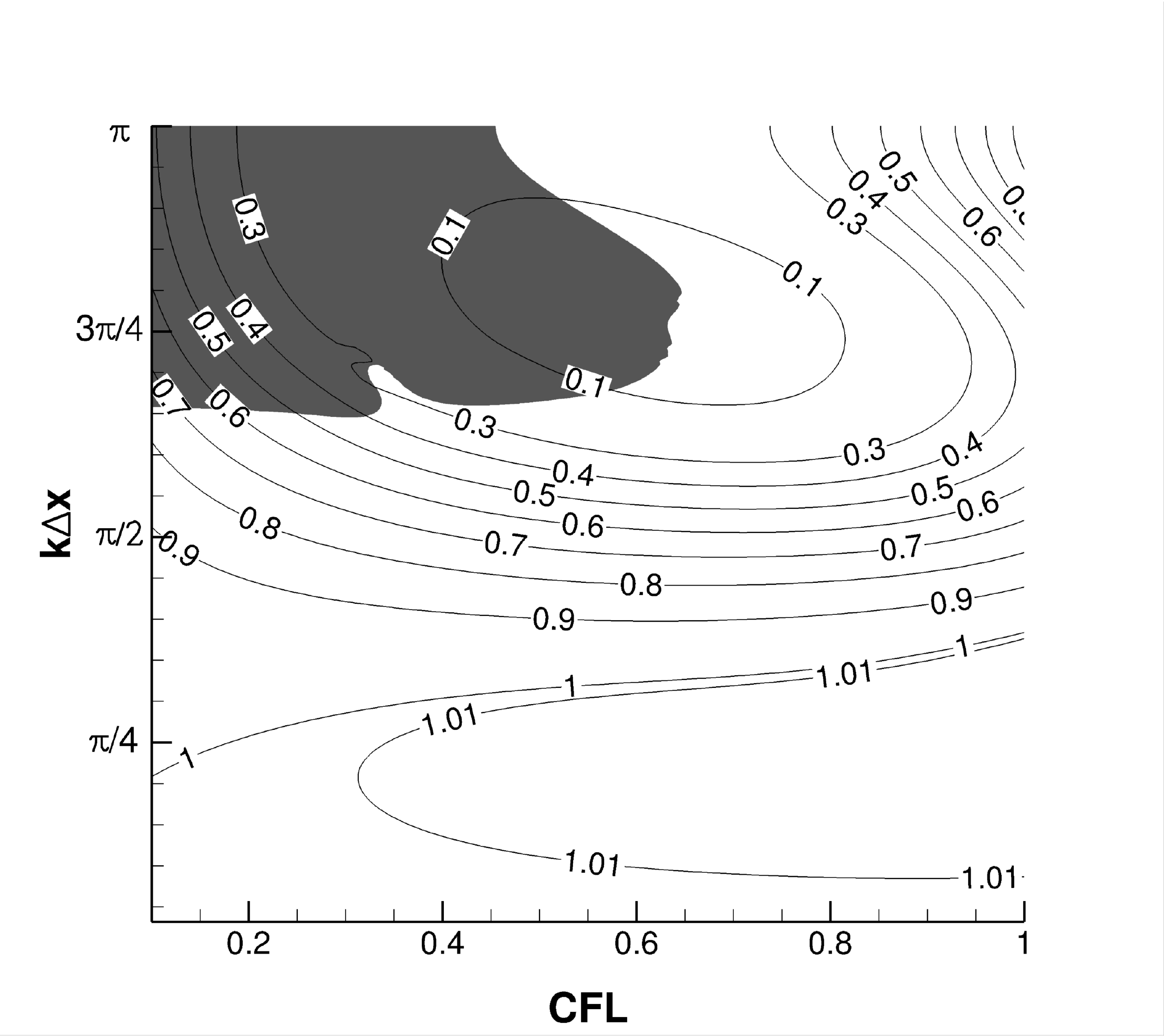}\\
\end{tabular}
\caption{ Isocontours of the dissipation $\mu_{\beta}$ and $\mu_{\gamma}$ with grey zone for negative group velocity\label{fig:qwaves_absG_AIONUP_DOWN}}
\end{center}
\end{figure}


According to Fig.~\ref{fig:qwaves_absG_AIONUP_DOWN}, it appears that in case of hybrid time integration,
the CFL limit of the negative-$V^{gN}$ zone corresponds to a dissipation $\mu>0.2$ for step UP and $\mu>0.1$ for step DOWN.
As a consequence, numerical dissipation can help in attenuating the $q-$waves.


The theoretical analysis provided in this paper is valid for regular grids only and it is of great importance
to analyse the time-adaptive AION scheme ability to keep global accuracy for irregular grid.
Several configurations are analysed in Sec.~\ref{sec:Validation}.

\section{Validation\label{sec:Validation}}

This section is devoted to the validation of the AION+TA scheme using several test cases of
increasing complexity, starting from 1D propagation problem to 2D Euler computation.

\subsection{Wave propagation problem}

The previous theoretical analysis highlighted the fact than Heun+TA and AION+TA do not behave similarly
at the interface between temporal classes for regular grid, considering that the theoretical analysis
for irregular grid is cumbersome (see Vichnevetsky~\cite{Vichnevetsky_MCS_23_1981}). Here the goal
is to perform the numerical analysis of Heun+TA and AION+TA time integration schemes
for an irregular mesh. To do so, a wave packet is propagated in a domain in order to introduce a frequency content inside the computational domain.
A non-periodic computational domain of length $L_x=270~m$ composed of $N=1024$ cells, is initialized with:
\begin{equation}
\begin{aligned}
y(x,0)=cos\bigg[{2\pi}{f_e}(x-x_c)\bigg]\exp\left(-\frac{x-x_c}{K}\right)
\end{aligned}
\end{equation}
with $K=200$, $f_e=1/\pi$ and $x_c=90~m$. The wave packet is advected at velocity $c=1~m/s$. Moreover,
while the theoretical analysis was performed using a fixed mesh size, here, two temporal classes of cells with different sizes are introduced.
Starting from the space size $\Delta x_{max}$ for the largest cells, the second class with
the refined mesh is defined by:
\begin{equation}
\Delta x_j=\frac{1}{2}\Delta x_{max}\text{ for } \frac{N}{2}-100 \leq j \leq \frac{N}{2}+100  \\
\end{equation}
and $\Delta x_{max}=0.29~m$. Step configurations are localised at $x=120.2~m$ (step DOWN) and $x=149.5~m$ (step UP).
Time integration is first performed with the standard Heun+TA scheme until $t=100~s$ at the local time
step $\Delta t$ in cells of class rank $1$ and $\Delta t/2$ in cells of class rank $0$. Here, $\Delta t$ corresponds to the time step associated with the largest cells, at CFL$=0.6$, which means that:
$$
\hbox{CFL}=\frac{\Delta t}{c\Delta x_{max}}.
$$

\begin{figure}[!htbp]\begin{center}
\begin{tabular}{cc}
\includegraphics[width=8cm]{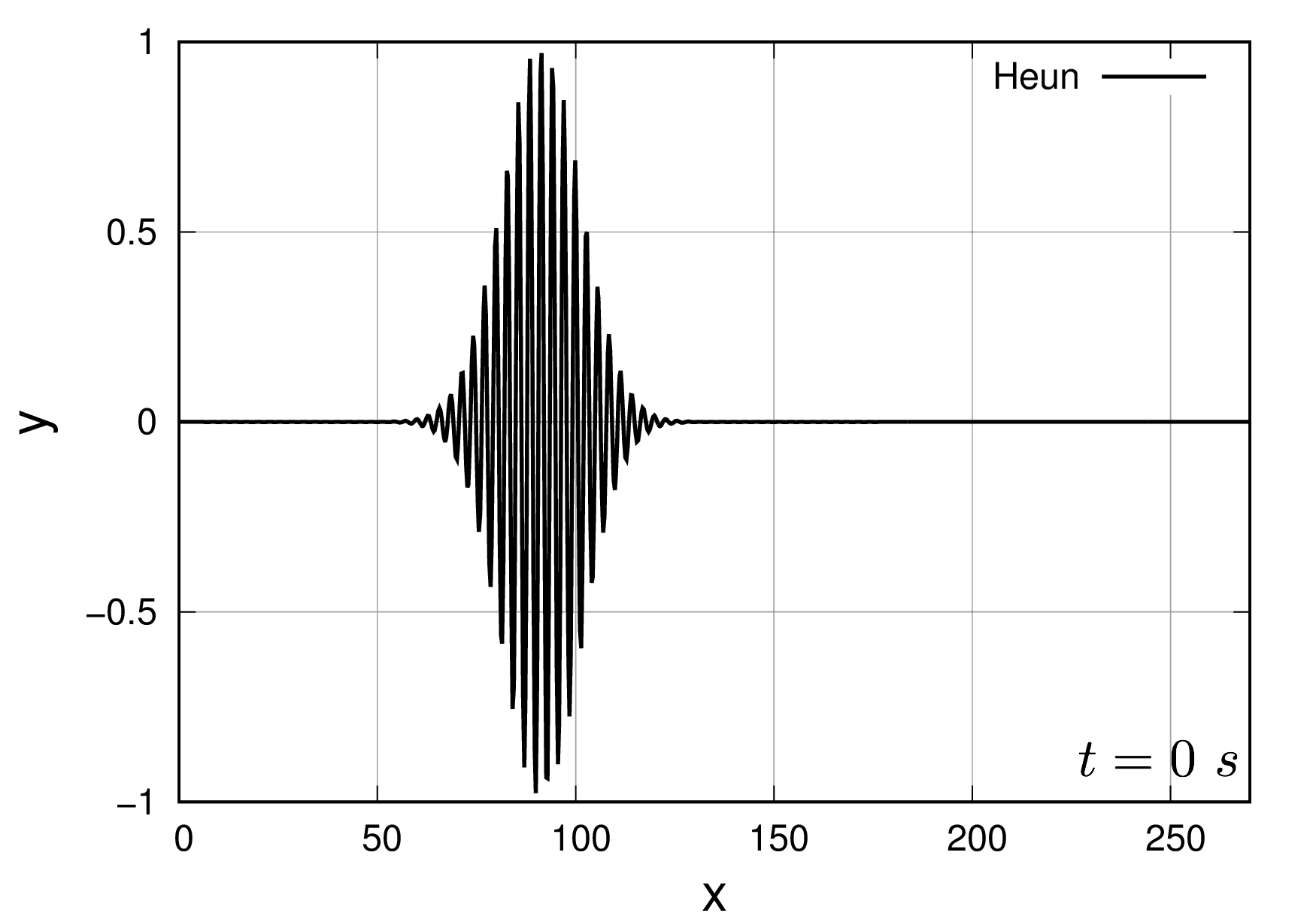} & \includegraphics [width=8cm]{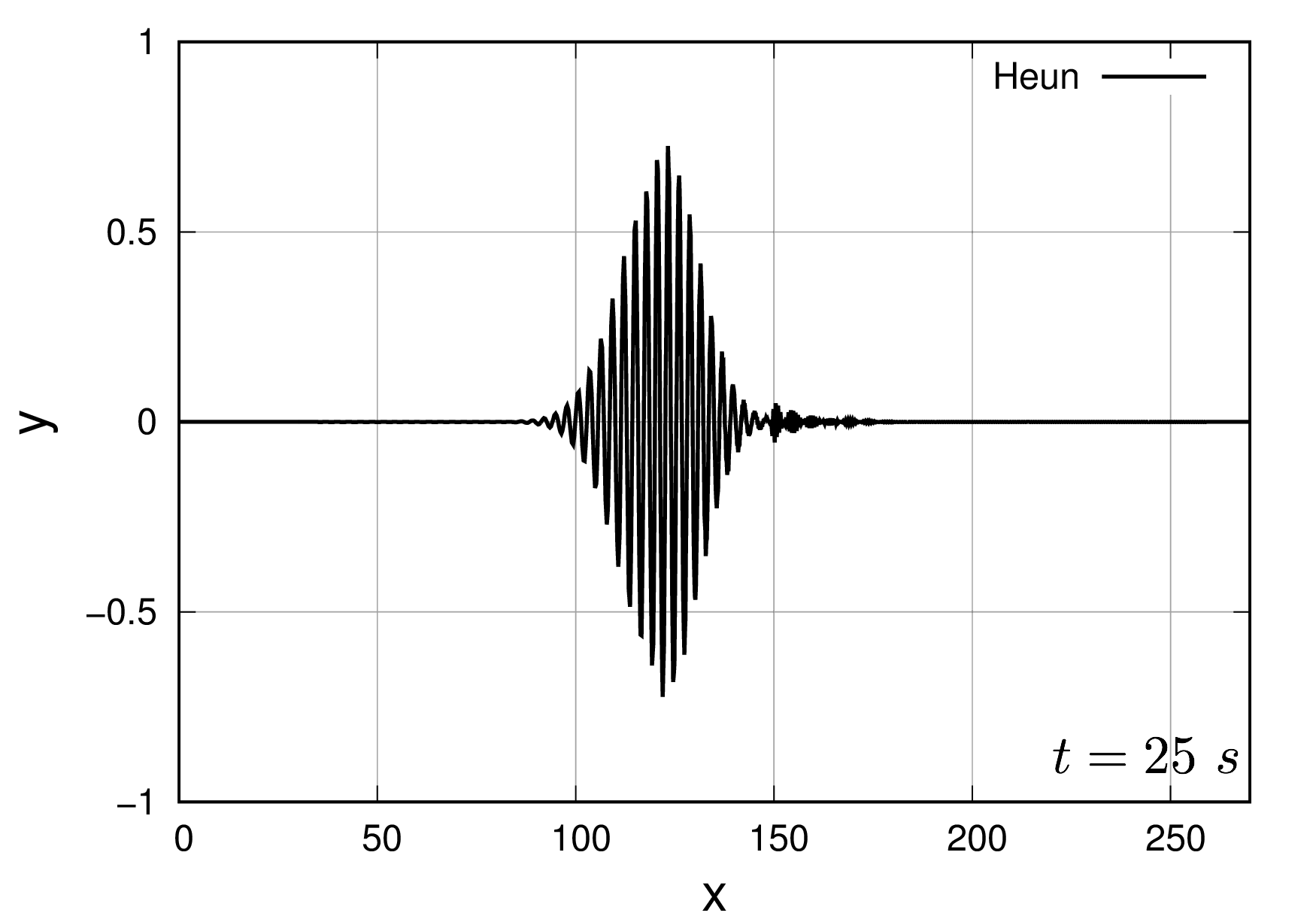}\\
\end{tabular}
\includegraphics [width=8cm]{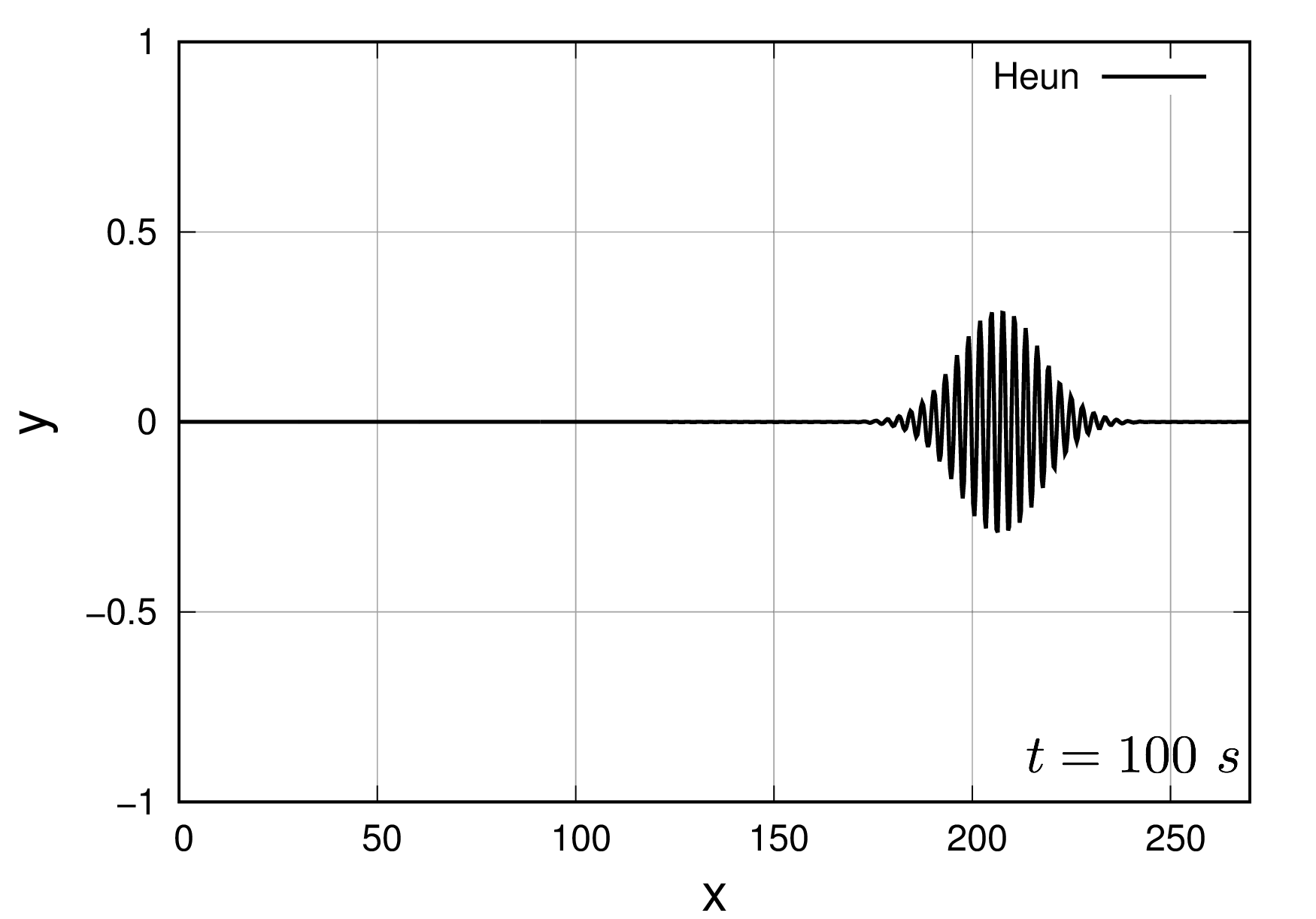}
\caption{Propagation of wavepacket with Heun+TA until $t=100s$ (CFL$=0.6$). \label{fig:waves_Heun_06}}
\end{center}
\end{figure}

Fig.~\ref{fig:waves_Heun_06} shows, as expected due to the theoretical analysis,
that a sinusoidal component of the wave-packet with a
certain wavenumber ($p-$waves) is amplified when the wave-packet goes through the step configurations
(at $t=25~s$ in Fig.~\ref{fig:waves_Heun_06}).
A Fast Fourier Transform (FFT) of the numerical solution $y$ is performed to highlight the phenomena, with the sampling frequency equal to $1~Hz$.

\begin{figure}[!htbp]\begin{center}
\begin{tabular}{cc}
\includegraphics[width=8cm]{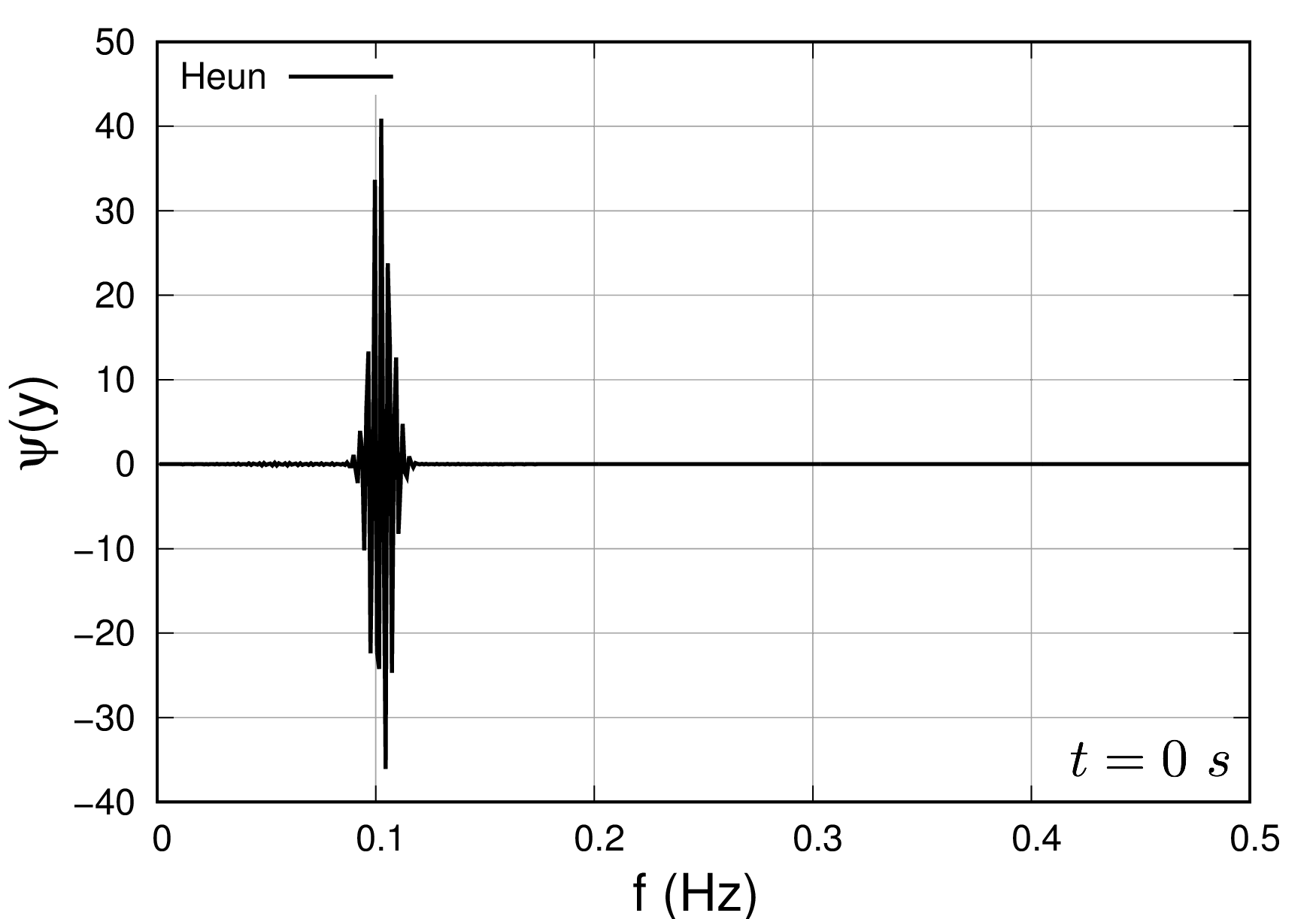} & \includegraphics [width=8cm]{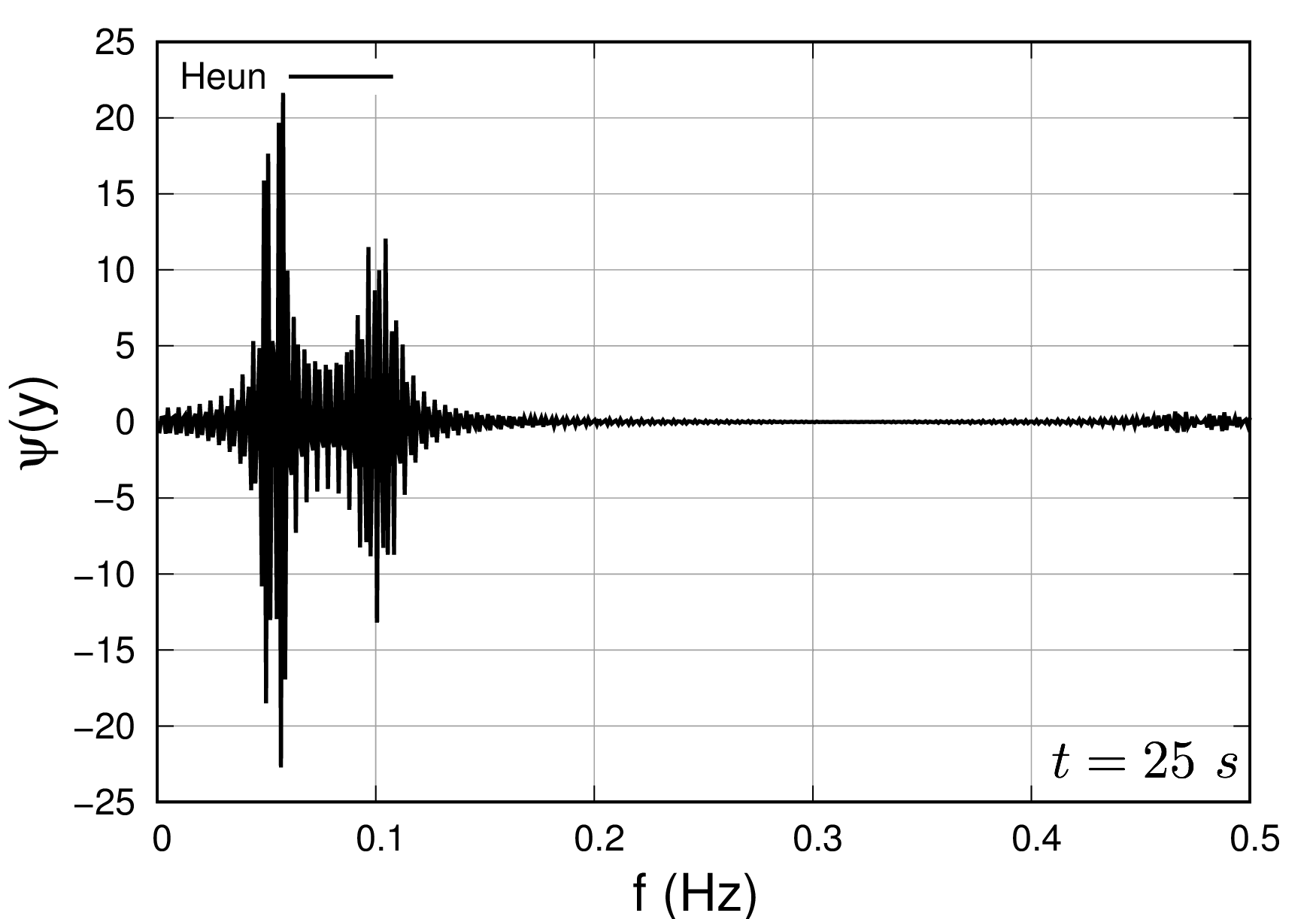}\\
\end{tabular}
\includegraphics [width=8cm]{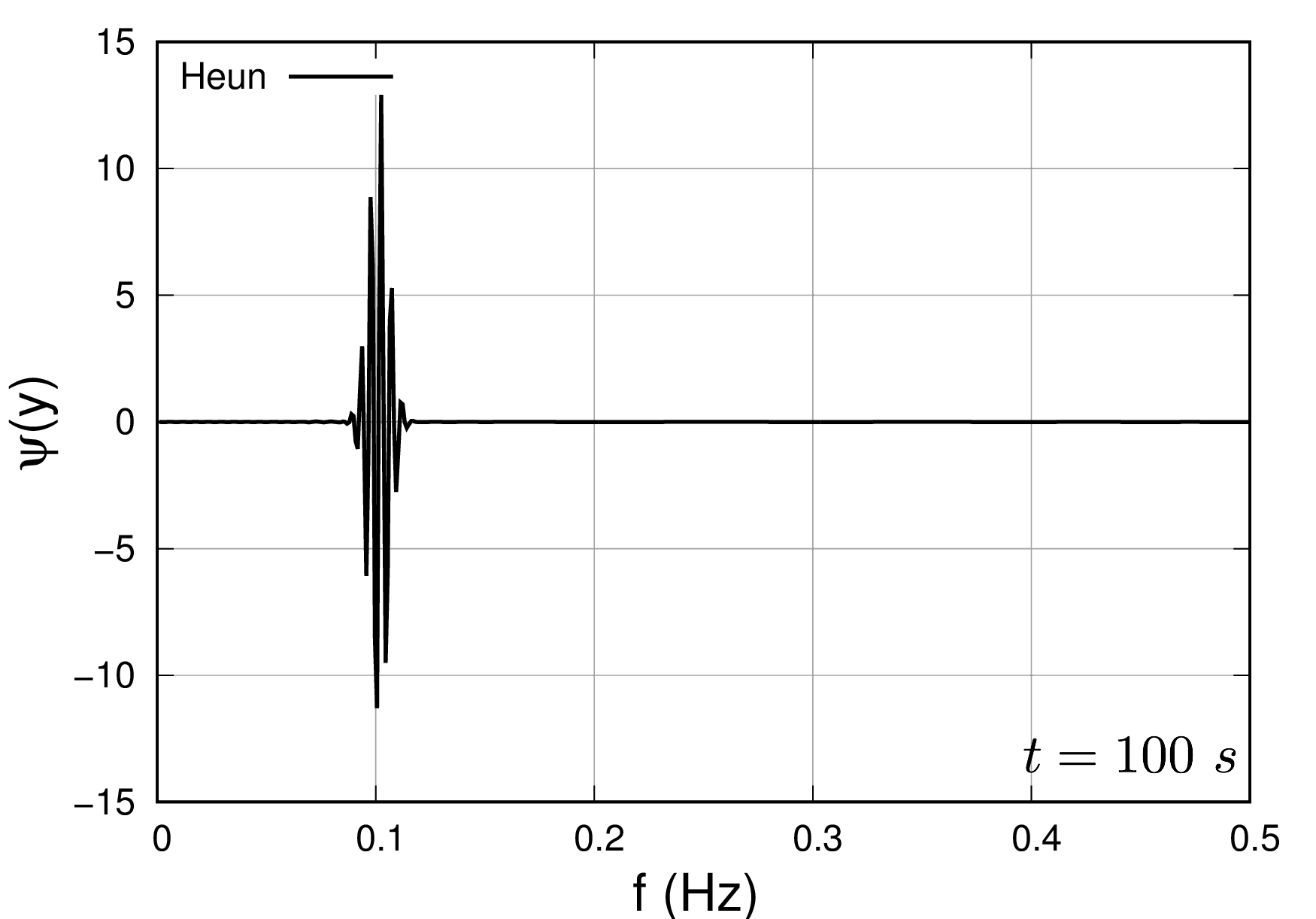}
\caption{Fast Fourier transform $\Psi(y)$ of the numerical solution obtained thanks to Heun+TA at $t=0~s,~25~s,~100~s$. \label{fig:waves_Heun_06_FFT}}
\end{center}
\end{figure}

The main frequency observed from the FFT of $y$ (named $\Psi(y)$ in the following) at time $t=0~s$
(Fig.~\ref{fig:waves_Heun_06_FFT}) is equal to $0.1~Hz$. This frequency is linked to the discretisation
of the signal according to $\Delta x_{max}$. The second dominant frequency obtained at $t=25~s$
is equal to $0.05~Hz$ and corresponds to the discretisation of the signal according to the space step of the
finest part of the domain ($\Delta x_{max}/2$). The amplified $p$-waves observed at time $t=25~s$ corresponds to the small
value of the FFT obtained at the frequency near $0.46~Hz$. This frequency is strongly higher than the main
frequencies that composed the signal as it is observed in Fig.~\ref{fig:waves_Heun_06}. Hence, Heun+TA is shown
to amplify some waves.

\begin{figure}[!htbp]\begin{center}
\begin{tabular}{cc}
\includegraphics[width=8cm]{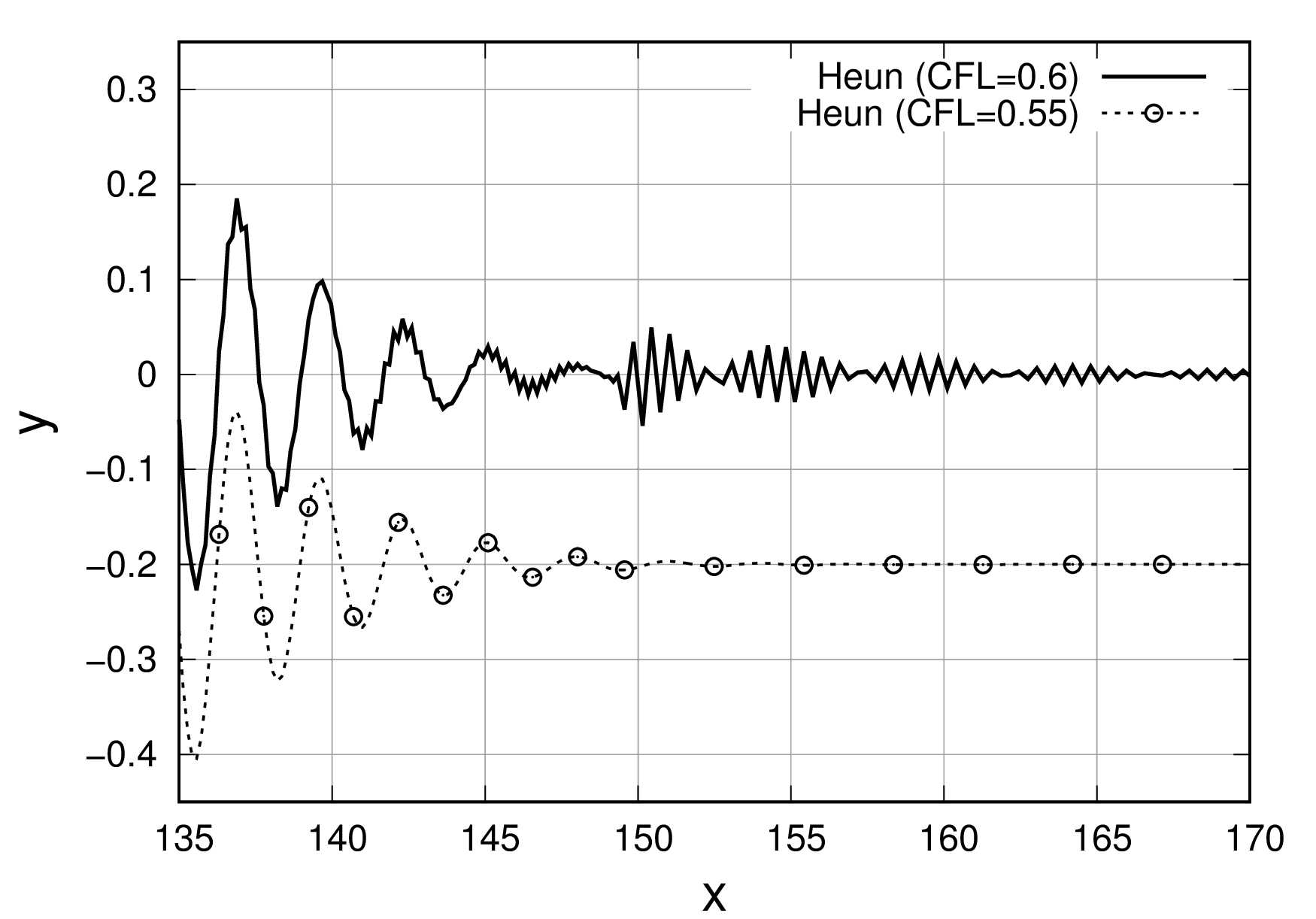} & \includegraphics [width=8cm]{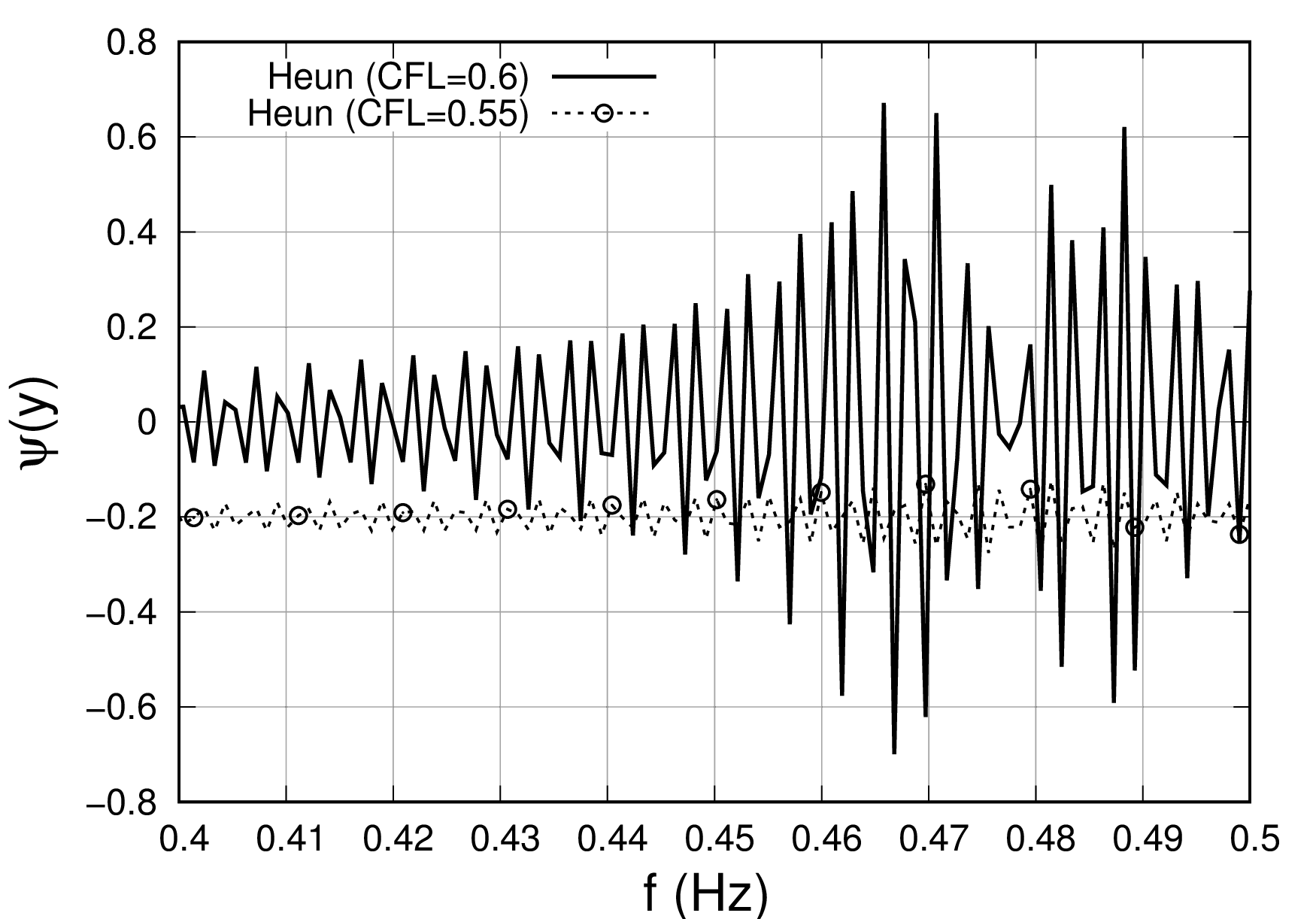}\\
\end{tabular}
\caption{Comparaison of numerical solution time integrated by Heun+TA scheme with CFL=0.55 and CFL=0.6 at $t=25~s$. \label{fig:waves_Heun055_06}}
\end{center}
\end{figure}

Amplification was shown to appear for a computation at CFL=0.6, near the stability limit
of Heun's scheme. If amplification occurs due to the local instability of the space/time scheme,
choosing a lower CFL value could reduce the amplified spectrum and in order to confirm our assumption,
the very same simulation is now performed at CFL=0.55 until $t=25~s$ with the Heun+TA scheme.
Fig.~\ref{fig:waves_Heun055_06} focuses on the wave packet advected until $t=25~s$. Two computations,
with CFL=0.6 and CFL=0.55 are compared and in order to increase readiness, the numerical solution for
CFL=0.55 is translated by -0.2 in y-axis. \MM{The amplification present at CFL=0.6 disappears at CFL=0.55.
The global stability properties of the Heun+TA
scheme involve amplification of $p-$waves at CFL=0.6}

The same simulation is time-integrated thanks to the AION+TA scheme.
For the hybrid part of the AION scheme, the parameter $\omega_j$ for cell $j$
is chosen such that:
\begin{equation}
\left\{
\begin{array}{ll}
  \displaystyle \omega_j=\alpha \, \omega_{j-1} &  \hbox{ for } \displaystyle\frac{N}{2}-102 \leq j \leq \frac{N}{2}  \\
  \vspace*{-3mm} &\\
  \displaystyle \omega_j=\frac{1}{\alpha} \, \omega_{j-1} & \hbox{ for } \displaystyle\frac{N}{2}+1 \leq j \leq \frac{N}{2}+102  \\
  \vspace*{-3mm} &\\
  \displaystyle \omega_j = 1 & \hbox{elsewhere,}
\end{array}
\right.
\end{equation}
with $\alpha=0.90$, and it is also used for the steps UP and DOWN introduced previously.


\begin{figure}[!htbp]\begin{center}
\begin{tabular}{cc}
\includegraphics[width=8cm]{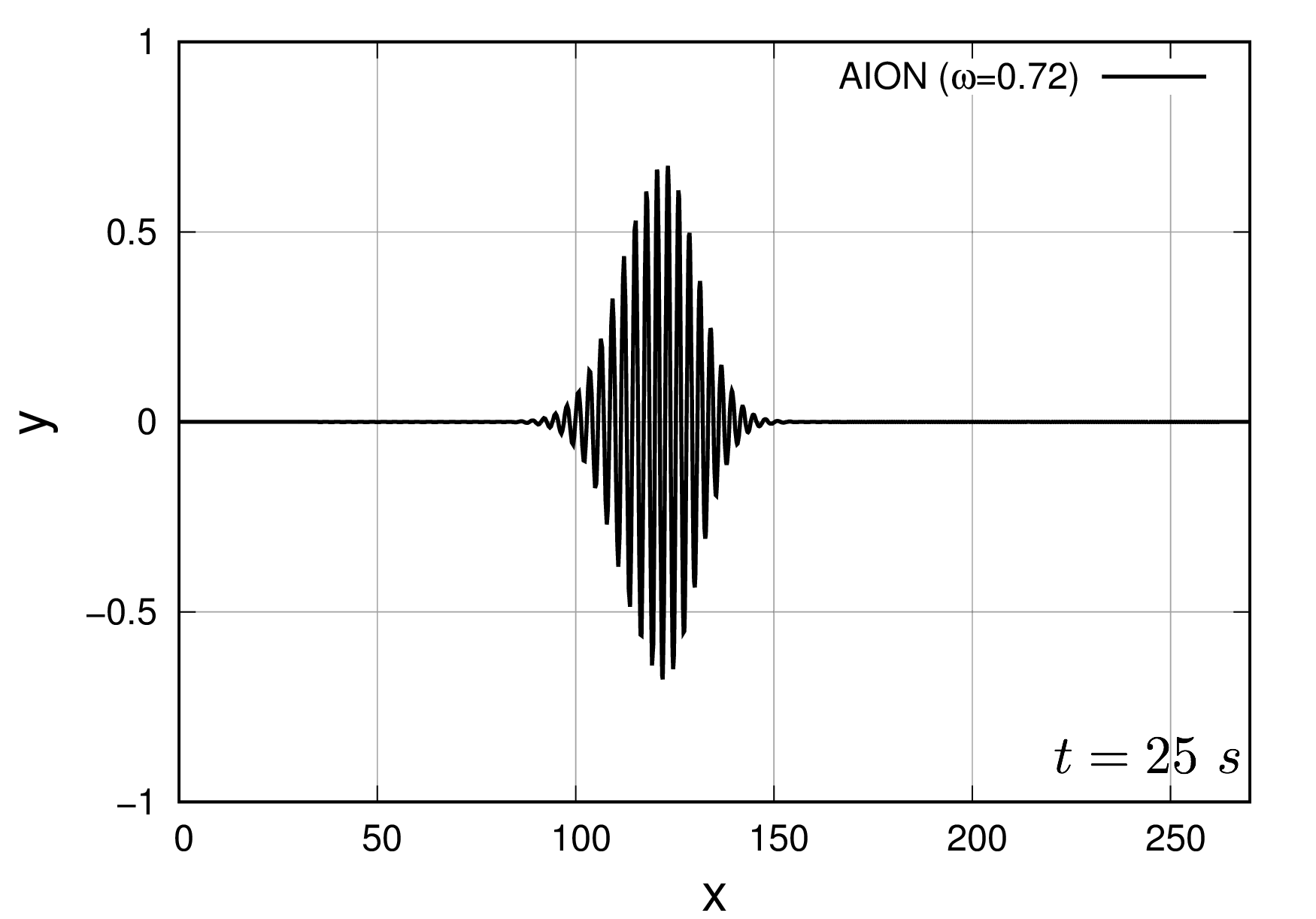} & \includegraphics [width=8cm]{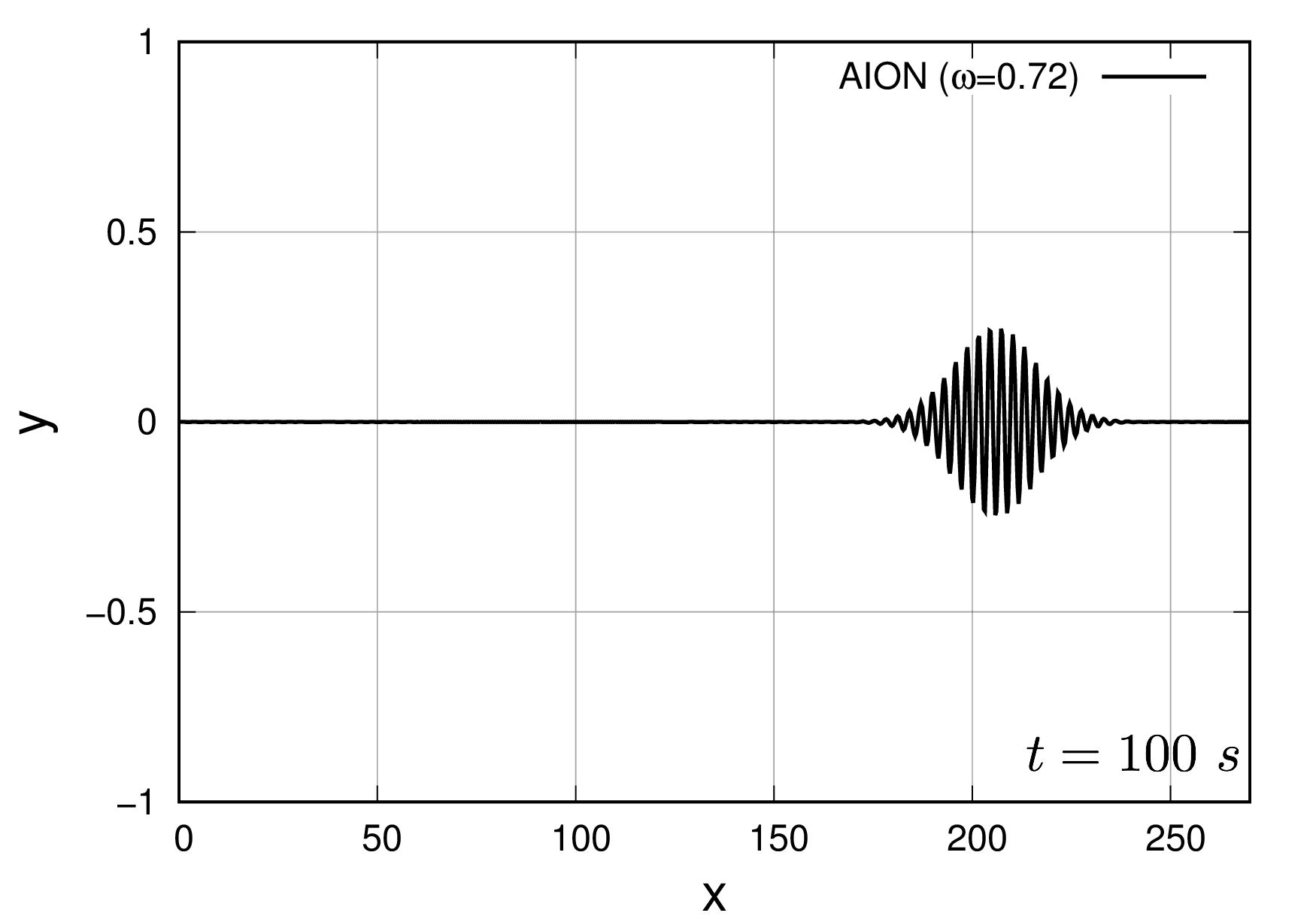}\\
\end{tabular}
\caption{Propagation of wavepacket with AION+TA until $t=100~s$ (CFL$=0.6$). \label{fig:waves_AION_06}}
\end{center}
\end{figure}

\begin{figure}[!htbp]\begin{center}
\begin{tabular}{cc}
\includegraphics[width=8cm]{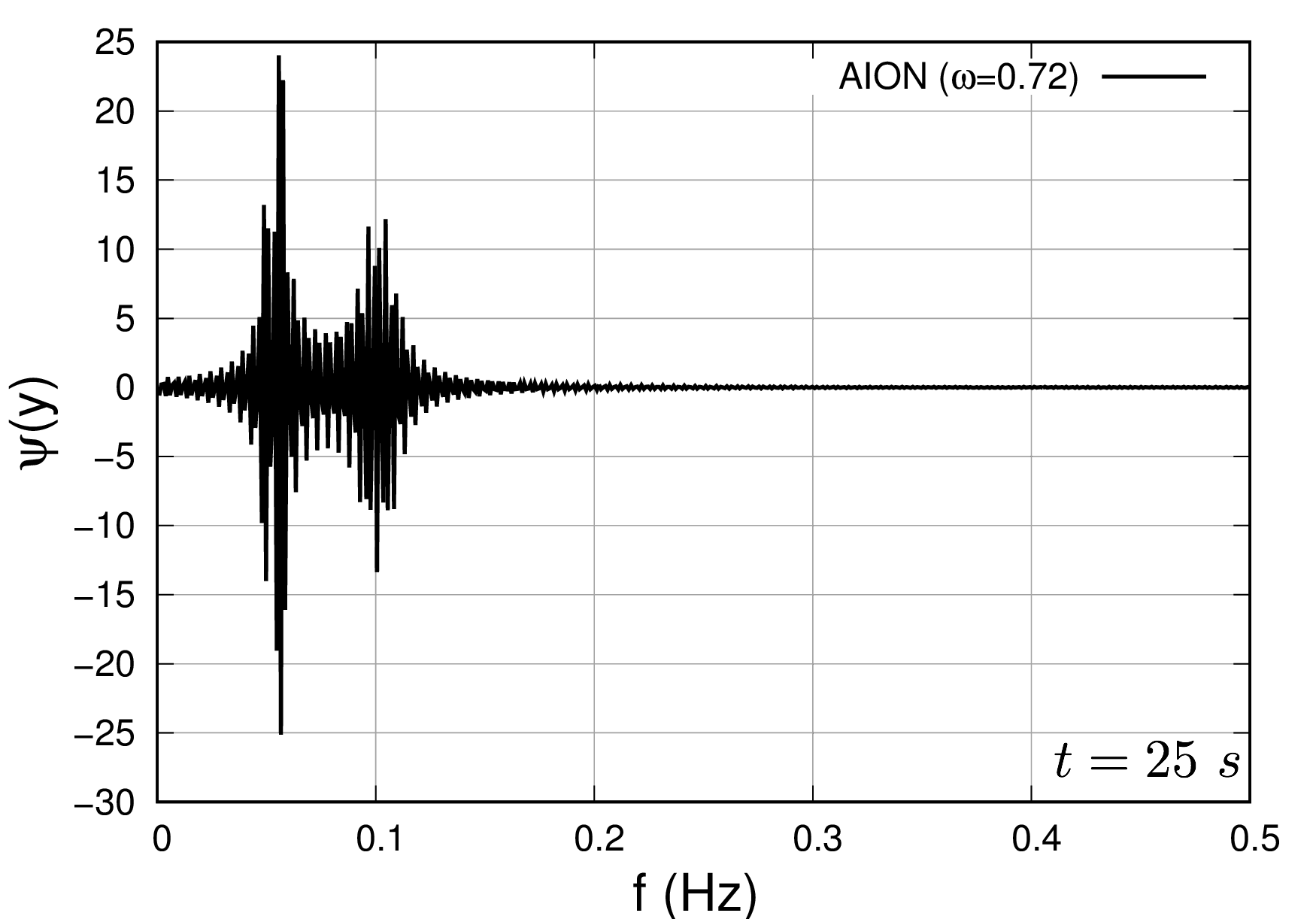} & \includegraphics [width=8cm]{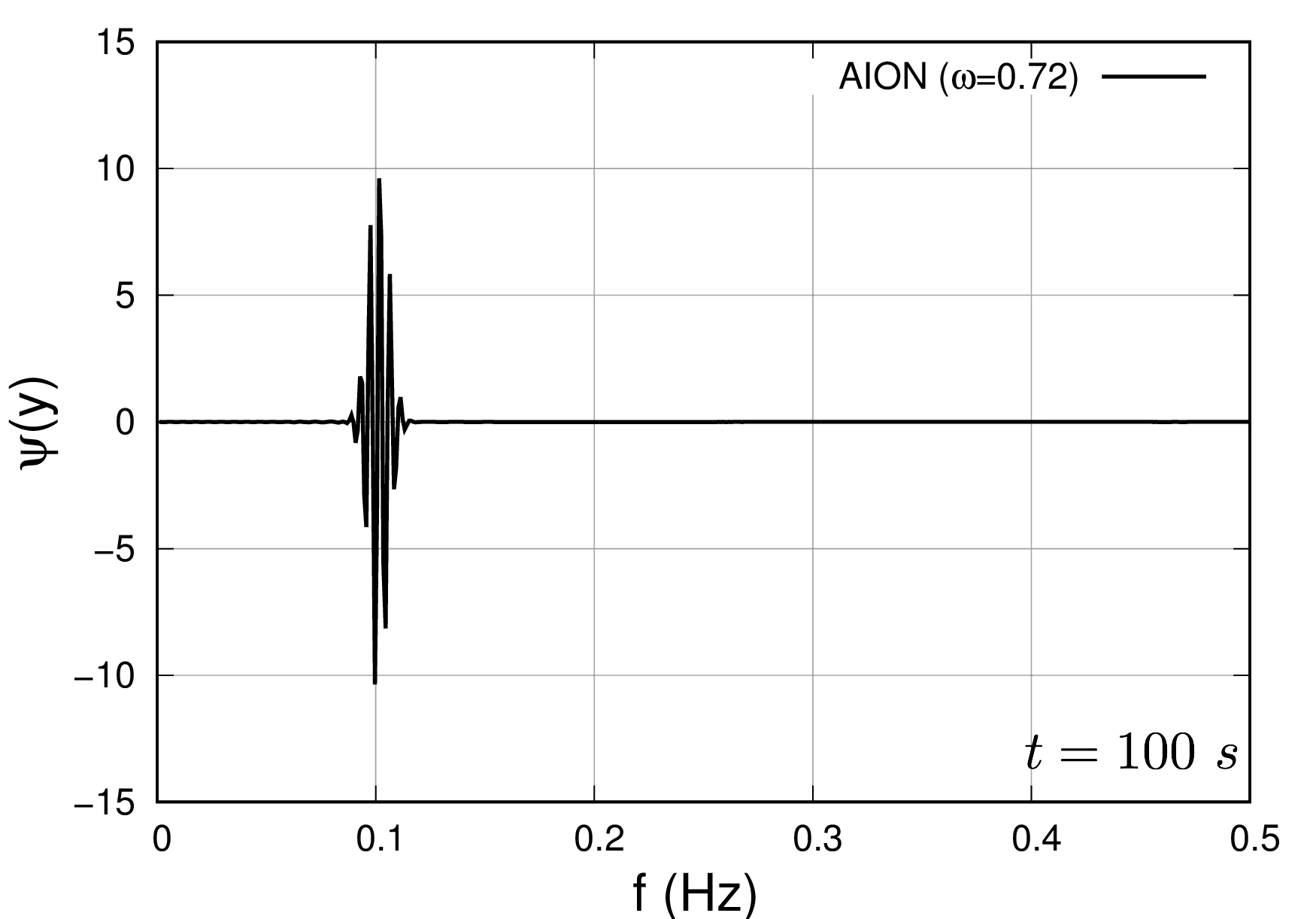}\\
\end{tabular}
\caption{Fast Fourier transform $\Psi (y)$ of the numerical solution obtained thanks to AION+TA at $t=25~s,~100~s$ (step configuration in hybrid part). \label{fig:waves_AION_06_FFT}}
\end{center}
\end{figure}

According to Fig.~\ref{fig:waves_AION_06}, no amplified $p-$waves appear in the computational domain using
the AION+TA time integration. This status is confirmed by the FFT at the frequency
$0.46~Hz$ (Fig.~\ref{fig:waves_AION_06_FFT}).
Hence, even if the AION+TA scheme uses many ingredients of the Heun+TA scheme,
it is shown to be more stable and to allow larger CFL values or stable time steps: AION+TA can be seen
as an enhancement of Heun+TA time integrator.


The next test case is dedicated to the analysis of the scheme ability to handle compressible effects.

\subsection{Sod's tube}

The coupled space/time analysis was always performed using the local regularity of the flow. Here,
our goal is to analyse the scheme behaviour on an academic case including discontinuities.
The Sod's tube is an unsteady inviscid one-dimensional test case where the computational domain of
length $L_x = 1~m$ is split into two parts separated by a membrane located
initially at $x=0.5L_x$. The initial flow is defined as:
\begin{equation}
\begin{aligned}
\begin{pmatrix}
\rho_L \\[3mm]
p_L \\[3mm]
U_L \\
\end{pmatrix}=\begin{pmatrix}
1.0\\[3mm]
1.0 \\[3mm]
0.0 \\
\end{pmatrix}
, \begin{pmatrix}
\rho_R \\[3mm]
p_R \\[3mm]
U_R \\
\end{pmatrix}=\begin{pmatrix}
0.125\\[3mm]
0.1 \\[3mm]
0.0 \\
\end{pmatrix},
\end{aligned}
\end{equation}
where $L$ refers to the left side and $R$ to the right side of the membrane. At $t=0$, the membrane is broken,
and waves move inside the computational domain. The solution is composed of a rarefaction wave, a contact discontinuity, and a shock at the final time $t=0.2~s$.

The Euler equations are solved using a 1-exact (second-order) space scheme using the Roe approximate Riemann solver.
In order to avoid spurious oscillations and to obtain a TVD solution, the minmod slope limiter~\cite{Roe_1986_ARFM} is used.
The computational domain is composed of $N=300$ cells located in regular parts with a uniform mesh size and
in irregular parts with non-uniform mesh size such that :
\begin{equation}
\begin{aligned}
\text{for} \text{ } \text{ } \frac{N}{2}-45 \leq j \leq \frac{N}{2} \text{ }: \Delta x_j=\alpha \, \Delta x_{j-1} \\
\text{for} \text{ } \text{ } \frac{N}{2}+1 \leq j \leq \frac{N}{2}+45 \text{ }: \Delta x_j=\frac{1}{\alpha} \, \Delta x_{j-1} \\
\end{aligned}
\end{equation}
with $\alpha=0.973$.
Time integration is performed using the temporal adaptive procedure, and
cells are associated to rank $K$ using:
\begin{equation}
K=int\bigg[\frac{\ln(\Delta t_j/\Delta t_{min})}{\ln(2)}\bigg]
\end{equation}
with
$$
\Delta t_j = CFL \, \frac{\Delta x_j}{\| \vec{v_j}\| + c_j}.
$$
Two temporal classes of cells are obtained such as the local time step of the computation is equal to
$\Delta t_{max}/2$ in cells of class $0$ and $\Delta t_{max}$ in cells of class $1$, where $\Delta t_{max}$
is the maximal allowed time step in the whole domain. For the AION time integration, the parameter
$\omega_j$ is controlled as:
\begin{equation}
\begin{aligned}
\text{for} \text{ } \text{ } \frac{N}{2}-37 \leq j \leq \frac{N}{2} \text{ }: & \text{ }\omega_j=\alpha.\omega_{j-1} \\
\text{for} \text{ } \text{ } \frac{N}{2}+1 \leq j \leq \frac{N}{2}+37 \text{ }: & \text{ }\omega_j=\frac{1}{\alpha}.\omega_{j-1} \\
\text{elsewhere} \text{ } \text{ } & \text{ }\omega_j=1
\end{aligned}
\end{equation}

\begin{figure}[!htbp]\begin{center}
\begin{tabular}{cc}
\includegraphics[width=8cm]{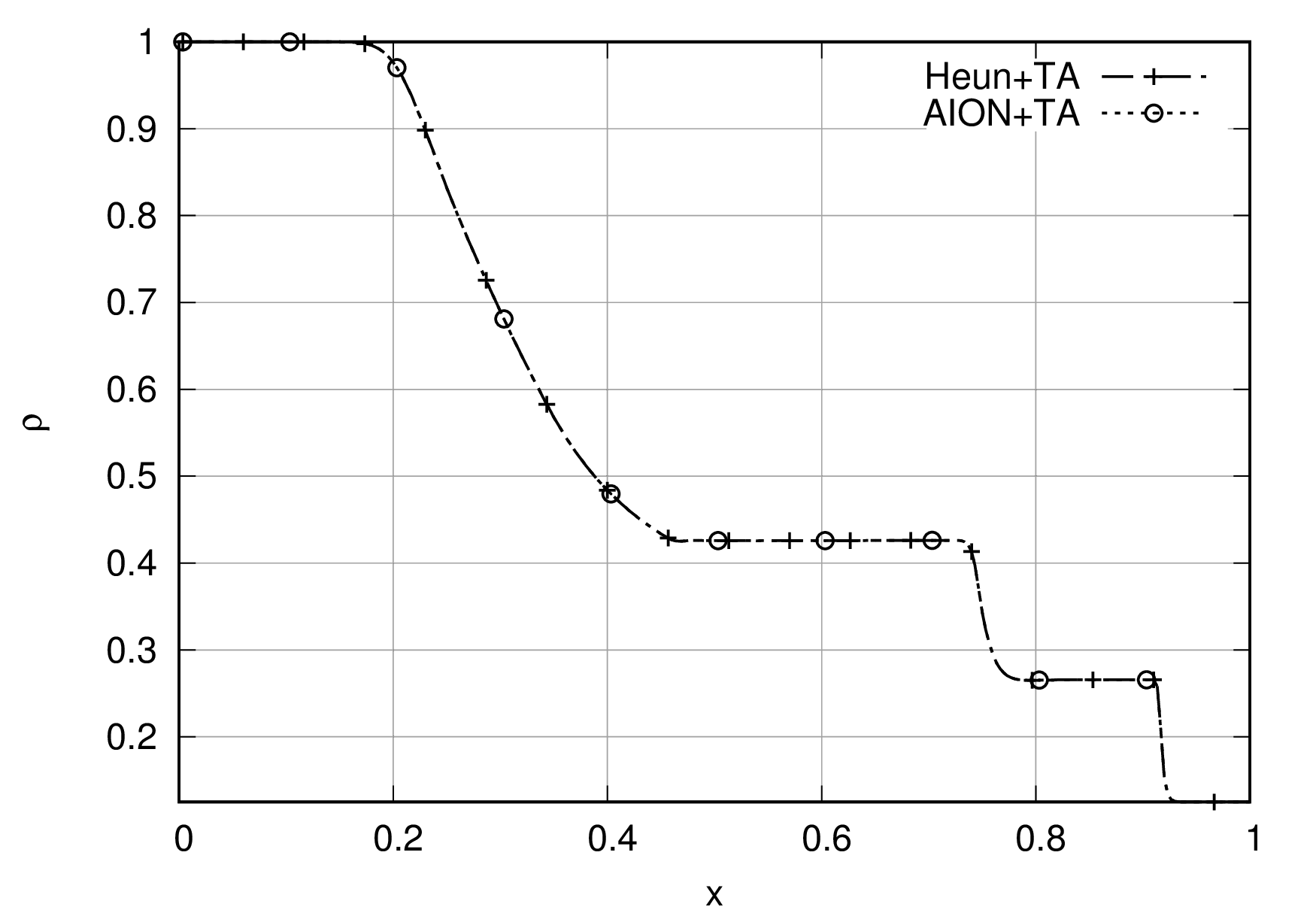} & \includegraphics [width=8cm]{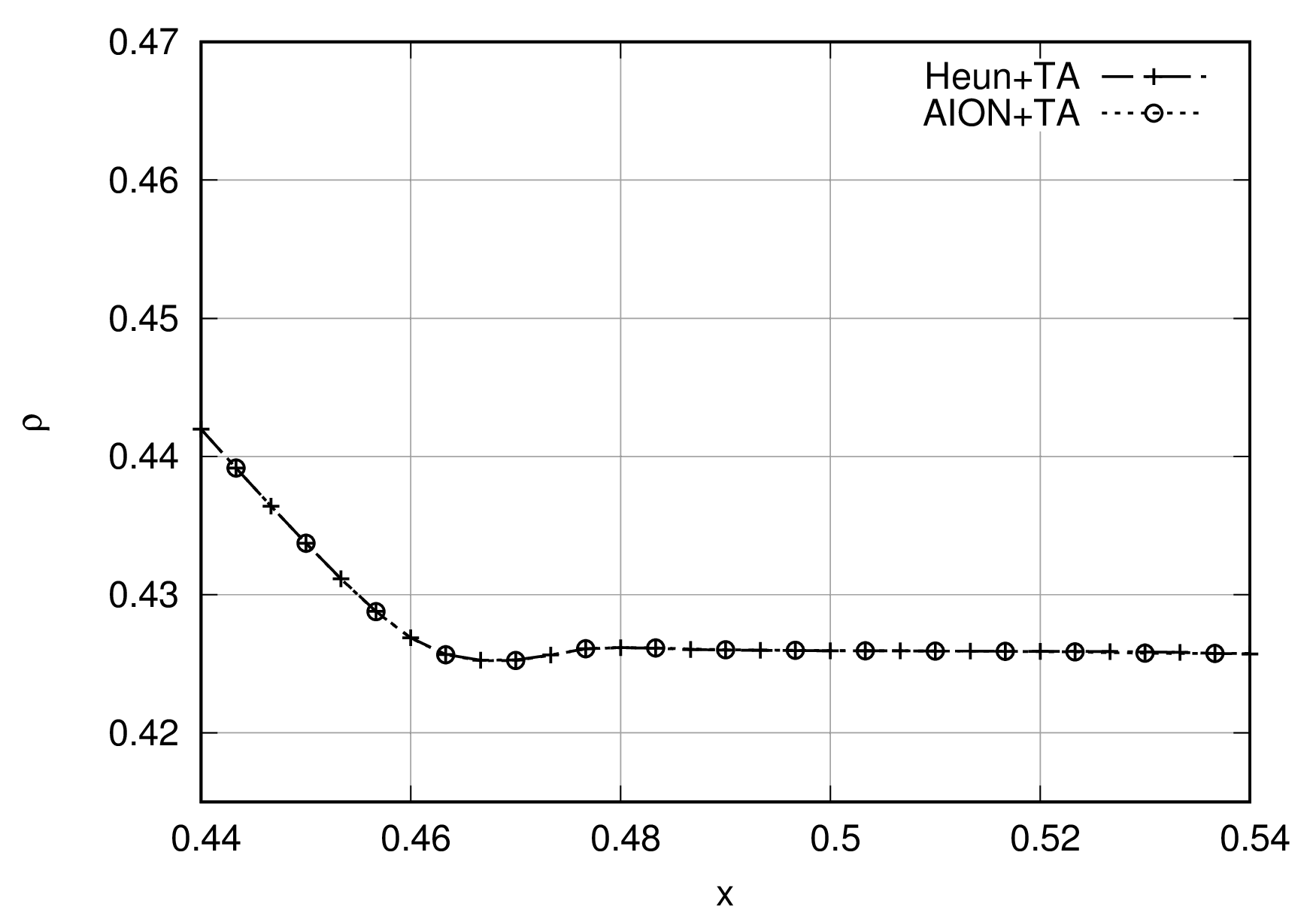}\\
\end{tabular}
\includegraphics [width=8cm]{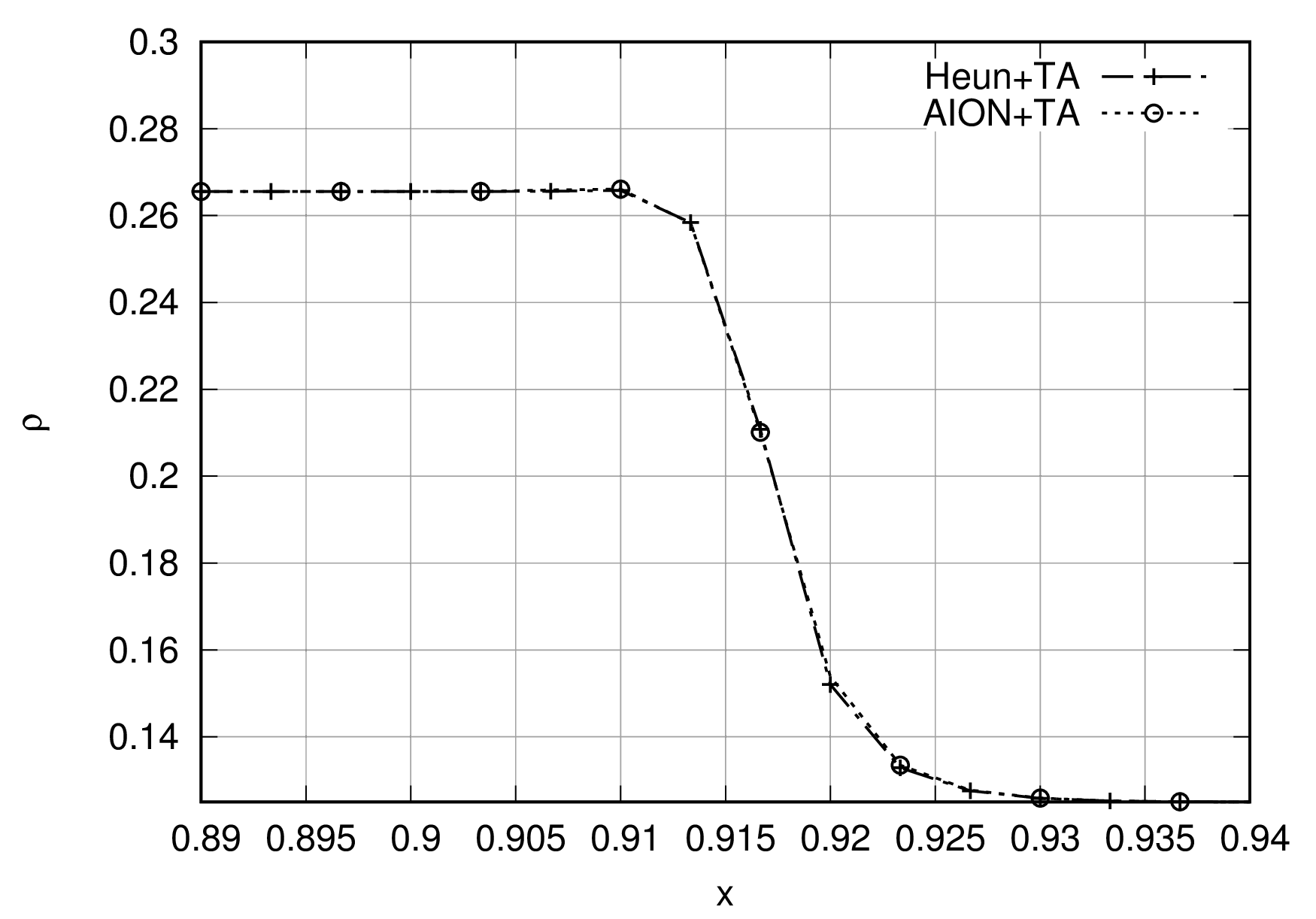}
\caption{Sod' shock tube at CFL$=0.1$. Global view of the density profile at $t=0.2s$ and close-up views
near the rarefaction wave and near the shock. \label{fig:rho_Sod_CFL0.1}}
\end{center}
\end{figure}

\begin{figure}[!htbp]\begin{center}
\begin{tabular}{cc}
\includegraphics[width=8cm]{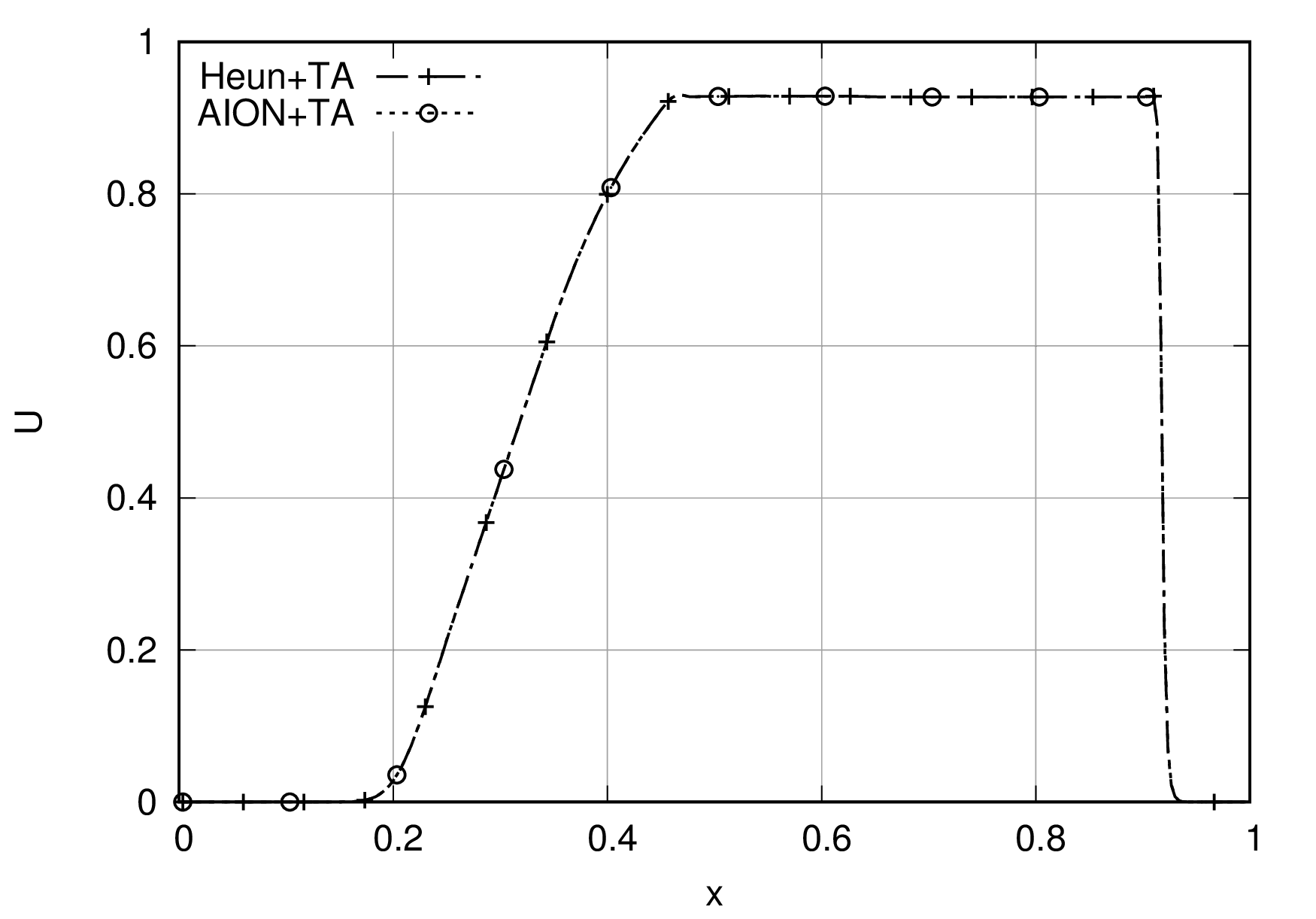} & \includegraphics [width=8cm]{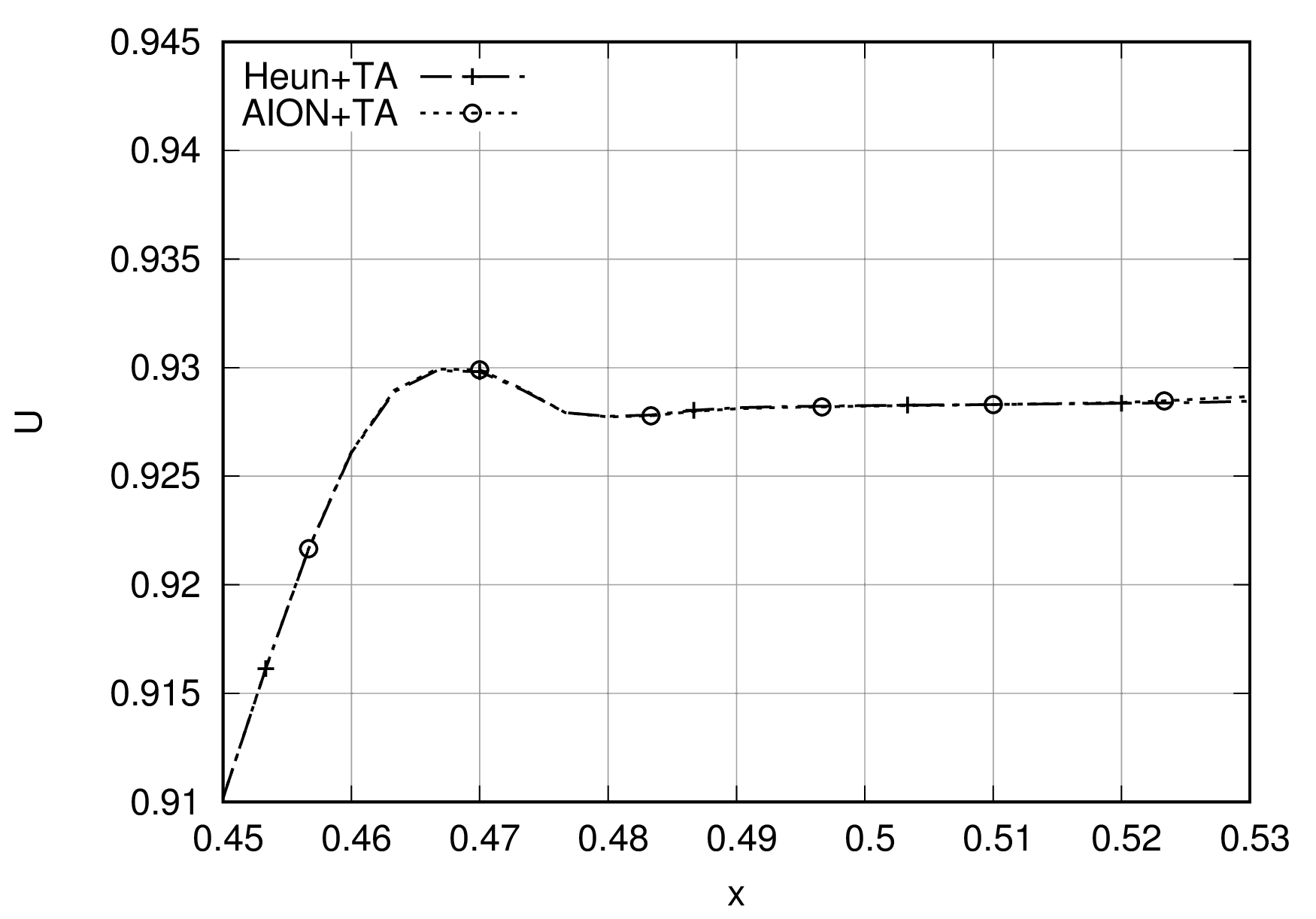}\\
\end{tabular}
\includegraphics [width=8cm]{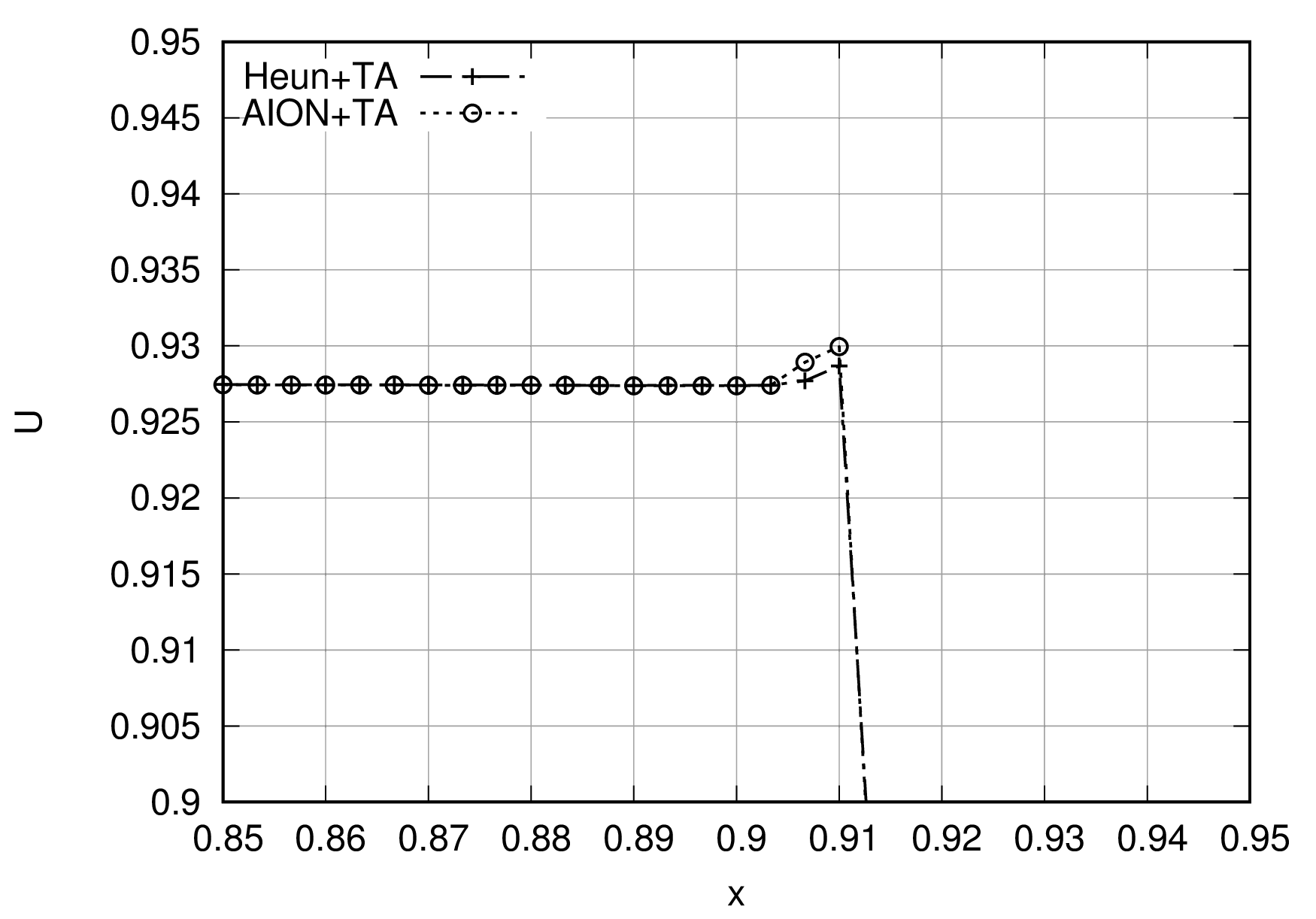}
\caption{Sod' shock tube at CFL$=0.1$. Global view of the density profile at $t=0.2s$ and close-up views
near the rarefaction wave and near the shock. \label{fig:U_Sod_CFL0.1}}
\end{center}
\end{figure}

For CFL$=0.1$, the Heun+TA scheme is compared to the AION+TA scheme.
Figs.~\ref{fig:rho_Sod_CFL0.1} and~\ref{fig:U_Sod_CFL0.1} show the density and velocity profiles,
with a focus in the region of the rarefaction wave and near the shock. The temporal adaptive
approaches (Heun+TA and AION+TA) lead to results in agreement with the theoretical behaviour.

\begin{figure}[!htbp]\begin{center}
\begin{tabular}{cc}
\includegraphics[width=8cm]{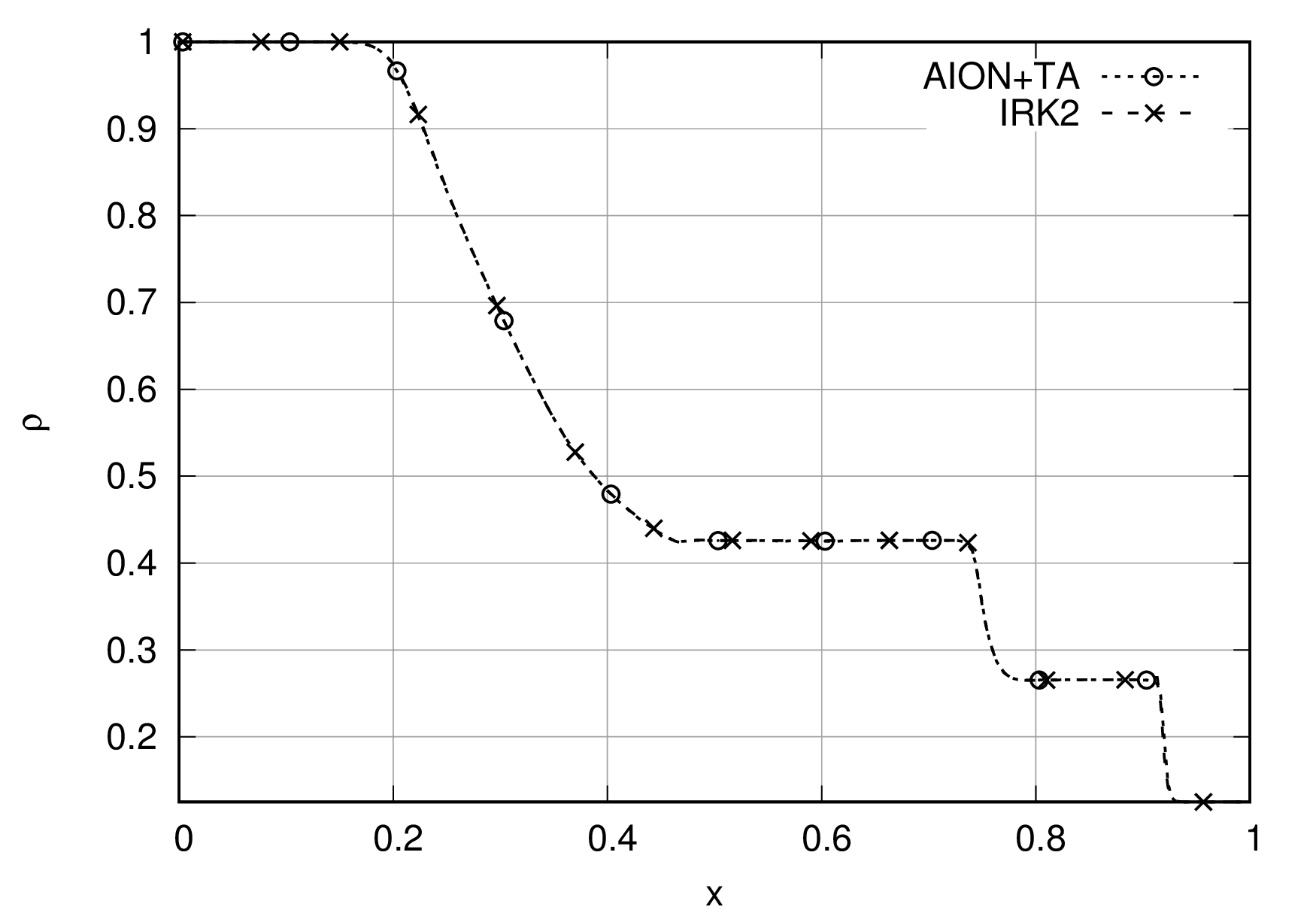} & \includegraphics [width=8cm]{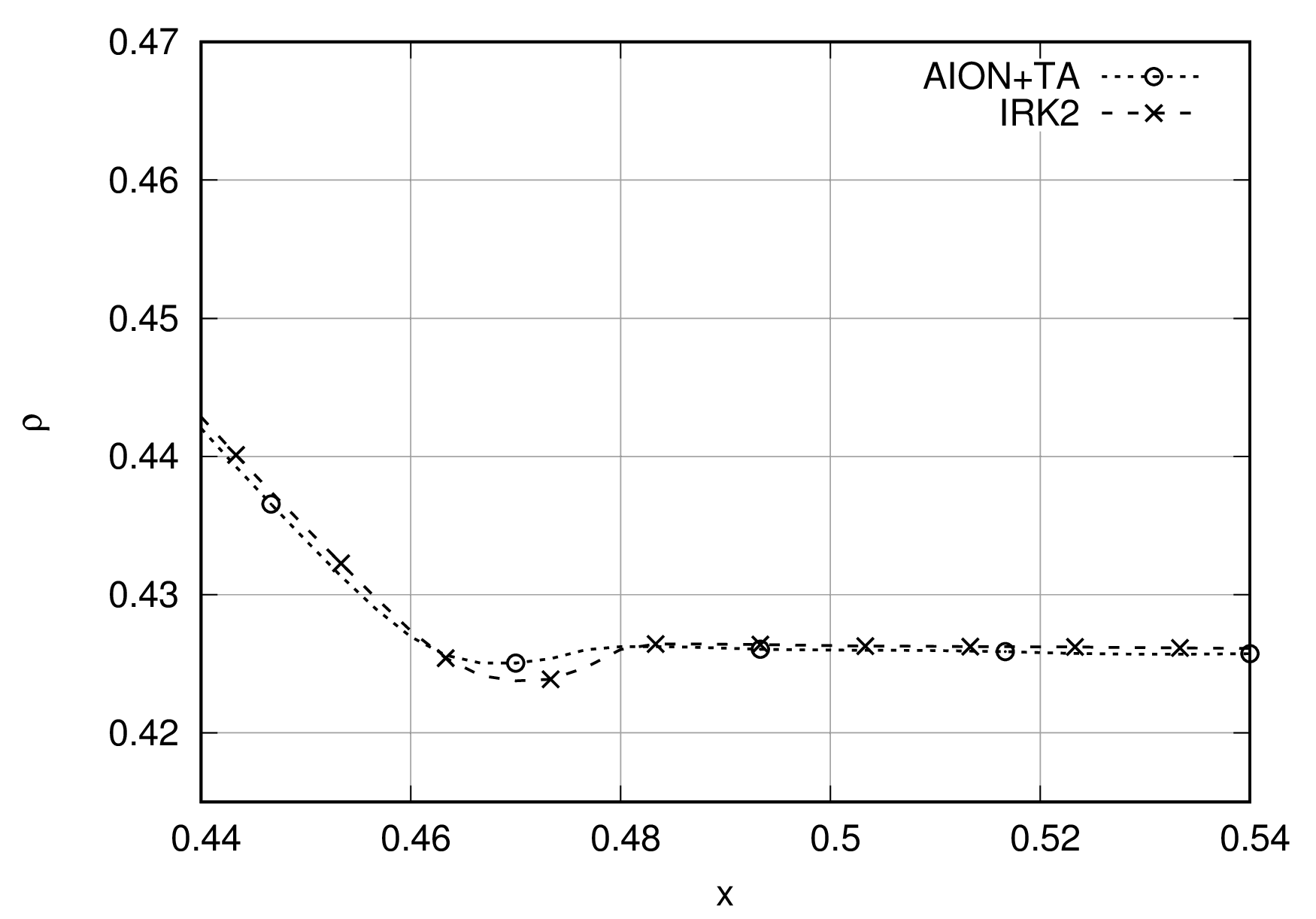}\\
\end{tabular}
\includegraphics [width=8cm]{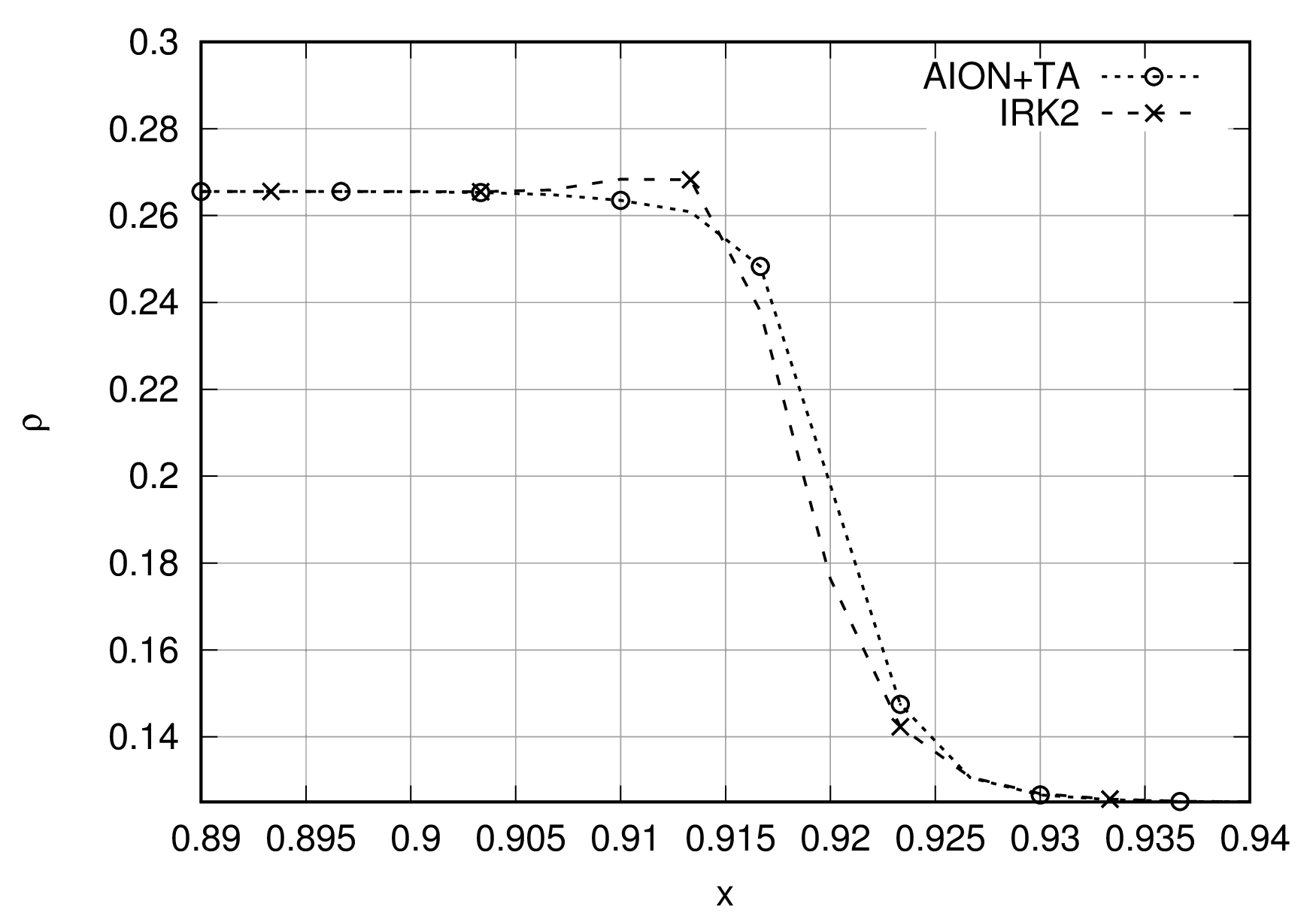}
\caption{Sod' shock tube at CFL$=0.45$. Global view of the density profile at $t=0.2s$ and close-up views
near the rarefaction wave and near the shock. \label{fig:rho_Sod_CFL0.45}}
\end{center}
\end{figure}

\begin{figure}[!htbp]\begin{center}
\begin{tabular}{cc}
\includegraphics[width=8cm]{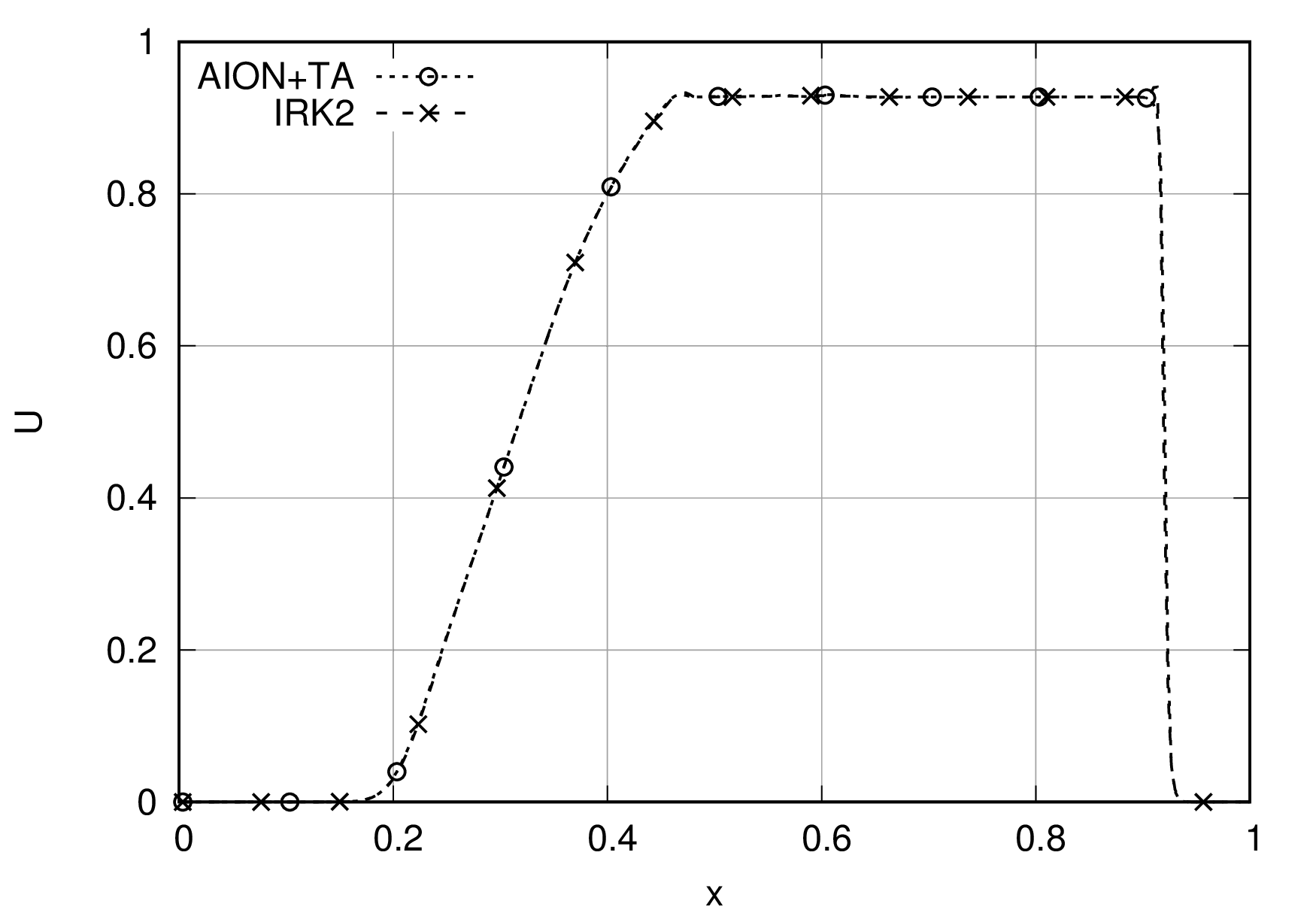} & \includegraphics [width=8cm]{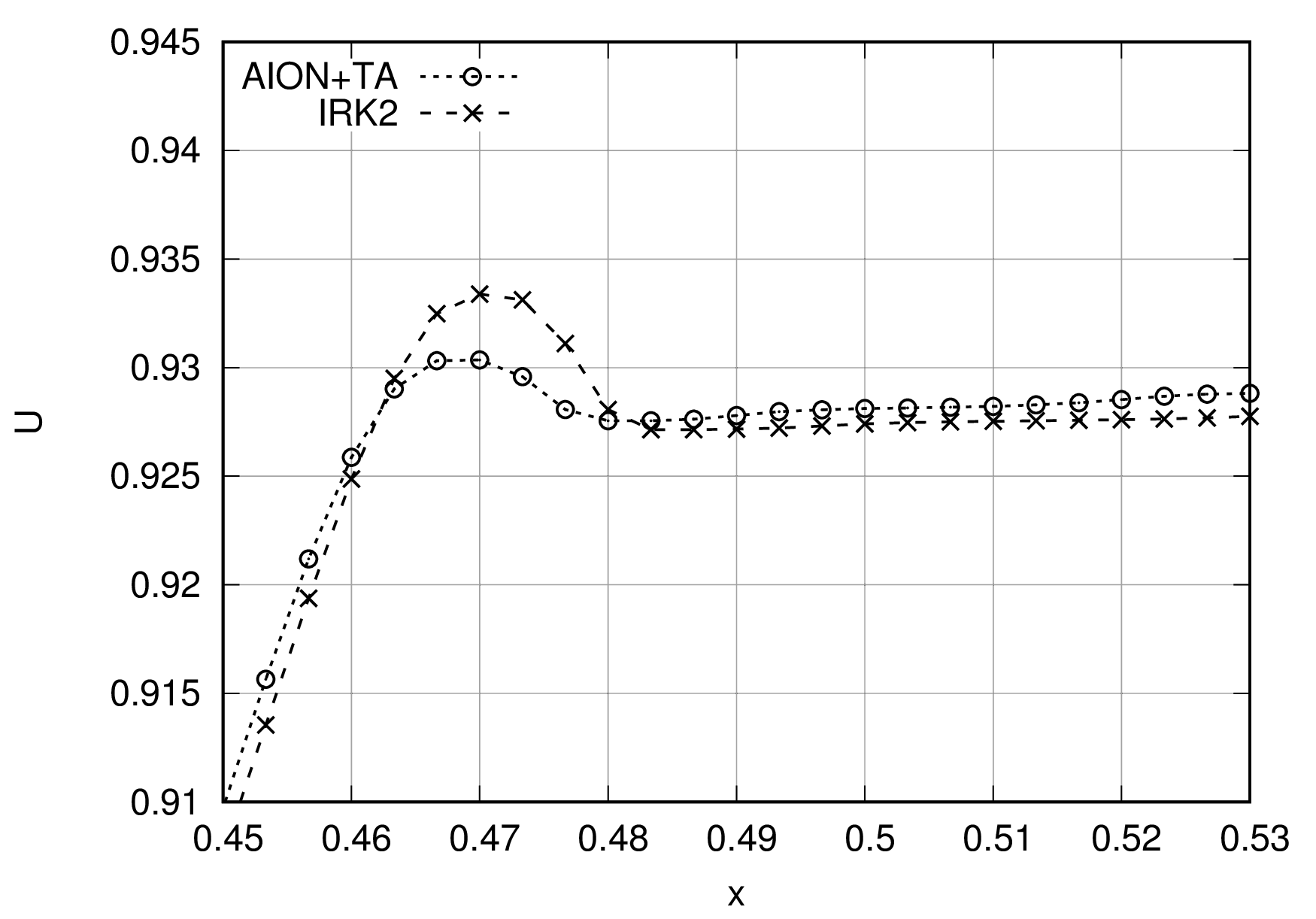}\\
\end{tabular}
\includegraphics [width=8cm]{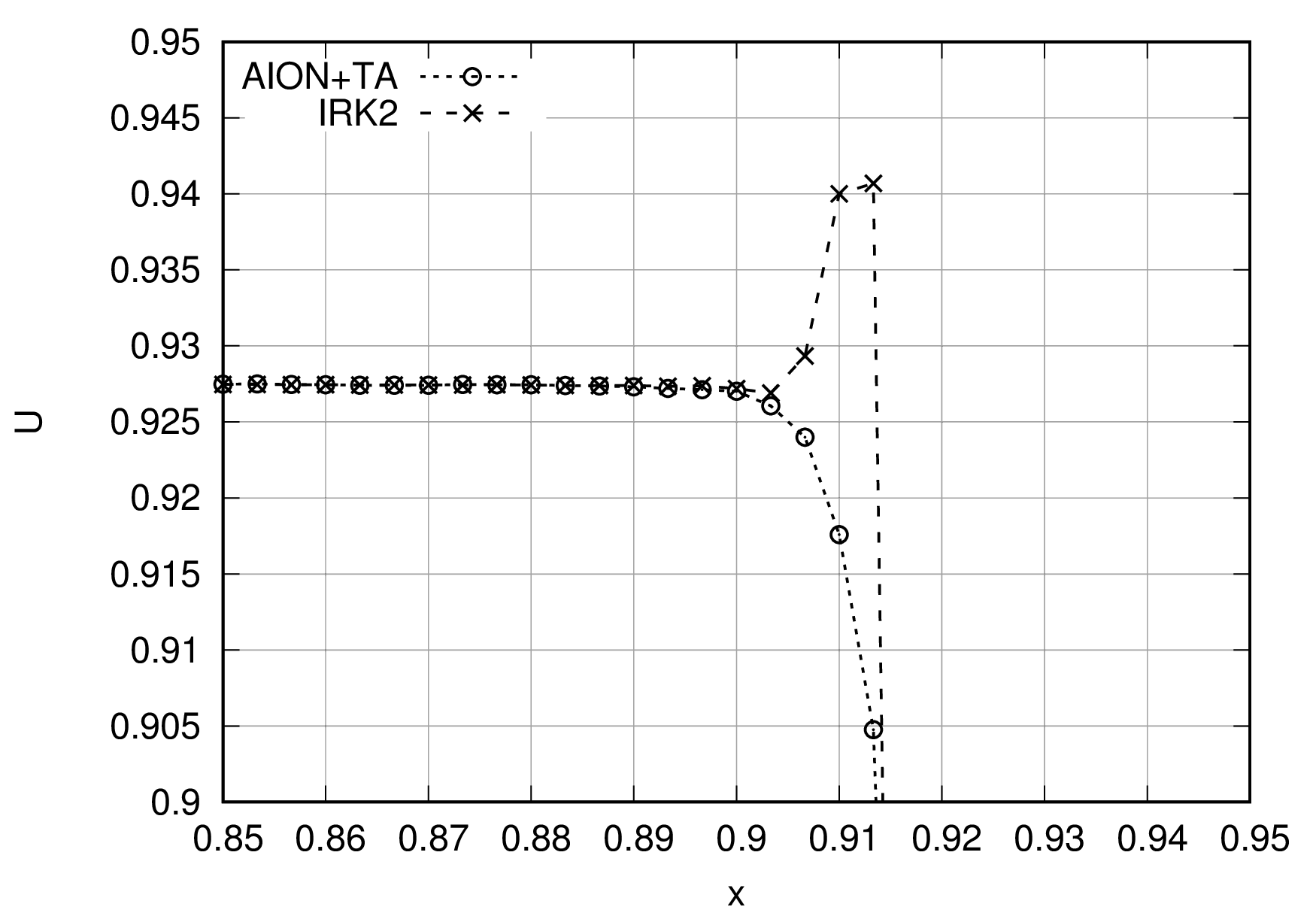}
\caption{Sod' shock tube at CFL$=0.45$. Global view of the density profile at $t=0.2s$ and close-up views
near the rarefaction wave and near the shock. \label{fig:U_Sod_CFL0.45}}
\end{center}
\end{figure}

A second set of computations is performed for CFL$=0.45$. Here, the Heun+TA scheme
is unstable, and AION+TA results are compared to those of the standard implicit IRK2 scheme
(Figs~\ref{fig:rho_Sod_CFL0.45} and~\ref{fig:U_Sod_CFL0.45}).
Both numerical solutions are very close.
Paying attention to the velocity and density near the shock, it seems that the AION+TA scheme leads
to smoother results than the IRK2 scheme, and overshoots are dissipated.

From these results, it can be concluded that the temporal adaptive approach AION+TA is in agreement with the requirements to handle shock, contact discontinuity and rarefaction waves. Moreover, the AION scheme improves the explicit temporal adaptive method with an enhanced stability.

The next two-dimensional test case is dedicated to the analysis of the AION+TA scheme accuracy.

\subsection{Two-dimensional linear advection of an isentropic vortex}

The transport of an isentropic vortex, solution of Euler's equations, is one of the standard
test cases of the High Order Workshop since quality of the results is directly linked with scheme accuracy.
Indeed, this problem allows to control the capability of the numerical scheme to preserve vorticity in an unsteady simulation and also to estimate the total order of accuracy.
The computational domain is composed of a square domain $[-\frac{L}{2}, \frac{L}{2}]^2$ ($L=0.1$)
with periodic boundary conditions.
An isentropic vortex defined by its characteristic radius $R$ and strength $\beta$ is imposed on a uniform flow of pressure $P_{\infty}$, temperature $T_{\infty}$ and
Mach number $M_{\infty}$.
The vortex is initialized in the center of the computational domain $(x_c,y_c)=(0,0)$. The initial state is defined by:
\begin{equation}
\begin{aligned}
\delta u&=- U_{\infty} \, \beta \, \frac{(y-y_c)}{R}e^{-\frac{r^2}{2}}, \\
\delta v&= U_{\infty} \, \beta \, \frac{(x-x_c)}{R}e^{-\frac{r^2}{2}},\\
\delta T&=\frac{(U_{\infty} \, \beta)^2}{2}.e^{-\frac{r}{2}}, \\
u_0&=U_{\infty}+\delta u, \\
v_0&=\delta v,
\end{aligned}
\end{equation}
with:
\begin{equation}
\begin{aligned}
r&=\frac{\sqrt{(x-x_c)^2+(y-y_c)^2}}{R}, \\
U_{\infty}&=M_{\infty} \, \sqrt{\gamma \, R_{gas} \, T_{\infty}}.
\end{aligned}
\end{equation}
$R_{gas}=287.15 \text{  }J/kg/K$ is the gas constant and a constant ratio of specific heats $\gamma =1.4$
is considered. The isentropic relation leads to the complete set of unknowns:
 \begin{equation}
\begin{aligned}
T_0&=T_{\infty}-\delta T, \\
\rho_0&=\frac{P_{\infty}}{R_{gas} \, T_{\infty}} \, \big(\frac{T_0}{T_{\infty}}\big)^{\frac{1}{\gamma-1}}, \\
P_0&=\rho_0 \, R_{gas} \, T_0
\end{aligned}
\end{equation}
The "fast vortex" test case is considered here and defined by
 \begin{equation}
P_{\infty} = 10^5 \text{  } N/m^2,
T_{\infty} = 300\text{  }K,
M_{\infty} = 0.5,
\beta =\frac{1}{5},
R = 0.005 .
\end{equation}

The solution is time-marched during three periods inside the periodic box.
The computation is performed with a Successive-Correction 2-exact formulation for the spatial scheme (order three) and time integrated by Heun+TA, standard IRK2 and AION+TA schemes.
The scheme is not designed
to be TVD, but many computations are performed for this case without the need for slope limiters at
any order of accuracy, as shown in results of the High Order Workshop.
An irregular domain of $260^2$ degrees of freedom (DOF) is considered, and the ratio between the largest and the smallest cell sizes is equal to $11$.
Reference~\cite{Muscat_JCP_XX_2018} highlights that the standard Heun's scheme is not able to perform such test case configuration
for a time step $\Delta t$ imposed according to the CFL of the biggest cells contrary to the AION scheme.
Here, the computation is designed to obtain four temporal classes of cells, with a $\Delta t_{min}$ of the temporal classes $0$
corresponding to CFL$=\frac{U_{\infty}\Delta t_{min}}{\underset{j}{\text{min}}(h_j)}=0.9$.
The parameter $\omega_j$ is defined according to cell size as
\begin{equation}
\begin{aligned}
\omega_j =\frac{|\Omega_j|}{\underset{j}{\text{max}}(|\Omega_j|)},
\end{aligned}
\end{equation}
with $|\Omega_j|$ the volume of cell $j$.
\begin{figure}[!ht]
\begin{center}
\includegraphics [width=10cm]{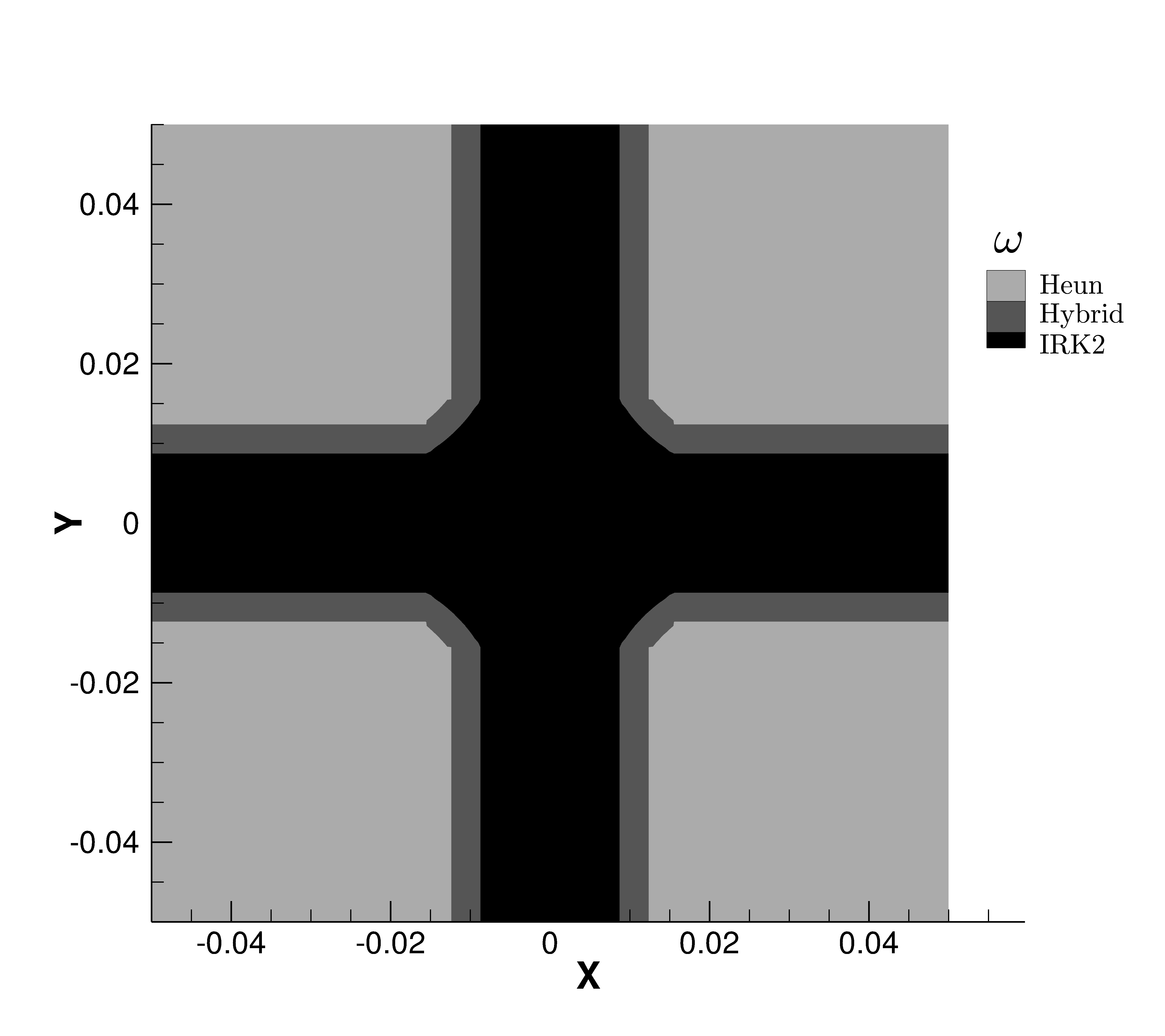}
\caption{Value of $\omega_j$
\label{fig:omega_j_260}}
\end{center}
\end{figure}
In this case, the proportion of cells with $\omega_j<1$ corresponding to hybrid and implicit cells is equal to $73\%$ (see Fig.~\ref{fig:omega_j_260}).

\begin{figure}[!htbp]\begin{center}
\begin{tabular}{cc}
\includegraphics[width=8cm]{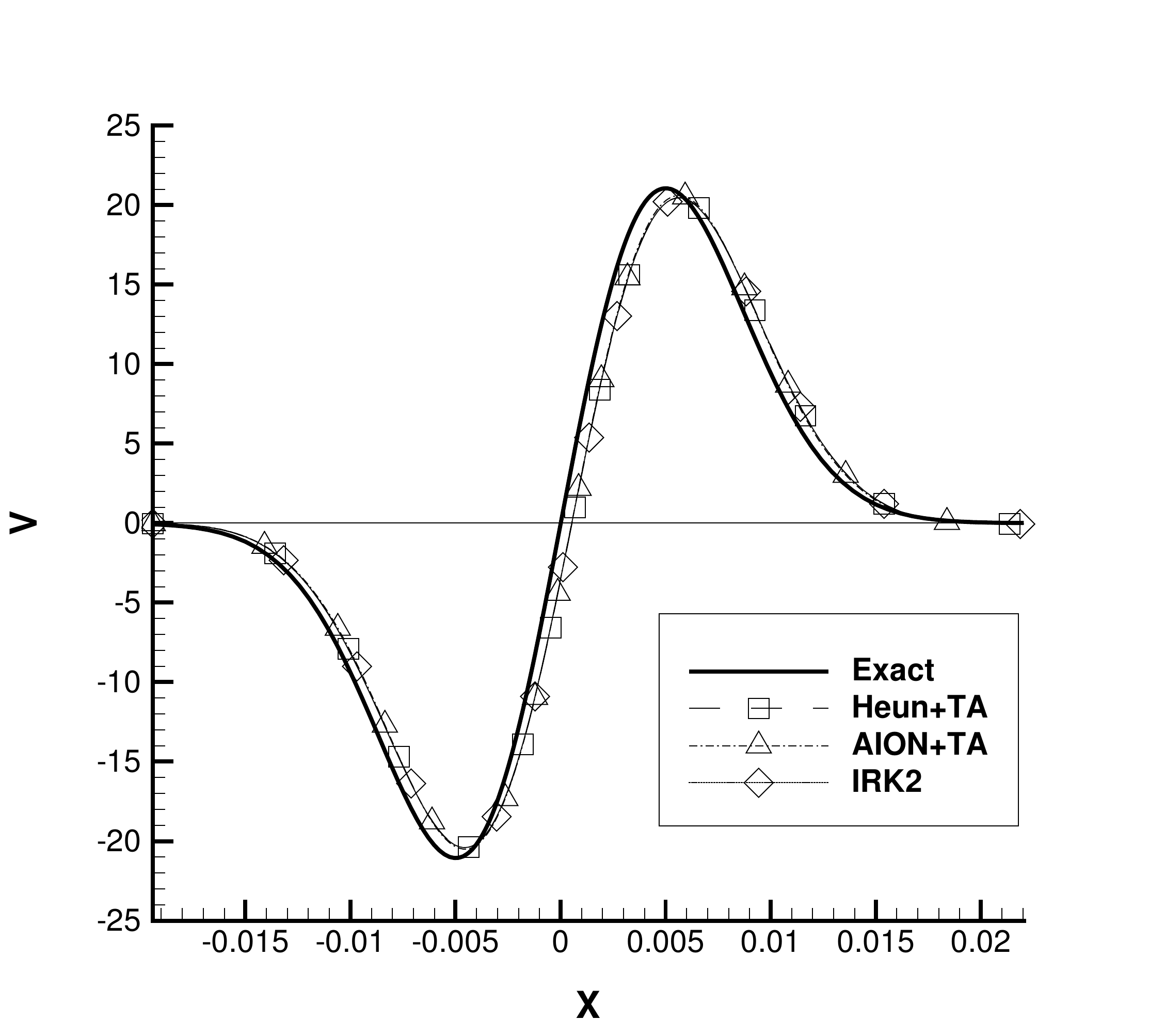} & \includegraphics [width=8cm]{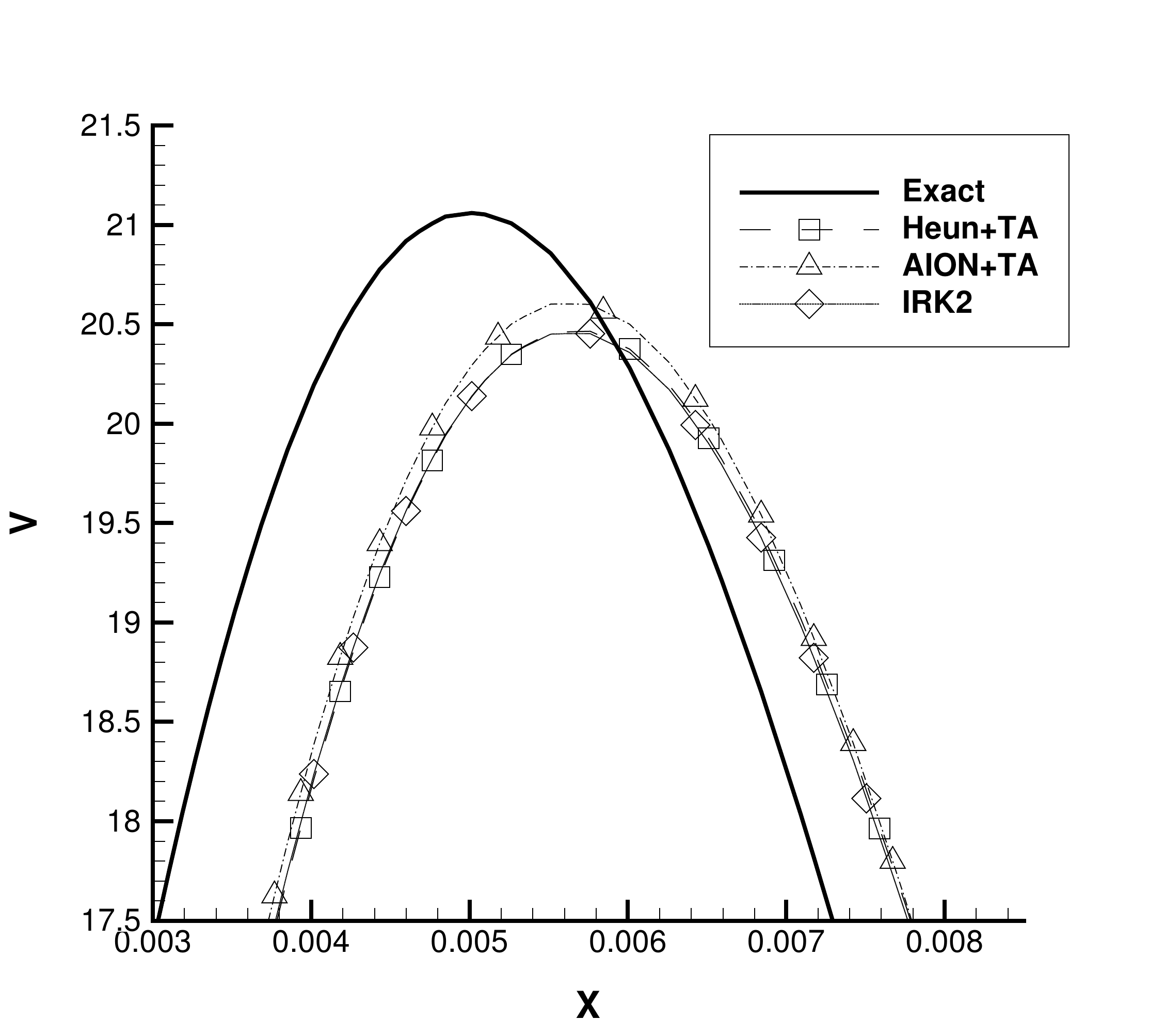}\\
\end{tabular}
\caption{ Velocity field (CFL=0.1) \label{fig:compar_260_temp_V_TA}}
\end{center}
\end{figure}

\newpage
\begin{figure}[!htbp]\begin{center}
\begin{tabular}{cc}
\includegraphics[width=8cm]{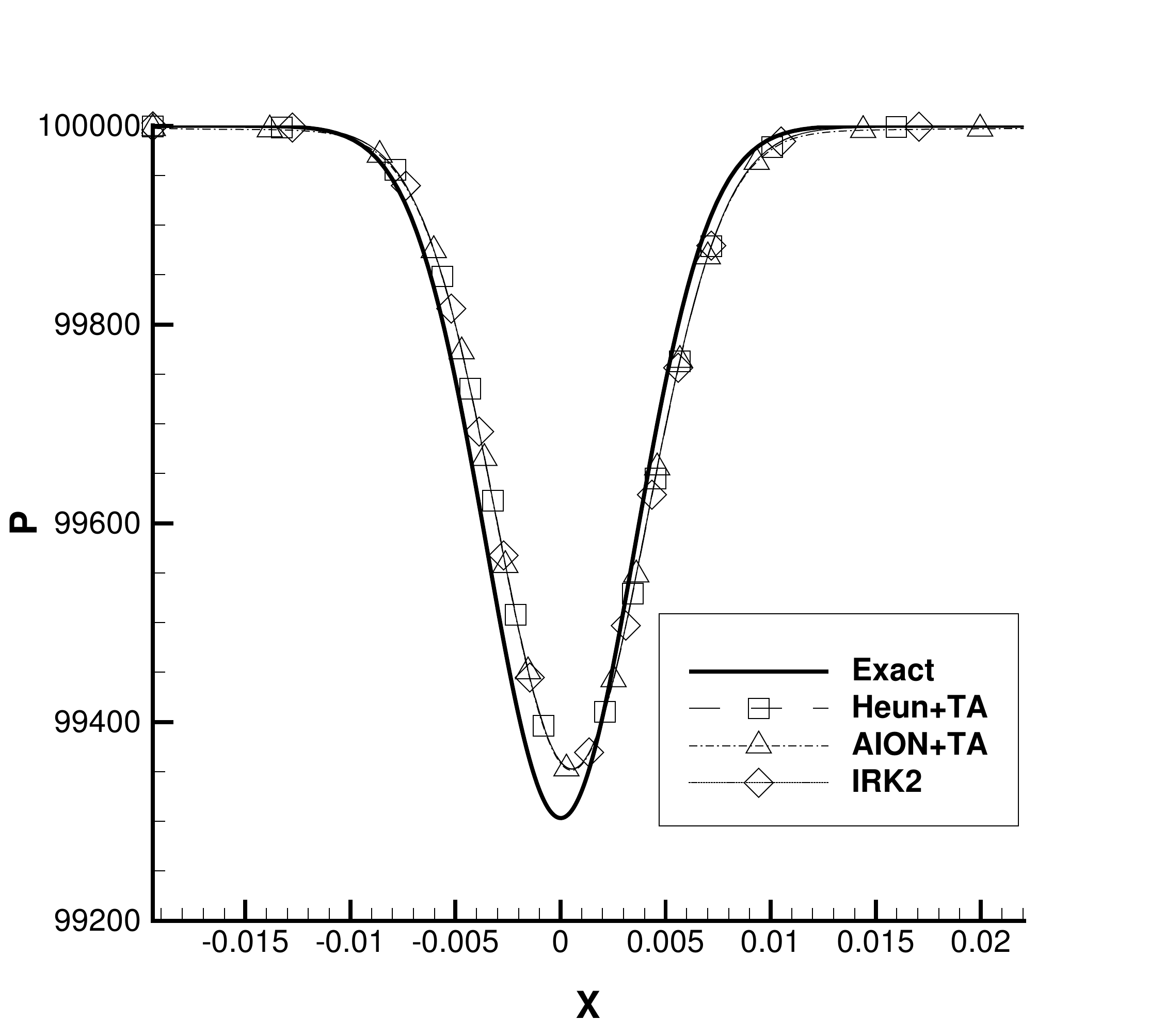} & \includegraphics [width=8cm]{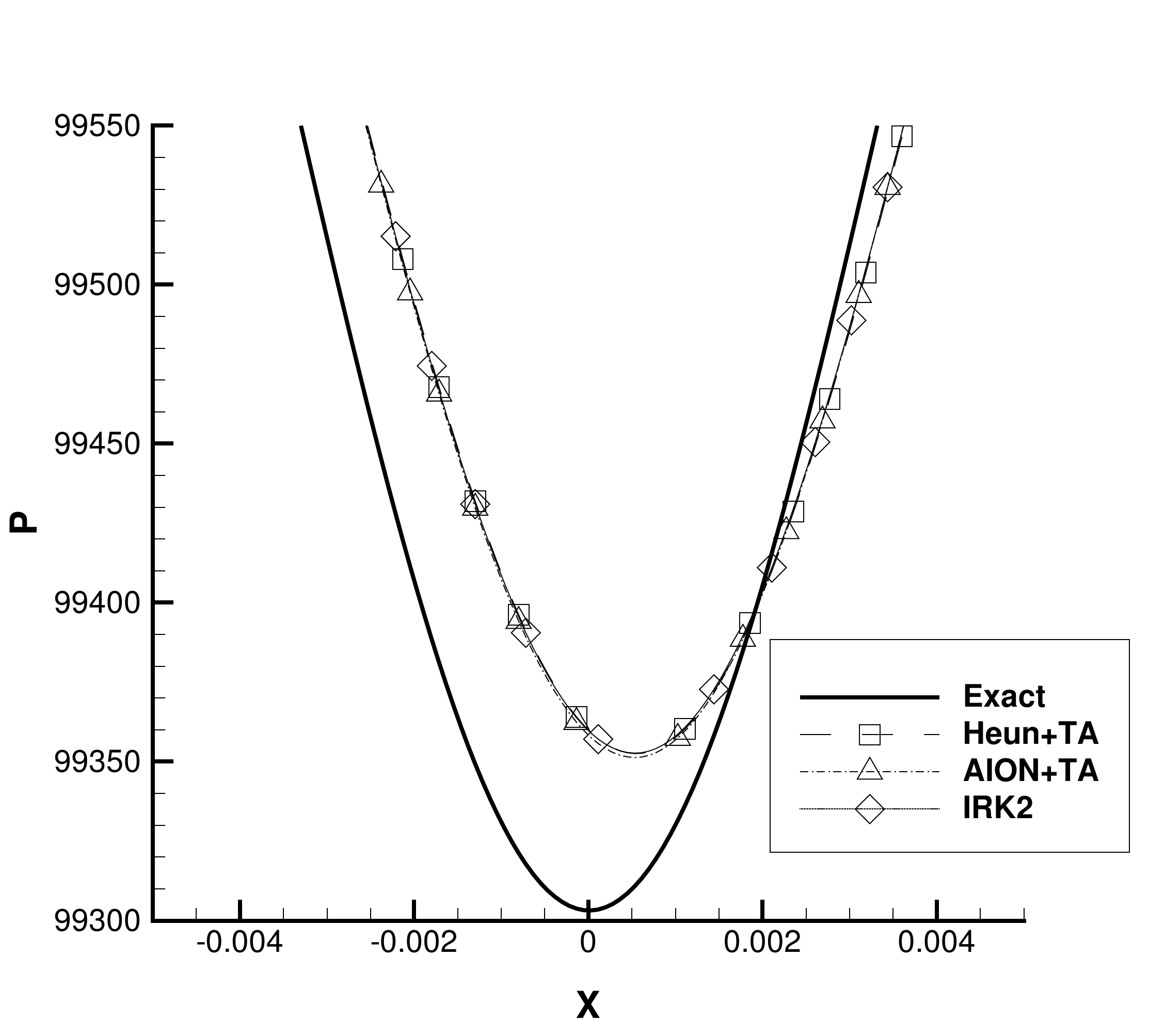}\\
\end{tabular}
\caption{Pressure field (CFL=0.1) \label{fig:compar_260_temp_P_TA}}
\end{center}
\end{figure}

According to the pressure and velocity fields in Figs.~\ref{fig:compar_260_temp_V_TA} and~\ref{fig:compar_260_temp_P_TA},
it appears that all time integrators have slightly the same properties of dissipation and dispersion.
The AION+TA scheme tends to less dissipate than the other second-order time integrators.
The temporal adaptive approach keep dispersive and dissipative properties of standard time integrators (Heun and AION) in multi-dimension.

\MM{
Hence the AION+TA scheme maintain the accuracy of the standard AION scheme.}
\newpage

\section{Conclusion}

Today industrial complex simulations require at the same time the computation of unsteady effects of
turbulence, while maintaining a low CPU cost by optimizing the discretization in the boundary layer
and by switching to the RANS model. The work presented in this paper is a first step towards
an efficient coupling of RANS and LES simulations in an industrial context.
LES and RANS model do not require the same numerical ingredients: LES is generally performed using
an explicit time integration in order to easily control the spectral properties of the scheme. In
contrary, RANS simulations require to converge fast to the steady solution by means of
an unconditionally stable implicit formulation.

In a previous paper, the coupling procedure to handle an explicit time integrator
and an implicit formulation in a cell-centered finite volume formulation was presented.
The new scheme, called AION, couples the Heun's scheme and the
second-order implicit Runge-Kutta scheme using a transition scheme called hybrid for which
a specific definition of the interface flux is required.

Today, standard unsteady industrial simulations with FLUSEPA use a specific time adaptive formulation
based on Heun's scheme. This procedure was not described in details and the current paper focuses first
on the description of this scheme currently used in the industrial solver FLUSEPA. This is motivated by
the need for a clear description of the scheme and by the need to demonstrate basic mathematical properties.
Actually, it is demonstrated that the extended scheme remains second-order accurate, is conservative,
but can produce amplification of some waves at the interface between cell classes.

In a second step, the same analysis is extended to the AION scheme coupled with time adaptation.
As the time adaptive version of Heun's scheme, the AION scheme with time adaptation needs to gather
cells in temporal classes according to their own maximal (stable) time step and uses sub-cycling
for time integration of temporal classes until the biggest time step. Several key ingredients were
analysed. First, it was demonstrated that the scheme remains conservative using a space/time analysis.
Then, the local time analysis revealed that the AION scheme coupled with time adaptation was second-order
accurate on regular grids. Finally, the Fourier analysis of the time adaptive version of AION scheme
revealed that dissipation and dispersion behaviours are not really influenced by the time adaptive
procedure.

The last part of the paper is devoted to the solutions for a set of canonical applications. One- and
two-dimensional simulations with exact solutions confirm numerically {\it a posteriori} the theoretical
properties demonstrated previously. The procedure is shown capable to handle shock, rarefaction wave and
a contact discontinuity.


The next improvement concerns the application of the proposed technique to RANS/LES industrial computations.
The smallest cells are time integrated implicitly at high CFL number while the extended temporal adaptive
time integration is performed elsewhere. The main corner stone will be to match the hybrid parameter
$\omega$ with the shielding function that switches between RANS and LES models.


\bibliographystyle{abbrv}
\bibliography{biblio}

\end{document}